\documentclass[twocolumn]{aastex631}

\usepackage{ gensymb }
\usepackage{textcomp}
\usepackage{mathtools}
\usepackage{booktabs}
\usepackage{amsmath}	
\usepackage{comment}

\accepted{}
\submitjournal{ApJ}

\shorttitle{Inside out Quenching around Spider Web Protocluster at $z \sim 2$}
\shortauthors{Laishram et al.}

\begin{document}

\title{Spider-Webb: Spatially-Resolved Evidence of Inside-Out Quenching in the Spiderweb Protocluster at $z \sim 2$}

\correspondingauthor{Ronaldo Laishram}
\email{ronaldo.laishram@nao.ac.jp}

\author [0000-0002-0322-6131]{Ronaldo Laishram}
\affiliation{National Astronomical Observatory of Japan,
2-21-1 Osawa, Mitaka, Tokyo 181-8588, Japan}

\author[0000-0002-0479-3699]{Yusei Koyama}
\affiliation{National Astronomical Observatory of Japan,
2-21-1 Osawa, Mitaka, Tokyo 181-8588, Japan}

\author[0000-0001-7713-0434]{Abdurrahman Naufal}
\affiliation{Department of Astronomical Science,
The Graduate University for Advanced Studies,
2-21-1 Osawa, Mitaka,
Tokyo 181-8588, Japan}
\affiliation{National Astronomical Observatory of Japan,
2-21-1 Osawa, Mitaka,
Tokyo 181-8588, Japan}

\author[0000-0002-2993-1576]{Tadayuki Kodama}
\affiliation{Astronomical Institute, Tohoku University, 6-3, Aramaki, Aoba, Sendai, Miyagi 980-8578, Japan}

\author[0000-0003-4442-2750]{Rhythm Shimakawa}
\affiliation{Waseda Institute for Advanced Study (WIAS), Waseda University, 1-21-1, Nishi-Waseda, Shinjuku, Tokyo 169-0051, Japan}

\author[0000-0002-9509-2774]{Kazuki Daikuhara}
\affiliation{Astronomical Institute, Tohoku University, 6-3, Aramaki, Aoba, Sendai, Miyagi 980-8578, Japan}

\author[0000-0001-7147-3575]{Helmut Dannerbauer}
\affiliation{Instituto de Astrofísica de Canarias (IAC), E-38205 La Laguna, Tenerife, Spain}
\affiliation{Universidad de La Laguna, Dpto. Astrofísica, E-38206 La Laguna, Tenerife, Spain}

\author[0000-0002-5963-6850]{Jose Manuel Pérez-Martínez}
\affiliation{Instituto de Astrofísica de Canarias (IAC), E-38205 La Laguna, Tenerife, Spain}
\affiliation{Universidad de La Laguna, Dpto. Astrofísica, E-38206 La Laguna, Tenerife, Spain}

\author[0000-0003-4528-5639]{Pablo G. P\'erez-Gonz\'alez}
%\email{pgperez@cab.inta-csic.es}
\affiliation{Centro de Astrobiolog\'{\i}a (CAB), CSIC-INTA, Ctra. de Ajalvir km 4, Torrej\'on de Ardoz, E-28850, Madrid, Spain}

\begin{abstract} 
We present a spatially-resolved analysis of galaxy quenching within the Spiderweb Protocluster at $z \sim 2.16$, combining deep imaging from the James Webb Space Telescope (JWST) and the Hubble Space Telescope (HST). Utilizing pixel-by-pixel spectral energy distribution fitting, we derive maps of stellar mass, star formation rate (SFR), specific SFR (sSFR), and rest-frame UVJ colors. Quiescent galaxies, predominantly found at $\log(M_*/M_{\odot}) \geq 10.5$, exhibit clear mass-dependent inside-out quenching, with central sSFR approximately an order of magnitude lower than outer regions, while lower-mass star-forming galaxies show flat sSFR profiles. Central star formation activity fundamentally anti-correlates with Sérsic index, indicating reduced activity in bulge-dominated systems. Spatially resolved UVJ colors reveal heterogeneous internal star formation, distinguishing star-forming regions in quiescent hosts from those in globally star-forming systems. These findings demonstrate that quenching mechanisms were effectively operating by $z \sim 2$, with the observed inside-out patterns and morphological correlations consistent with AGN-driven feedback processes. Our study provides key observational constraints on galaxy evolution during this critical epoch.
\end{abstract}

\keywords{Protoclusters (1297); Galaxy evolution (594); Quenched galaxies (2016); Galaxy structure (622); Galaxy properties (615), High-redshift galaxy clusters (2007)}

\section{Introduction} \label{sec:intro}

Understanding the formation and evolution of galaxies across cosmic time is a central goal of modern astrophysics. Galaxies exhibit a diverse range of properties, shaped by a complex interplay of processes including gas accretion, star formation, mergers, and feedback mechanisms \citep{Somerville_et_al_1999, Cole_et_al_2000}. A key aspect of this quest involves studying galaxy evolution in the early universe, particularly at high redshifts (z $>$ 2) when galaxies were undergoing their most active phases of star formation and assembly \citep{Steidel_et_al_2004,Madau_et_al_2014}. These early epochs are crucial for understanding the origin of the diverse galaxy populations observed today.

Protoclusters, the progenitors of present-day galaxy clusters, represent unique environments for studying galaxy evolution in the early universe \citep{Chiang_et_al_2013, Overzier_2016}. These overdense regions are believed to play a critical role in the formation and quenching of massive galaxies at early epochs \citep{Steidel_et_al_2005, Muldrew_et_al_2018}. Observations of protoclusters at cosmic noon (z $\sim$ 1.5–3) have revealed a wide range of evolutionary stages, from those actively forming stars to those beginning to exhibit a red sequence of quiescent galaxies \citep{Cucciati_et_al_2018, Shi_et_al_2021, Ito_et_al_2023, Perez_et_al_2024,Daikuhara_et_al_2024, Laishram_et_al_2024}. The high density of galaxies within protoclusters can enhance interactions and mergers, potentially fueling starbursts and active galactic nuclei (AGN) \citep{Steidel_et_al_2005}. Furthermore, the environmental conditions within protoclusters can influence the efficiency of star formation and the quenching of star formation, leading to accelerated galaxy evolution \citep{Steidel_et_al_2005, Shimakawa_et_al_2018}. Theoretical models indicate that protocluster environments are responsible for a substantial fraction (20\%--50\%) of the cosmic star formation rate density during the epoch $z = 2$--10 \citep{Chiang_et_al_2017}. This enhanced star formation activity requires continuous gas replenishment through both cold gas accretion via cosmic web filaments and episodic merger events \citep{Dekel_et_al_2006, Genel_et_al_2008}, providing the necessary fuel for the rapid assembly of the most massive galactic systems during this critical period of cosmic evolution \citep{Bethermin_et_al_2014}. Although some of this star formation can be traced through rest-frame ultraviolet and optical observations \citep{Toshikawa_et_al_2016, Ito_et_al_2020, Huang_et_al_2022,Laishram_et_al_2024}, a considerable portion remains hidden behind dense dust clouds, with the bulk of the emission emerging in the far-infrared and submillimeter wavelengths \citep{Smail_et_al_1997, Greve_et_al_2005}.

A crucial aspect of studying galaxy evolution is understanding the processes that regulate star formation within galaxies and ultimately lead to their quenching. Observations have revealed a tight correlation between the integrated star formation rate (SFR) and stellar mass, known as the star-forming main sequence (SFMS) \citep{2007ApJ...660L..43N,2007ApJ...670..156D}. This relation indicates that galaxies grow in mass over cosmic time in a state of self-regulated semi-equilibrium. However, the mechanisms that cause galaxies to deviate from the SFMS and transition to a quiescent state remain a key question in galaxy evolution \citep{Whitaker_et_al_2012, 2014MNRAS.440..889S,2016ApJ...832...79P}. Spatially resolved studies of galaxy structure and star formation can provide valuable insights into the processes that drive the quenching of star formation in galaxies.
While our understanding of star formation cessation remains incomplete, observational evidence demonstrates that the transition from active to quiescent states depends critically on both intrinsic galaxy properties and external environmental factors. The probability of quenching correlates strongly with internal structural characteristics including the prominence of bulge components, the degree of central light concentration as quantified by sérsic indices, the surface density of stellar mass in galaxy cores, and central velocity dispersion measurements \citep{Cameron_et_al_2009, Mendel_et_al_2013, Wuyts_et_al_2011, Fang_et_al_2013, Barro_et_al_2017}. Simultaneously, the surrounding cosmic environment exerts significant influence through the local galaxy number density and the morphological characteristics of large-scale structure \citep{Balogh_et_al_2004, Weinmann_et_al_2006, Peng_et_al_2010}. These observational correlations suggest that star formation quenching operates through multiple channels, with both galaxy-scale physical processes and environmental mechanisms playing complementary roles in driving the transformation of actively star-forming systems into passive galaxies.

Distinguishing between these competing quenching mechanisms requires detailed analysis of the spatial distribution of stellar populations within individual galaxies, as different physical processes predict distinct radial patterns of quenching \citep{Abdurro_Akiyama_2018,Nelson_et_al_2021}. Internal mechanisms such as AGN feedback and merger-induced gas inflows typically produce ``inside-out'' quenching where star formation ceases first in galaxy centers due to central gas consumption and powerful outflows \citep{Zolotov_et_al_2015,Tacchella_et_al_2016}. In contrast, environmental processes such as ram pressure stripping and tidal interactions preferentially remove gas from galaxy outskirts, potentially producing ``outside-in'' quenching patterns \citep{Masters_et_al_2010,Chen_et_al_2019, Bluck_et_al_2020, Shimakawa_et_al_2022}.

These spatial signatures can only be reliably detected through spatially resolved analysis techniques that probe galaxy properties on sub-kiloparsec scales. Spatially resolved studies of galaxy structure can provide valuable insights into the processes that drive the quenching of star formation in galaxies.
Spatially resolved spectral energy distribution (SED) fitting has emerged as a powerful technique for investigating galaxy evolution at sub-galactic scales \citep{Wuyts_et_al_2011,Cibinel_et_al_2015,2021ApJS..254...15A}.  By applying stellar population synthesis models to individual pixels, this technique provides a detailed view of galaxy structure and evolution compared to traditional integrated SED fitting \citep{Cibinel_et_al_2015,Tacchella_et_al_2015,Nelson_et_al_2016}. The advent of JWST has revolutionized spatially resolved stellar population studies, enabling unprecedented depth and spatial resolution for analyzing galaxy structure and evolution at both low and high redshifts \citep{Gimenez_Arteaga_et_al_2023,Abdurrof_et_al_2023,Perez_Gonzalez_et_al_2023,Liu_et_al_2024,Iglesias_Navarro_et_al_2025,Haryana_et_al_2025}. This approach enables us to probe the radial distribution of star formation and stellar mass, understand the inside-out nature of galaxy growth, and investigate the environmental effects on galaxy evolution \citep{Whitaker_et_al_2012,Davies_et_al_2021}. Furthermore, studies have shown a strong connection between central stellar mass surface density and the quenching of star formation in galaxies \citep{Barro_et_al_2017,Fang_et_al_2013b,Cheung_et_al_2012,Whitaker_et_al_2017}.

This paper presents a comprehensive pixel-by-pixel SED fitting analysis of the Spiderweb protocluster at z $\sim$ 2.16 \citep{Miley_et_al_2006,kurk_search_2000}. 

The Spiderweb protocluster PKS 1138-262, centered on the radio galaxy \citep{Bolton_et_al_1979, Roettgering_et_al_1994, Roettgering_et_al_1997, Carilli_et_al_1997, Pentericci_et_al_1997}, represents one of the most extensively studied protoclusters in the early universe. The protocluster was discovered through the identification of a significant overdensity of Ly$\alpha$ emitters around the central radio galaxy \citep{kurk_search_2000, Pentericci_et_al_2000,Miley_et_al_2006}, with subsequent spectroscopic confirmation of member galaxies \citep{pentericci_search_2000}. Multi-wavelength observations have identified a rich variety of galactic systems throughout the protocluster, encompassing H$\alpha$ emitters \citep{kurk_search_2004,kurk_search_2004-1, Koyama_et_al_2013,Shimakawa_et_al_2014, Shimakawa_et_al_2018,Jin_et_al_2021,Perez_et_al_2023,Naufal_et_al_2023,Daikuhara_et_al_2024}, multiple active galactic nucleus (AGN) candidates detected via X-ray observations \citep{Pentericci_et_al_2002,Croft_et_al_2005,Tozzi_et_al_2022b, Lepore_et_al_2024,shimakawaJWSTNIRCamNarrowband2024}, the detection of a developing red sequence in the cluster center \citep{kurk_search_2004-1, Kodama_et_al_2007, Zirm_et_al_2008, Doherty_et_al_2010, Tanaka_et_al_2010, naufalRevealingQuiescentGalaxy2024}, an ensemble of far-infrared Herschel sources \citep{Rigby_et_al_2014}, dust-obscured submillimeter galaxies (SMGs; \citep{Dannerbauer_et_al_2014, Zeballos_et_al_2018}) and gas-rich CO emitters \citep{Dannerbauer_et_al_2017, Emonts_et_al_2018, Tadaki_et_al_2019, Chen_et_al_2024, Perez_et_al_2025}, with detailed mapping of the protocluster's extended structure through CO(1–0) observations using the Australian Telescope Compact Array \citep{Jin_et_al_2021}, and via 1.1mm dust continuum imaging with ALMA \citep{Zhang_et_al_2024}. This extensive multi-wavelength dataset establishes the Spiderweb protocluster as an exceptional laboratory for investigating galaxy evolution within high-density environments at $z >$ 2.

Recent ALMA observations have detected the thermal Sunyaev-Zeldovich effect, providing direct evidence for a developed intracluster medium \citep{Di_Mascolo_et_al_2023}, confirming the Spiderweb protocluster as a transitional structure evolving toward a massive galaxy cluster.
Specifically, we focus on the spatially resolved star formation properties of galaxies. This paper is structured as follows: Section 2 describes the observational data and sample selection. Section 3 outlines our methodology for pixel-by-pixel SED fitting. Section 4 presents the results of our analysis. Section 5 discusses these results in the context of galaxy formation and evolution. Finally, Section 6 summarizes the main findings of this work.

\section{Data} \label{sec:data}

\subsection{The Spiderweb Protocluster Dataset} \label{sec:dataset}

We utilize deep multi-wavelength imaging from the \textit{James Webb Space Telescope} (\textit{JWST}) and the \textit{Hubble Space Telescope} (\textit{HST}) to investigate galaxies in the Spiderweb protocluster at $z \approx 2.16$. The Spiderweb protocluster represents one of the most extensively studied high-redshift protoclusters and provides an exceptional laboratory for examining galaxy evolution processes during the epoch of peak cosmic star formation. Our comprehensive dataset spans rest-frame ultraviolet through near-infrared wavelengths, enabling robust spectral energy distribution (SED) fitting across the critical spectral features that trace stellar populations and star formation activity.

The photometric dataset comprises six broad-band filters spanning a wavelength range from 4750 \AA{} to 4.1 $\mu$m in the observer frame, corresponding to rest-frame wavelengths from approximately 1500 \AA{} to 1.3 $\mu$m at the protocluster redshift. Table \ref{tab:data} summarizes the instrumental configurations, depth characteristics, and literature sources for each photometric band employed in our analysis.

Deep optical imaging in the rest-frame ultraviolet regime was obtained using the \textit{HST} Advanced Camera for Surveys (ACS) in the F475W and F814W filters \citep{mileySpiderwebGalaxyForming2006}. The ACS imaging achieves 5$\sigma$ limiting magnitudes of 28.68 and 28.40 AB magnitudes in F475W and F814W, respectively, measured within 0.3-arcsecond diameter apertures.

Near-infrared imaging was obtained through the \textit{JWST} Near Infrared Camera (NIRCam) as part of Cycle 1 General Observer program 1572 (P.I. Dannerbauer and Co P.I. Koyama). The NIRCam observations encompass three filters: F115W, F182M, and F410M. Data were processed using the JWST Science Calibration pipeline version 1.10.2 under Calibration Reference Data System context 1140.pmap, with additional processing using the Rainbow JWST pipeline \citep{Perez_Gonzalez_et_al_2024}. For details about the observations and data reduction, refer to the \citet{shimakawaJWSTNIRCamNarrowband2024} and \citet{perez-martinezJWSTNIRCamPav2024}.

\begin{deluxetable}{lllc}
\label{tab:data}
    \tablecaption{Photometric bands used for SED fitting with CIGALE.}
    \scriptsize
    \tablehead{\colhead{Instrument} & \colhead{Filter} & \colhead{PSF FWHM} & \colhead{Reference} \\
               \colhead{} & \colhead{} & \colhead{(arcsec)} & \colhead{}}
    \startdata
    HST/        &           &        & \\
    ACS         & F475W     & 0.126  & \citet{mileySpiderwebGalaxyForming2006} \\
    \ldots      & F814W     & 0.125  & \ldots \\
    WFC3        & F160W     & 0.194  & \citet{naufalRevealingQuiescentGalaxy2024} \\
    JWST/       &           &        &\\
    NIRCam      & F115W     & 0.084  & \citet{shimakawaJWSTNIRCamNarrowband2024} \\
    \ldots      & F182M     & 0.090  & \ldots \\
    \ldots      & F410M     & 0.158  & \ldots \\
    \enddata
    \tablecomments{PSF FWHM values measured using photutils for original (unconvolved) images. For pixel-by-pixel SED fitting, all images are PSF-matched to F160W (FWHM = 0.194 arcsec) and resampled to a common pixel grid. Detailed photometric depths and sensitivity maps are provided in the referenced papers.}
\end{deluxetable}

Our analysis employs the spectroscopically confirmed protocluster member catalog compiled by \citet{naufalRevealingQuiescentGalaxy2024}, which provides robust redshift measurements and photometric properties for galaxies within the protocluster environment. This catalog utilizes \textit{HST} WFC3 G141 grism slitless spectroscopy to achieve spectroscopic confirmation of protocluster membership. \citet{naufalRevealingQuiescentGalaxy2024} identified 40 galaxies as robust members of the protocluster with $H_{160} < 25$ mag, of which 19 were previously classified as H$\alpha$ emitters (HAEs) through narrowband selection by \citet{Shimakawa_et_al_2018}. Additionally, \citet{naufalRevealingQuiescentGalaxy2024} spectroscopically identified new [O\,\textsc{iii}] emitters as protocluster members, along with previously unidentified quiescent galaxies selected based on the strength of the 4000\,\AA\ break in their spectra. The spectroscopic approach employed in this catalog effectively eliminates potential contamination from foreground and background sources that could otherwise affect purely photometric selection methods, thereby ensuring a clean sample of confirmed protocluster members for our spatially resolved analysis.

\section{Methods} \label{sec:methods}

\subsection{Image Processing and Pixel Binning } \label{sec:image_processing}

Accurate spectral energy distribution (SED) fitting requires careful treatment of instrumental effects and systematic differences in spatial resolution between observations obtained with different telescopes and instruments. We begin our analysis by establishing a common reference frame for all photometric observations across the six filters spanning the observed wavelength range from optical to near-infrared.

A critical step in multi-wavelength analysis involves homogenizing both the pixel sampling and spatial resolution across all photometric bands to enable meaningful pixel-by-pixel comparisons. We first resample all images to a common pixel scale of 0.06 arcsec/pixel to ensure consistent spatial sampling across all filters.  The spatial resolution varies significantly between our HST and JWST observations, with original PSF FWHM values ranging from 0.084 arcsec (JWST/NIRCam F115W) to 0.194 arcsec (HST/WFC3 F160W), as listed in Table~\ref{tab:data}. The HST/WFC3 F160W filter provides the poorest spatial resolution in our dataset. We therefore degrade all higher-resolution images to match the F160W PSF characteristics (final FWHM = 0.194 arcsec) through convolution with appropriately constructed kernels (see Appendix~\ref{app:psf} for detailed methodology).

PSF characterization is performed using unsaturated stars identified within our imaging area. We select stars that are not saturated in each filter and construct empirical PSF models by stacking these stellar profiles together. Convolution kernels are then derived to match all images to the worst resolution filter (F160W). This approach ensures that all final images possess consistent spatial resolution characteristics, enabling robust pixel-by-pixel photometric measurements.

Following pixel scale matching and PSF homogenization, we extract postage stamp images of $91 \times 91$ pixels (5.46 × 5.46 arcsec) centered on each target galaxy. This field size provides adequate coverage of the galaxy and surrounding background while maintaining computational tractability for the pixel-by-pixel fitting procedures.

Traditional pixel-by-pixel analysis of high-redshift galaxies faces inherent limitations imposed by photometric signal-to-noise constraints, particularly in the low surface brightness regions that dominate galaxy outskirts. These limitations can severely compromise the reliability of derived physical parameters, especially for faint stellar populations with complex star formation histories. To address these challenges while preserving crucial spatial information, we employ the adaptive binning algorithm implemented in the \texttt{piXedFit} package \citep{abdurraufIntroducingPixedfitSpectral2021}.

The \texttt{piXedFit} binning algorithm represents a significant advancement over conventional spatial binning schemes by incorporating spectral similarity as a fundamental constraint in the binning process. This approach optimizes the signal-to-noise ratio of the resulting SEDs while maintaining spatial resolution through careful consideration of the intrinsic SED shape variations across galaxy regions. The algorithm operates under three primary constraints that collectively ensure both photometric reliability and preservation of genuine spatial structure.

For our analysis, we adopt a maximum reduced $\chi^2$ threshold of 5.0 for determining acceptable SED shape similarity between neighboring pixels, following the methodology established for high-redshift JWST observations by \citet{Abdurrof_et_al_2023}. This threshold value has been demonstrated to work effectively for high-redshift galaxy analysis, representing a compromise between the competing requirements of achieving adequate photometric signal-to-noise ratios and preserving genuine spectral diversity across different galaxy regions. The minimum bin diameter is set to 3.5 pixels, which exceeds the PSF FWHM size of our lowest-resolution observations (3.23 pixels) and ensures that spatial structure is not artificially degraded below the intrinsic resolution limit of our dataset.

Signal-to-noise requirements are established independently for each photometric band, reflecting the varying depth and sensitivity characteristics of our multi-wavelength dataset. We require minimum detection significance of $3\sigma$ in the filters F160W, F410M, F115W, and F182M for each binned pixel group, with F160W and F410M serving as our primary detection bands due to their optimal combination of depth and spatial resolution. We do not apply signal-to-noise thresholds to the HST filters F475W and F814W, as the inherently lower signal-to-noise ratios in these bands at high redshift would impose excessive constraints on the pixel-binning process, resulting in coarser binning maps that would compromise the spatial resolution and lose critical structural information from the original images \citep{Abdurrof_et_al_2023}.

\subsection{Spectral Energy Distribution Fitting with CIGALE} \label{sec:sed_fitting}

We perform SED fitting on each bin within the galaxy using the Code Investigating GALaxy Emission (\texttt{CIGALE}) package \citep{boquien_cigale_2019}, which employs Bayesian inference techniques to derive physical parameters from observed photometric measurements. 

Our SED fitting configuration incorporates several key physical components designed to capture the stellar populations and interstellar medium properties of protocluster galaxies at $z \sim 2.1$. For stellar population synthesis, we adopt the \citet{bruzualStellarPopulationSynthesis2003} models with a Chabrier initial mass function, providing robust mass estimates consistent with modern stellar evolution theory. For stellar metallicity, we allow CIGALE to fit for three values: $Z = 0.004$, $0.008$, and $0.02$ (approximately $0.2$, $0.4$, and $1.0~Z_{\odot}$), spanning sub-solar to solar metallicities. CIGALE selects the best-fit metallicity for each spatial bin based on the photometric data, allowing metallicity to vary spatially within individual galaxies and mitigating systematic biases from imposing a single fixed metallicity value.

The star formation history is modeled using delayed exponential functions with $e$-folding times ($\tau_{\text{main}}$) spanning 10, 30, 50, 100, 300, 500, 1000, 3000, 5000, and 10000 Myr. This parameterization takes the form:
\begin{equation}
\text{SFR}(t) \propto t \exp(-t/\tau_{\text{main}})
\end{equation}
where $\tau_{\text{main}}$ represents the $e$-folding time and $t$ is the age of the stellar population. This functional form naturally incorporates an initial rise in star formation activity followed by exponential decline which is predicted by gas infall models.

The age of the main stellar population ($\text{age}_{\text{main}}$) is sampled up to 3.1 Gyr, which is the upper limit corresponding to the age of the universe at $z \sim 2.1$.

Dust attenuation is modeled using the modified starburst attenuation law with the nebular line color excess $E(B-V)_{\text{lines}}$ sampled from 0 to 1.0 magnitudes in steps of 0.1 magnitudes. We adopt a total-to-selective extinction ratio $R_V = 4.05$, appropriate for starburst environments, and apply the Milky Way extinction law for emission line attenuation.

The infrared dust emission component utilizes the \citet{dale2014ModelsStarforming} templates with a fixed $\alpha$ slope parameter of 2.0. We explicitly exclude any active galactic nucleus contribution by setting $\text{fracAGN} = 0.0$. We note that the inclusion of AGN parameters for the central bins of galaxies does not significantly alter our results. However, we caution that SED fitting results can change when including X-ray photometry for AGN identification, as demonstrated by \citet{shimakawaJWSTNIRCamNarrowband2024} who showed that approximately one-third of previously considered star-forming galaxies in the Spiderweb protocluster are more likely to be passively-evolving galaxies with low-luminosity AGNs when multiwavelength SED fitting includes AGN components. Nevertheless, our SED fitting approach, which covers 6 space-based HST and JWST filters spanning observed wavelengths from 4750 \AA{} to 4.1 $\mu$m and lacks X-ray imaging, is less susceptible to these biases as AGN contamination becomes more dominant at longer infrared wavelengths where AGN-heated dust emission contributes significantly to the observed flux.

A fundamental challenge in photometric SED fitting is the well-established degeneracy between stellar population age, metallicity, and dust extinction, where older, less dusty stellar populations can produce spectral energy distributions nearly indistinguishable from younger, more heavily attenuated systems \citep{Worthey_et_al_1994, Conroy_et_al_2013}. This degeneracy becomes even more pronounced when relying on photometric redshift estimates, which introduce additional parameter space uncertainty \citep{Salim_et_al_2007}.
Our analysis incorporates several methodological advantages that help mitigate, though not completely eliminate, these degeneracies. First, we employ spectroscopically confirmed redshifts from \textit{HST} WFC3 G141 grism observations, removing the redshift uncertainty that would otherwise expand the allowed parameter space for stellar population fits \citep{kriek2013dust, boquien2019cigale}. Second, our comprehensive wavelength coverage from rest-frame far-ultraviolet ($\sim$1500~\AA, observed F475W) through rest-frame near-infrared ($\sim$1.3~$\mu$m, observed F410M) spans spectral regions with distinct sensitivities to stellar populations and dust attenuation. This wavelength baseline, particularly the inclusion of rest-frame near-infrared data from JWST, provides critical leverage for distinguishing between age and dust effects \citep{takagi1999age,pozzetti2000extremely}. Third, we include three metallicity values (Z = 0.004, 0.008, 0.02) in the SED fitting rather than imposing a single fixed metallicity. While the age-metallicity degeneracy cannot be completely broken with broad-band photometry alone \citep{Worthey_et_al_1994}, allowing metallicity to vary as a free parameter prevents systematic biases that arise when the assumed metallicity differs from the intrinsic value. A mismatch between the fixed and true metallicity would force compensating changes in other parameters (particularly age and dust) to achieve acceptable fits, thereby increasing uncertainties in the derived star formation histories and stellar population properties. Recent studies have demonstrated that similar multi-wavelength coverage extending into the rest-frame near-infrared can provide sufficient constraints to distinguish between these competing effects in high-redshift galaxies (see Appendix B of \citealt{Abdurrof_et_al_2023}). While residual degeneracies remain inherent to photometric SED fitting, our analysis framework is designed to minimize their impact on the primary conclusions regarding spatially resolved quenching patterns, which rely primarily on relative trends rather than absolute parameter determinations.

\subsection{Derived Physical Parameters} \label{sec:derived_parameters}

From the SED fitting analysis, we extract several key physical parameters that characterize the spatially resolved properties of protocluster galaxies. The primary outputs include stellar mass surface densities ($\Sigma_*$), SFR surface densities ($\Sigma_{\text{SFR}}$), stellar ages, dust extinction values, and sSFR derived from the ratio of SFR to stellar mass. To validate our spatially resolved methodology, we performed independent SED fitting analysis using aperture-integrated photometry for our combined sample of 38 galaxies. Total stellar masses derived from summing individual spatially resolved fits show excellent agreement with masses obtained from integrated photometry, with a median offset of $+0.006$ dex, scatter of $0.142$ dex, and Pearson correlation coefficient of $r = 0.982$. Total SFR estimates from summed resolved fits show a median offset of $-0.13$ dex relative to integrated photometry with a scatter of $0.436$ dex and correlation of $r = 0.624$. This demonstrates overall consistency between the two approaches, noting that the larger scatter in SFR reflects the ability of spatially resolved fitting to capture local variations in dust attenuation and star formation history that are averaged out in integrated measurements.

\begin{figure*}

\includegraphics[width=0.99\textwidth]{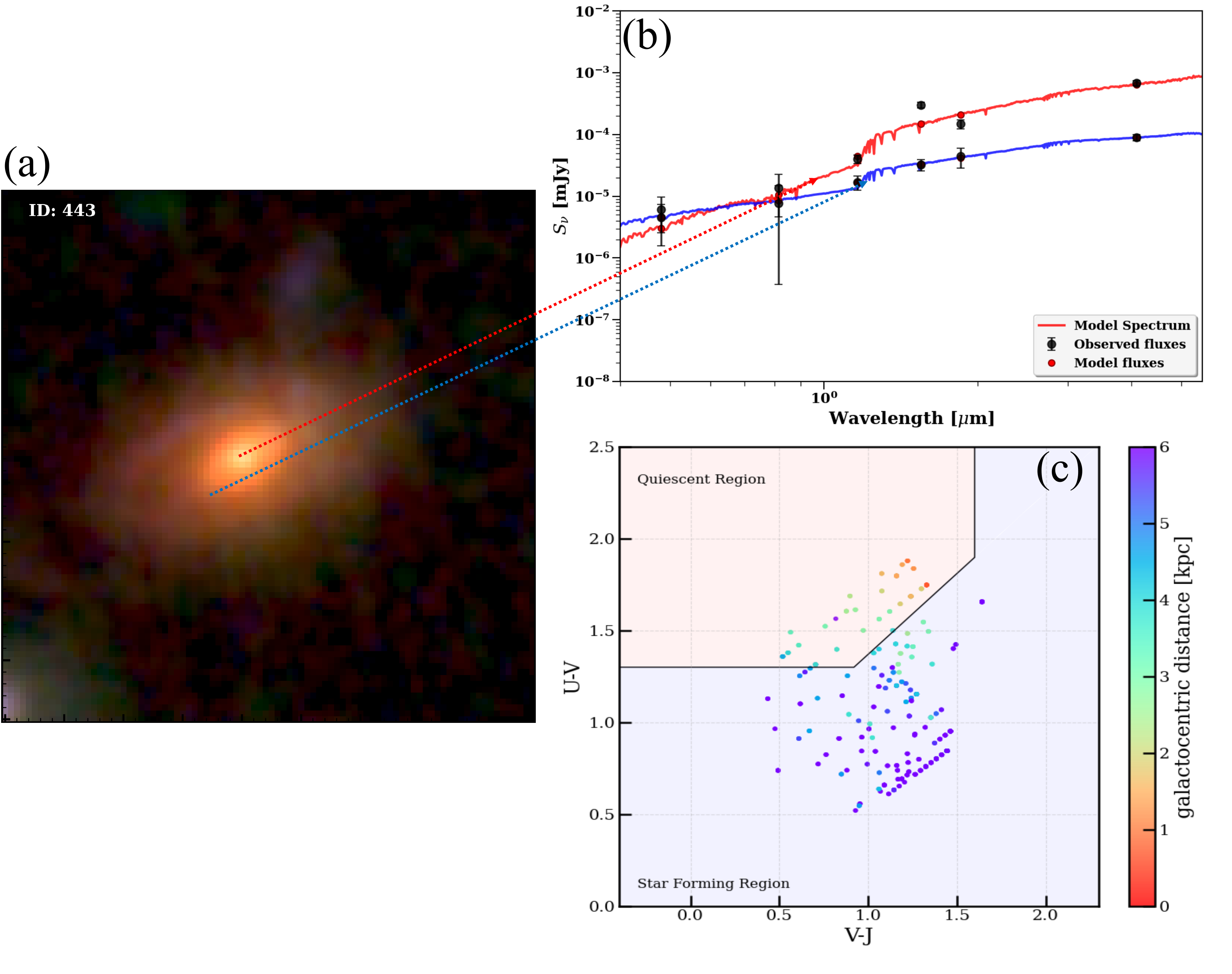}

\caption{Multi-wavelength analysis of galaxy ID 443 demonstrating our pixel-by-pixel methodology. \textbf{(a)} RGB composite image constructed using Red: F410M, Green: F160W+F182M, Blue: F475W+F814W+F115W, revealing the galaxy's morphological structure and spatial distribution of stellar light across different wavelengths. Red and blue dotted arrows indicate the spatial locations of two specific bins analyzed in panel (b). \textbf{(b)} SED fitting results from CIGALE for the two spatial bins corresponding to the regions indicated by the arrows in panel (a), showing observed photometry (black circles with error bars) and corresponding best-fit model spectra (red and blue lines). The agreement between observations and models across all six photometric bands demonstrates the reliability of our spatially resolved SED fitting approach for constraining stellar population properties at sub-galactic scales. \textbf{(c)} Spatially resolved UVJ color-color diagram with individual bins color-coded by galactocentric distance, illustrating the spatial distribution of star-forming and quiescent regions within the galaxy according to \citet{williamsRestFrameUltravioletColors2009} selection criteria and revealing the heterogeneous nature of stellar populations across the galaxy.}

\label{fig:galaxy_443_analysis}

\end{figure*}

Figure~\ref{fig:galaxy_443_analysis} illustrates our spatially resolved analysis methodology using galaxy ID 443 as an example. All galaxy IDs referenced in this work correspond to the catalog presented in \citet{naufalRevealingQuiescentGalaxy2024}. The RGB composite image (panel a) constructed from HST and JWST photometry reveals the galaxy's morphological structure and stellar light distribution. The SED fitting results (panel b) demonstrate the quality of CIGALE fits to spatially binned photometry, comparing two different spatial regions to highlight variations in stellar populations across the galaxy. The spatially resolved UVJ diagram (panel c) shows the distribution of star-forming and quiescent regions within the galaxy according to \citet{williamsRestFrameUltravioletColors2009} criteria, with individual bins color-coded by galactocentric distance measured from the GALFIT \citep{Peng_et_al_2002} morphological center. This analysis reveals the heterogeneous nature of stellar populations and star formation activity across different galactic regions, providing observational constraints on the spatial patterns of galaxy evolution processes.

\begin{figure*}
\includegraphics[width=\textwidth]{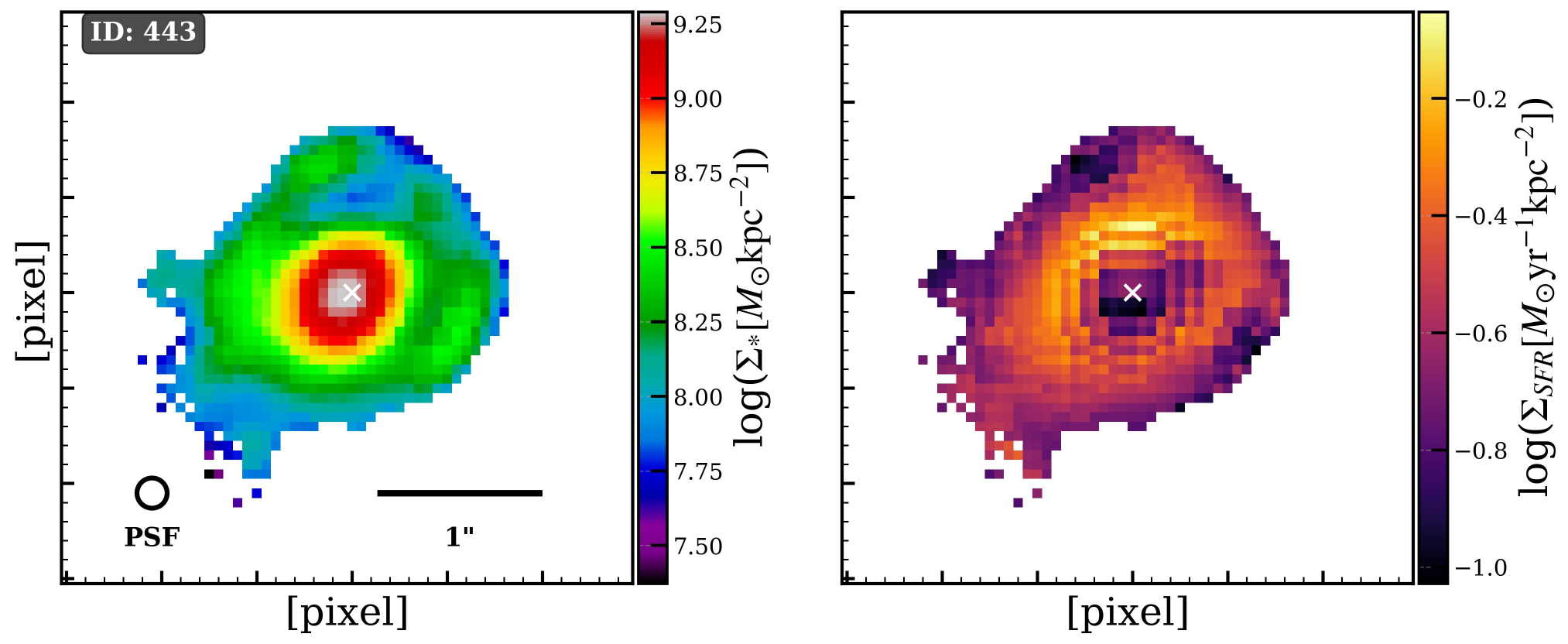}
\caption{Spatially resolved maps of $\Sigma_*$ (left panel) and $\Sigma_{\rm SFR}$ (right panel) for one of the quiescent galaxies in our sample (ID 443). The maps show a zoomed view centered on the galaxy, displaying stellar mass surface density in units of $\log(\Sigma_*[M_{\odot}\,\mathrm{kpc}^{-2}])$ and star formation rate surface density in units of $\log(\Sigma_{\mathrm{SFR}}[M_{\odot}\,\mathrm{yr}^{-1}\,\mathrm{kpc}^{-2}])$. These pixel-by-pixel measurements demonstrate the spatial resolution and detail achieved through our SED fitting methodology, revealing the internal structure and distribution of stellar properties within the galaxy.}
\label{fig:map_plot}
\end{figure*}

Figure~\ref{fig:map_plot} illustrates representative maps of stellar mass surface density (left panel) and SFR surface density (right panel) for one galaxy in our sample (ID 443). The galaxy center is marked with a white cross, determined from GALFIT morphological fitting. The $\Sigma_*$ map shows values ranging from approximately $\log(\Sigma_*) = 7.5$ to $9.25$ $M_{\odot}$ kpc$^{-2}$, with the highest concentrations observed in the central regions. The $\Sigma_{\rm SFR}$ map displays values spanning $\log(\Sigma_{\text{SFR}}) = -1.0$ to $-0.3$ $M_{\odot}$ yr$^{-1}$ kpc$^{-2}$, demonstrating spatial variations in current star formation activity across the galaxy. The complete set of stellar mass and $\Sigma_{\rm SFR}$ maps for all 11 quiescent galaxies is presented in the appendix (Figure~\ref{fig:quiescent_galaxy_maps}), while the corresponding maps for all 27 star-forming galaxies are shown in the appendix (Figures~\ref{fig:sf_galaxy_maps}). Individual pixel uncertainties account for both SED fitting errors and additional uncertainties from scaling pixel fluxes to derive spatially resolved parameters. Typical uncertainties are 0.31 dex for stellar mass surface density and 0.57 dex for SFR surface density based on analysis of 19,981 pixels across 38 galaxies.
These spatially resolved maps reveal the diversity in galaxy structures and star formation patterns within the protocluster environment.

Specific star formation rates (sSFR = SFR/$M_*$) provide insights into galaxy evolution by normalizing current star formation activity relative to existing stellar mass. The spatially resolved sSFR distributions derived from these maps enable identification of regions experiencing different evolutionary phases and provide observational constraints on the spatial patterns of star formation regulation within individual galaxies.

We additionally derive mass-to-light (M/L) ratios from the spatially resolved stellar mass and observed flux measurements to probe the age structure of stellar populations within individual galaxies. The M/L ratio provides a complementary diagnostic to sSFR for characterizing evolutionary state, as it is primarily sensitive to the stellar population age distribution rather than current star formation activity. We calculate M/L ratios using the F160W filter (HST/WFC3, observed wavelength $\sim$1.6~$\mu$m), which samples rest-frame optical wavelengths ($\sim$5000~\AA{} or V-band) at $z \sim 2.16$. This wavelength regime traces the bulk stellar mass while remaining relatively sensitive to age differences in stellar populations, making it suitable for identifying age gradients within galaxies.

For each spatial bin, we compute the monochromatic luminosity from the observed F160W flux using the luminosity distance derived from our adopted cosmology (H$_0$ = 70 km s$^{-1}$ Mpc$^{-1}$, $\Omega_M$ = 0.3, flat $\Lambda$CDM). The luminosity is expressed in solar units using the solar monochromatic luminosity density at V-band wavelengths ($L_{\lambda,V,\odot} \approx 2.4 \times 10^{29}$ erg s$^{-1}$ \AA$^{-1}$). The M/L ratio is then calculated as the ratio of stellar mass surface density to luminosity surface density within each spatial bin, expressed in units of M$_{\odot}$/L$_{\odot}$. These M/L ratios are subsequently analyzed as a function of galactocentric radius using the same radial binning procedure as applied to the stellar mass and sSFR profiles.

\section{Results} \label{sec:results}

\subsection{Spatially Resolved Radial Profiles} \label{sec:radial_profiles}

We present the spatially resolved stellar mass surface density and specific SFR profiles for 38 protocluster galaxies, comprising 27 star-forming and 11 quiescent systems as classified by \citet{naufalRevealingQuiescentGalaxy2024}.

Radial profiles are constructed using elliptical annuli . Galactocentric distances are measured from galaxy centers determined through GALFIT morphological fitting \citep{Peng_et_al_2002}, with elliptical annuli defined by the axis ratio and position angle from the structural fits. We assume intrinsic radial gradients without additional deprojection corrections, as our analysis focuses on relative trends within individual galaxies and systematic differences between populations. Within each annulus, we compute mean surface densities by summing the total stellar mass (or SFR) within valid galaxy pixels and dividing by the projected area. For sSFR profiles, we compute the ratio of SFR and stellar mass surface densities within each annulus.

All radial distances are normalized by the half-mass radius ($R_{\mathrm{e}}$) to enable direct comparison across galaxies. Population-averaged profiles are constructed by interpolating individual galaxy profiles onto a common radial grid, then computing mean values with uncertainties derived from bootstrap resampling (1000 iterations). Profiles extend reliably to approximately 3--4 $R_{\mathrm{e}}$ for most galaxies in our sample.

Two galaxies are excluded from the original sample of 40 robust members for methodological reasons. First, we exclude the central spiderweb galaxy (ID 577), the radio galaxy at the center of the protocluster, due to its fundamentally different nature as an active galactic nucleus with extended radio jets and complex multi-component morphology that distinguishes it from the normal galaxy population we aim to characterize. Second, we exclude one star-forming galaxy (ID 650) due to unreliable GALFIT fitting results. This galaxy showed poor convergence during profile fitting, with highly unstable center determination and unrealistic structural parameters that precluded robust radial profile analysis.
Quiescent galaxies are defined following \citet{naufalRevealingQuiescentGalaxy2024} based on 4000~\AA\ break strength $D_{\mathrm{n}}4000 > 1.1$ (as defined by \citealt{Balogh_et_al_1999}).
While our sample segregates naturally into two stellar mass regimes, the primary distinction is between star-forming and quiescent evolutionary states rather than stellar mass. For display purposes, we present the results separated by mass bins: $\log(M_*/\mathrm{M}_{\odot}) < 10.5$ containing 26 galaxies (25 star-forming and 1 quiescent), and $\log(M_*/\mathrm{M}_{\odot}) \geq 10.5$ containing 12 galaxies (2 star-forming and 10 quiescent), which reflects the strong correlation between stellar mass and evolutionary state in our sample. All radial profiles are normalized by the half-mass radius ($R_{\mathrm{e}}$) derived from the stellar mass surface density distributions to enable direct comparison across the sample.

\begin{figure*}
\includegraphics[width=\textwidth]{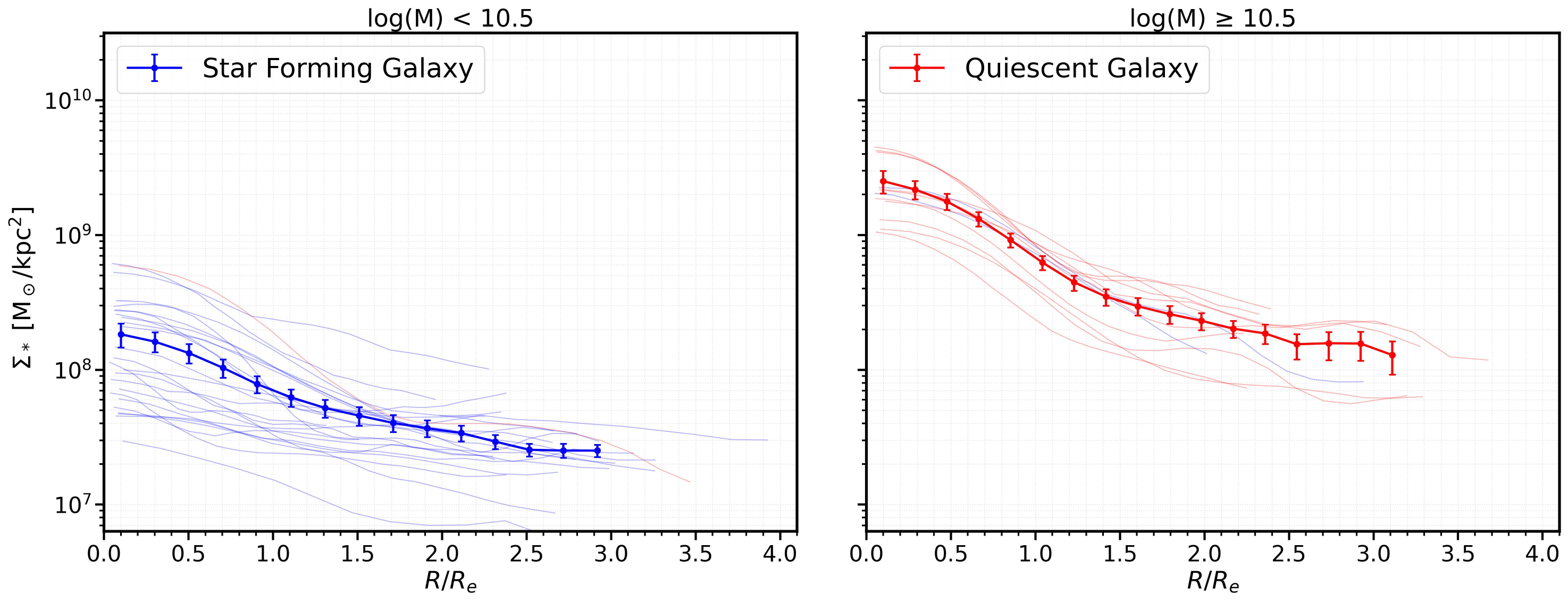}
\caption{Stellar mass surface density profiles as a function of normalized radius for protocluster galaxies in two stellar mass bins. Left panel shows galaxies with $\log(M_*/\mathrm{M}_{\odot}) < 10.5$ (blue), right panel shows $\log(M_*/\mathrm{M}_{\odot}) \geq 10.5$ (red). The higher mass bin is dominated by quiescent galaxies. Thin lines represent individual galaxies, thick lines show mean profiles with bootstrap-derived error bars (1000 iterations) representing 68\% confidence intervals. Radii are normalized by the half-mass radius $R_{\mathrm{e}}$.}
\label{fig:mass_profiles}
\end{figure*}

Figure~\ref{fig:mass_profiles} displays the stellar mass surface density profiles ($\Sigma_*$) as a function of normalized radius for both mass bins. Individual galaxy profiles are shown as thin colored lines, with mean profiles and their associated uncertainties represented by thick lines with error bars. We calculate uncertainties using bootstrap resampling with 1000 iterations. For the lower mass bin ($\log(M_*/\mathrm{M}_{\odot}) < 10.5$), which is dominated by star-forming galaxies, the mean stellar mass surface density decreases from $1.3 \times 10^8$~M$_{\odot}$~kpc$^{-2}$ at $R/R_{\mathrm{e}} = 0.5$ to $2.5 \times 10^7$~M$_{\odot}$~kpc$^{-2}$ at $R/R_{\mathrm{e}} = 3.0$. The higher mass bin ($\log(M_*/\mathrm{M}_{\odot}) \geq 10.5$), which is predominantly composed of quiescent galaxies, exhibits significantly elevated central surface densities, with mean values reaching $1.8 \times 10^9$~M$_{\odot}$~kpc$^{-2}$ at $R/R_{\mathrm{e}} = 0.5$ and declining to $1.6 \times 10^8$~M$_{\odot}$~kpc$^{-2}$ at $R/R_{\mathrm{e}} = 3.0$. The profiles extend reliably to approximately 3--4~$R_{\mathrm{e}}$ for the majority of galaxies in both mass bins.

The stellar mass surface density profiles demonstrate substantially higher central concentrations in quiescent galaxies compared to star-forming systems, with quiescent galaxies showing central surface densities that are approximately 14 times greater than those observed in star-forming galaxies in lower mass systems. This enhanced central mass concentration is consistent with the prevalence of quiescent galaxies \citep{Szomoru_et_al_2012, Tacchella_et_al_2015}, which exhibit more centrally concentrated stellar distributions characteristic of early-type morphologies.

\begin{figure*}
\includegraphics[width=\textwidth]{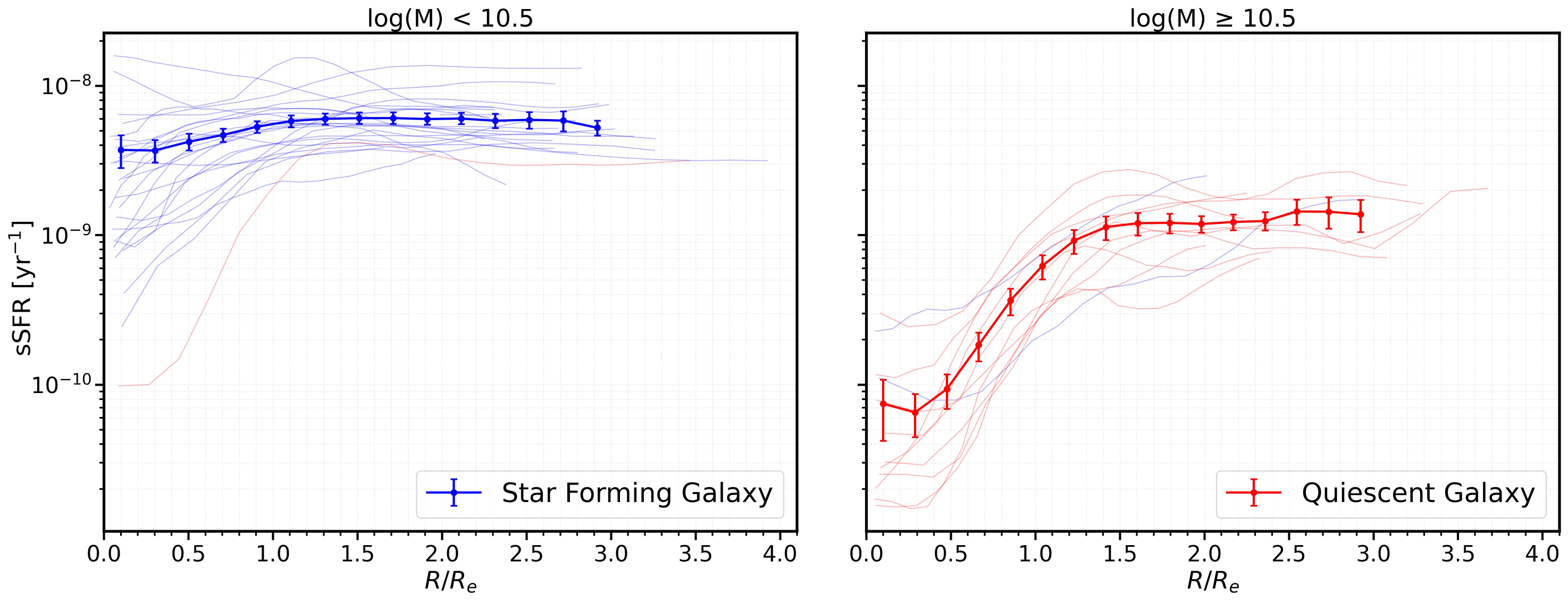}
\caption{Specific star formation rate (sSFR) profiles as a function of normalized radius for the same galaxy sample and mass bins as Figure \ref{fig:mass_profiles}. Left panel: Lower mass galaxies (mostly star-forming galaxies) show relatively flat sSFR profiles with gradual decline toward outer regions. Right panel: Higher mass galaxies (mostly quiescent galaxies) exhibit systematic increases in sSFR with radius. The central suppression of star formation activity in massive galaxies contrasts with the sustained activity observed in lower mass systems. Error bars represent 68\% confidence intervals derived from bootstrap resampling (1000 iterations).}
\label{fig:ssfr_profiles}
\end{figure*}

The sSFR profiles reveal markedly different behavior between star-forming and quiescent populations, as shown in Figure~\ref{fig:ssfr_profiles}.
For the lower mass bin, dominated by star-forming galaxies, the mean sSFR shows a relatively flat distribution, with values of $4.2 \times 10^{-9}$ yr$^{-1}$ at $R/R_{\mathrm{e}} = 0.5$ and $5.2 \times 10^{-9}$ yr$^{-1}$ at $R/R_{\mathrm{e}} = 3.0$.  This nearly constant profile is characteristic of actively star-forming systems with sustained activity across their disks. In contrast, the higher mass bin, dominated by quiescent galaxies, demonstrates a pronounced radial gradient in sSFR consistent with inside-out quenching patterns. The mean sSFR increases systematically from $9.4 \times 10^{-11}$~yr$^{-1}$ at $R/R_{\mathrm{e}} = 0.5$ to $1.4 \times 10^{-9}$~yr$^{-1}$ at $R/R_{\mathrm{e}} = 3.0$, representing a substantial increase across the measured radial range. Notably, two star-forming galaxies in this mass bin (ID 335 with Sérsic index $n=1.15$ and ID 551 with n$ = 2.44$) exhibit clear signatures of inside-out quenching despite their overall star-forming classification, indicating they represent systems in the early phases of the quenching process. The compact nature of ID 551 particularly exemplifies the structural characteristics associated with quenching onset. The central regions ($R/R_{\mathrm{e}} < 1.0$) exhibit consistently low sSFR values indicative of suppressed star formation activity, while the outer regions
show elevated star formation activity. This radial sSFR gradient demonstrates that star formation quenching proceeds from the galaxy centers outward, with central regions already quenched while outer regions maintain residual star formation. Individual galaxy profiles within the quiescent population display significant diversity, with some systems maintaining low sSFR across all radii while others exhibit steep positive gradients toward larger radii.

To quantify the radial variations in star formation activity, we compute the ratio of outer to inner sSFR using the half-mass radius as the boundary ($R_{\mathrm{e}}$). Star-forming galaxies in the lower mass bin (n=25) exhibit modest radial variation between inner ($R \leq R_{\mathrm{e}}$) and outer ($R > R_{\mathrm{e}}$) regions, with median sSFR values of $\log(\text{sSFR}) = -8.42 \pm 0.05$ and $-8.28 \pm 0.03$~yr$^{-1}$, respectively (errors represent 68\% bootstrap confidence intervals), corresponding to a factor of $1.4$ enhancement in outer regions. In marked contrast, quiescent galaxies in the higher mass bin (n=10) demonstrate a pronounced inside-out quenching signature. The median sSFR of the inner regions ($R \leq R_{\mathrm{e}}$) is $\log(\text{sSFR}) = -10.08 \pm 0.10$~yr$^{-1}$, while the outer regions ($R > R_{\mathrm{e}}$) exhibit a substantially elevated median of $-8.97 \pm 0.11$~yr$^{-1}$. This corresponds to an enhancement by a factor of $\approx 13$, confirming that star formation quenching proceeds from galaxy centers outward, with central regions already quenched while outer regions maintain residual star formation activity. This systematic pattern provides direct evidence for inside-out quenching as the dominant mechanism governing star formation cessation in massive galaxies within the Spiderweb protocluster.

\subsection{Mass-to-Light Ratio Radial Profiles} \label{sec:ml_profiles}

\begin{figure*}
\includegraphics[width=\textwidth]{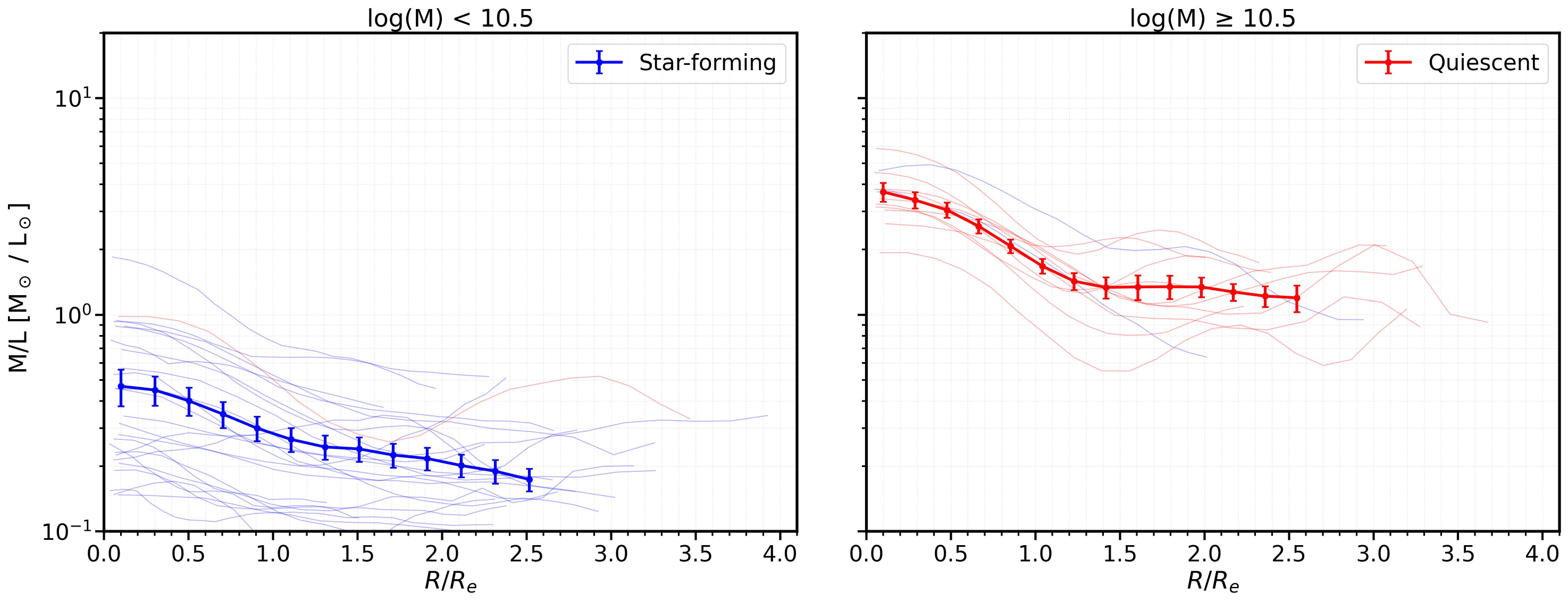}
\caption{Mass-to-light (M/L) ratio profiles as a function of normalized radius for protocluster galaxies, computed using the F160W filter (rest-frame optical wavelengths at $z \sim 2.16$). The left panel shows star-forming galaxies with $\log(M_*/M_{\odot}) < 10.5$ (blue), while the right panel displays quiescent galaxies with $\log(M_*/M_{\odot}) \geq 10.5$ (red). Thin lines represent individual galaxies, thick lines show mean profiles with bootstrap-derived error bars representing 68\% confidence intervals. Quiescent galaxies exhibit systematically higher M/L ratios and steeper negative gradients compared to star-forming systems.}
\label{fig:ml_profiles}
\end{figure*}

The mass-to-light (M/L) ratio provides a complementary diagnostic of stellar population properties within galaxies. The M/L ratio is sensitive to both stellar population age and ongoing star formation activity, as young stellar populations with actively forming massive stars exhibit systematically lower M/L ratios due to their high luminosity relative to stellar mass. Figure~\ref{fig:ml_profiles} presents the M/L radial profiles for our sample, computed using the F160W band which samples rest-frame optical wavelengths at $z \sim 2.16$.

Quiescent galaxies in the higher mass bin ($\log(M_*/M_{\odot}) \geq 10.5$) display systematically elevated M/L ratios compared to star-forming systems. The mean M/L profile exhibits a negative radial gradient, decreasing from approximately $3.4~M_{\odot}/L_{\odot}$ at $R/R_{\mathrm{e}} = 0.5$ to $1.9~M_{\odot}/L_{\odot}$ at $R/R_{\mathrm{e}} \sim 3.0$, consistent with observations of negative M/L gradients in local galaxies \citep{Garcia-Benito_et_al_2019, Goddard_et_al_2017} and at intermediate redshift \citep[e.g., $z \sim 1.4$;][]{Chan_et_al_2016}. This negative M/L gradient provides independent confirmation of the inside-out quenching pattern identified through the sSFR analysis in Section~\ref{sec:radial_profiles}, and explains the negative color gradients ($r_{\mathrm{mass}} < r_{\mathrm{light}}$) often observed in quiescent populations \citep{Suess_et_al_2019a}. The evolutionary pathway connecting these high-redshift observations to the local universe has been traced through fossil record analysis by \citet{Avila-Reese_et_al_2023}, who used MaNGA data to infer the evolution of M/L profiles in quiescent elliptical galaxies. They found that the M/L ratio of massive galaxies increases over time (until $z \sim 0.2$) at a faster rate in the inner regions than in the outer ones (see their Fig. 7), providing strong evidence for inside-out quenching in quiescent galaxies \citep[see also][]{Ibarra-Medel_et_al_2022}. This evolutionary trend is consistent with the idea that quiescent galaxies in protoclusters accelerated their formation process due to the dense environment.

The physical origin of the M/L gradient in quiescent galaxies can be understood through its relationship with the observed sSFR gradient. As shown in Section~\ref{sec:radial_profiles}, quiescent galaxies exhibit strong positive sSFR gradients, with central regions displaying lower sSFR ($\sim10^{-11}$ yr$^{-1}$) compared to outer regions ($\sim10^{-10}$ yr$^{-1}$). The central regions, having ceased star formation earlier and more completely, lack young luminous stars, resulting in elevated M/L ratios. The outer regions maintain residual low-level star formation activity, which contributes young stellar populations that increase luminosity relative to stellar mass, thereby reducing the M/L ratio. While the absolute sSFR levels in these quiescent systems are low compared to actively star-forming galaxies, the spatial variation in residual star formation activity produces observable gradients in both stellar population properties and M/L ratios, providing a direct signature of the inside-out quenching process.

Star-forming galaxies in the lower mass bin ($\log(M_*/M_{\odot}) < 10.5$) exhibit substantially lower M/L ratios overall, with values ranging from approximately $0.37~M_{\odot}/L_{\odot}$ in the centers to $0.17~M_{\odot}/L_{\odot}$ in the outer regions. These low M/L values reflect stellar populations dominated by ongoing star formation activity, where young massive stars contribute significant luminosity. The moderate radial M/L gradient observed in star-forming galaxies traces inside-out stellar mass assembly rather than differential quenching. The mass-weighted age profiles (Section~\ref{sec:age_profiles}) show that central regions assembled their stellar mass earlier than outer regions, producing slightly older underlying stellar populations in galaxy centers despite ongoing star formation throughout these systems. Combined with the flat sSFR profiles indicating uniformly distributed current star formation, the M/L gradient in star-forming galaxies reflects the age structure established during the mass assembly process. This anti-correlation between sSFR gradients and M/L gradients in actively star-forming systems is well-established \citep[e.g.,][]{Ge_et_al_2021}, see also \citet{Garcia-Benito_et_al_2019}, where stronger sSFR gradients produce steeper negative M/L gradients.

The contrast between the M/L profiles of star-forming and quiescent galaxies, both in absolute values and in their physical interpretation, demonstrates fundamentally different evolutionary states. For quiescent galaxies, the M/L gradient arises from differential suppression of residual star formation following the main epoch of mass assembly, as evidenced by the combination of steep sSFR gradients and relatively uniform mass-weighted ages. For star-forming galaxies, the M/L gradient traces the inside-out assembly history, with ongoing star formation maintaining low M/L ratios throughout. The combination of M/L profiles with sSFR and age gradients provides a comprehensive picture of how stellar populations and star formation activity are spatially distributed within protocluster galaxies. These M/L gradients directly trace the underlying color gradients and structural differences between mass and light distributions ($r_{e,\mathrm{mass}} < r_{e,\mathrm{light}}$) observed in high-redshift galaxies \citep{Suess_et_al_2019a}. Detailed analysis of color gradients and $r_{e,\mathrm{mass}}/r_{e,\mathrm{light}}$ ratios will be presented in future work.

\subsection{Stellar Age Radial Profiles} \label{sec:age_profiles}

\begin{figure*}
\includegraphics[width=\textwidth]{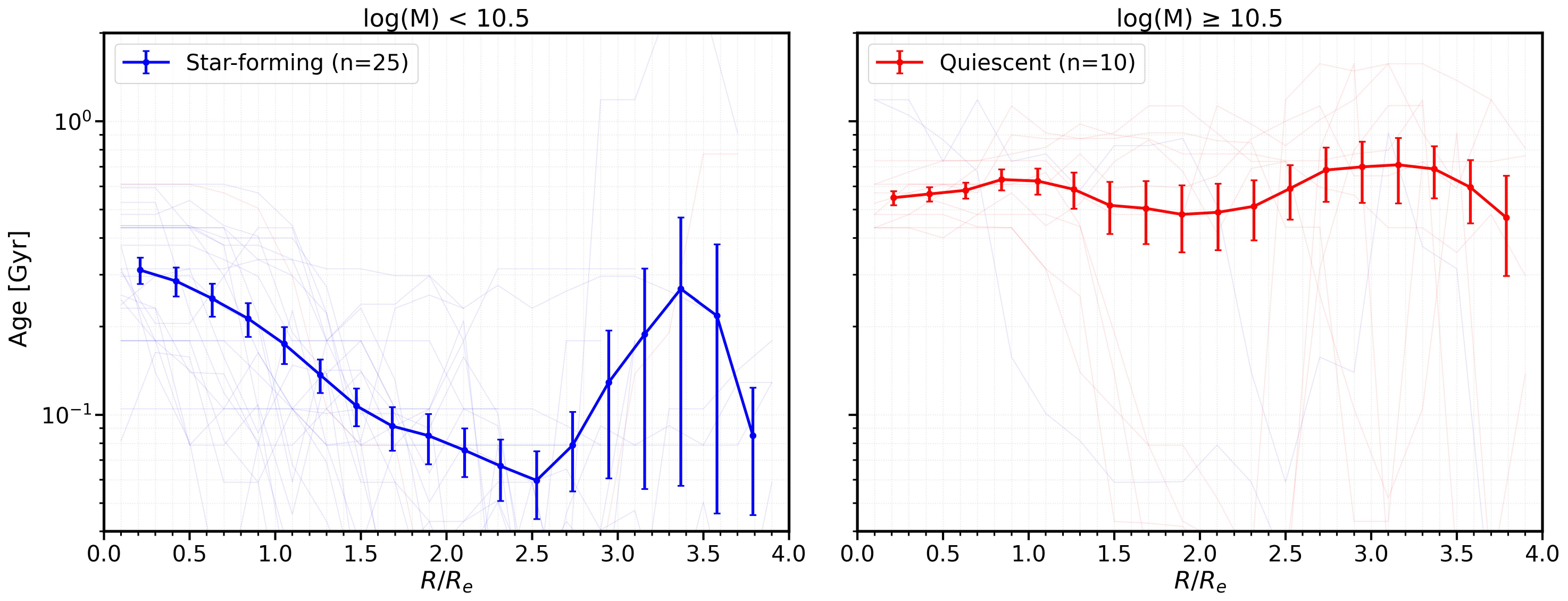}
\caption{Mass-weighted stellar age profiles as a function of normalized radius for protocluster galaxies. The left panel shows star-forming galaxies with $\log(M_*/M_{\odot}) < 10.5$ (blue), while the right panel displays quiescent galaxies with $\log(M_*/M_{\odot}) \geq 10.5$ (red). Thin lines represent individual galaxies, thick lines show mean profiles with bootstrap-derived error bars. Quiescent galaxies exhibit relatively flat age profiles at $\sim$0.4--0.5 Gyr across all radii, indicating uniform timing of the dominant stellar mass assembly phase. Star-forming galaxies show decreasing ages with radius in inner regions}
\label{fig:age_profiles}
\end{figure*}

To further characterize the stellar population properties and their spatial distribution, we examine the mass-weighted stellar age profiles derived from the CIGALE SED fitting. The mass-weighted age represents the mean age of the stellar population weighted by stellar mass contribution, providing a measure of when the bulk of stellar mass was assembled in each spatial region. This differs from luminosity-weighted age, which is more sensitive to recent star formation due to the high luminosity of young stars. Figure~\ref{fig:age_profiles} presents the mass-weighted age radial profiles for both star-forming and quiescent galaxy populations.

Quiescent galaxies in the higher mass bin exhibit relatively uniform mass-weighted ages across all radial bins consistent with the flat profiles reported by \citet{Abdurrof_et_al_2023}, with values of approximately $0.4$--$0.5$ Gyr from the central regions to the outer extent of the galaxies consistent with . This flat age profile indicates that the majority of stellar mass in these systems was assembled during a relatively concentrated epoch, with minimal radial variation in the timing of the dominant mass assembly phase. The absence of strong mass-weighted age gradients despite the presence of M/L and sSFR gradients (Sections~\ref{sec:ml_profiles} and \ref{sec:radial_profiles}) provides important constraints on the quenching timescale.

The combination of flat mass-weighted age profiles with steep sSFR gradients indicates that the spatial variations in current star formation activity do not reflect differences in the epoch of initial mass assembly, but rather differences in when and how completely star formation ceased in different regions. This pattern is consistent with a scenario in which the bulk of stellar mass was assembled relatively uniformly across these galaxies during a period of intense star formation, followed by inside-out quenching that preferentially and more completely suppressed star formation in the central regions. The residual low-level star formation in outer regions, while insufficient to significantly alter the mass-weighted age, produces observable gradients in sSFR and M/L ratios. The relatively flat age profiles suggest that the quenching process occurred recently and rapidly on galactic timescales, preventing the development of large age differences between regions that quenched at different times.

Star-forming galaxies in the lower mass bin show a trend of decreasing mass-weighted age with increasing radius, from approximately $0.3$--$0.4$ Gyr in the central regions to $\sim0.1$ Gyr at intermediate radii ($R/R_{\mathrm{e}} \sim 1.5$--$2.0$). However, the age measurements at larger radii ($R/R_{\mathrm{e}} > 2$) exhibit substantially larger uncertainties, limiting robust interpretation of the age structure in the outer regions of these galaxies. The age gradient in star-forming galaxies traces inside-out mass assembly, where central regions formed stars earlier in the galaxy's evolutionary history. The systematically younger mass-weighted ages in star-forming galaxies compared to quiescent systems ($\sim0.1$--$0.3$ Gyr vs. $\sim0.4$--$0.5$ Gyr) reflect their ongoing star formation activity, which continuously adds young stellar populations. The combination of age gradients with the relatively flat sSFR profiles (Section~\ref{sec:radial_profiles}) indicates that star formation remains active throughout these systems, with the age structure established during the earlier phases of inside-out growth.
We note that while spectroscopic follow-up observations would enable more detailed constraints on star formation histories and further refinement of age estimates, such observations are beyond the scope of this work. Nevertheless, our multi-band photometric analysis, incorporating multiple metallicity evolutionary tracks in the SED fitting, provides constraints on the stellar population gradients in these high-redshift protocluster galaxies.

\subsection{Spatially Resolved UVJ Color Analysis} \label{sec:uvj_analysis}

To further investigate the spatial distribution of star formation activity and quenching signatures, we perform spatially resolved UVJ color analysis on individual pixels within each galaxy. Rest-frame $U-V$ and $V-J$ colors are derived directly from the best-fit CIGALE SED models for each spatial bin. CIGALE computes rest-frame colors by integrating the best-fit SED through specified filter transmission curves, accounting for both the stellar population synthesis and dust attenuation properties determined from the SED fitting. We apply the UVJ selection criteria from \citet{williamsRestFrameUltravioletColors2009} to classify each pixel as either star-forming or quiescent based on these rest-frame colors. This approach enables identification of heterogeneous populations within individual galaxies and provides insight into the spatial distribution of different evolutionary states.

\begin{figure*}
\includegraphics[width=\textwidth]{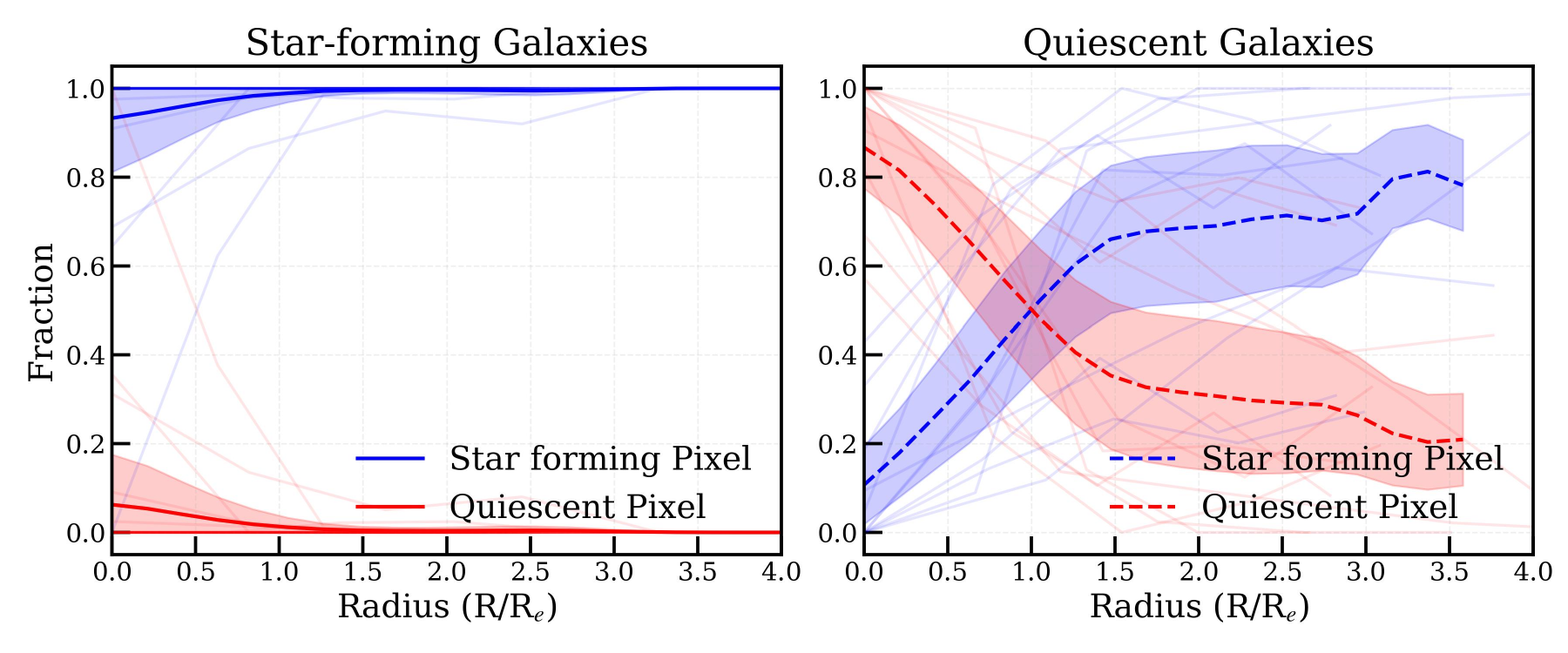}
\caption{Radial distribution of star-forming (blue) and quiescent (red) pixel fractions based on UVJ color selection criteria applied to spatially binned data. Left panel shows star-forming galaxies, right panel shows quiescent galaxies. Individual galaxy radial profiles are shown as thin semi-transparent lines, with population averages indicated by thick colored lines and shaded regions representing $\pm1\sigma$ uncertainties calculated from the scatter across individual galaxies. Note that for star-forming galaxies, the population averages approach the physical limits (SF fraction $\approx 1.0$, Q fraction $\approx 0.0$) due to uniformly active star formation across all radii in this population. Star-forming galaxies maintain predominantly star-forming classifications across all radii, with typical SF fractions $>$80\% and minimal radial variation. Quiescent galaxies exhibit systematic radial gradients with high quiescent fractions in centers ($\sim$90\%) decreasing to $\sim$40\% at large radii, consistent with inside-out quenching scenarios.}
\label{fig:uvj_fractions}
\end{figure*}

Figure~\ref{fig:galaxy_443_analysis}c demonstrates this approach for galaxy ID 443, where individual bins are color-coded by galactocentric distance measured from the GALFIT-determined center. The distribution of pixels within the UVJ diagram reveals the presence of both star-forming and quiescent regions within the same galaxy, with the color-coding illustrating how these different populations are spatially distributed from the galaxy center to its outskirts.
Figure \ref{fig:uvj_fractions} presents the radial distribution of star-forming and quiescent fractions for both star-forming and quiescent galaxy populations. Shaded regions represent $\pm1\sigma$ uncertainties calculated from the standard deviation of individual galaxy profiles at each radial bin. Star-forming galaxies demonstrate remarkably uniform pixel classifications across all radii, with star-forming pixels comprising approximately 87\%
of all pixels at all radial distances and quiescent pixels representing the remaining 13\%.
This flat radial distribution reflects the globally active star formation state of these systems and the absence of significant quenching gradients.

In contrast, quiescent galaxies exhibit pronounced radial gradients in pixel classifications that mirror the sSFR trends observed in the previous analysis. The fraction of quiescent pixels reaches more than 90\% in the central regions and decreases systematically to 60\% at radii beyond 8 kpc. Conversely, the star-forming pixel fraction increases from 10\% in galaxy centers to almost 50\% in the outer regions. This radial variation provides independent confirmation of the inside-out quenching pattern identified through direct sSFR measurements, demonstrating that central regions have transitioned to quiescent colors while outer regions retain star-forming characteristics.

\begin{figure*}
\centering
\includegraphics[width=0.49\textwidth]{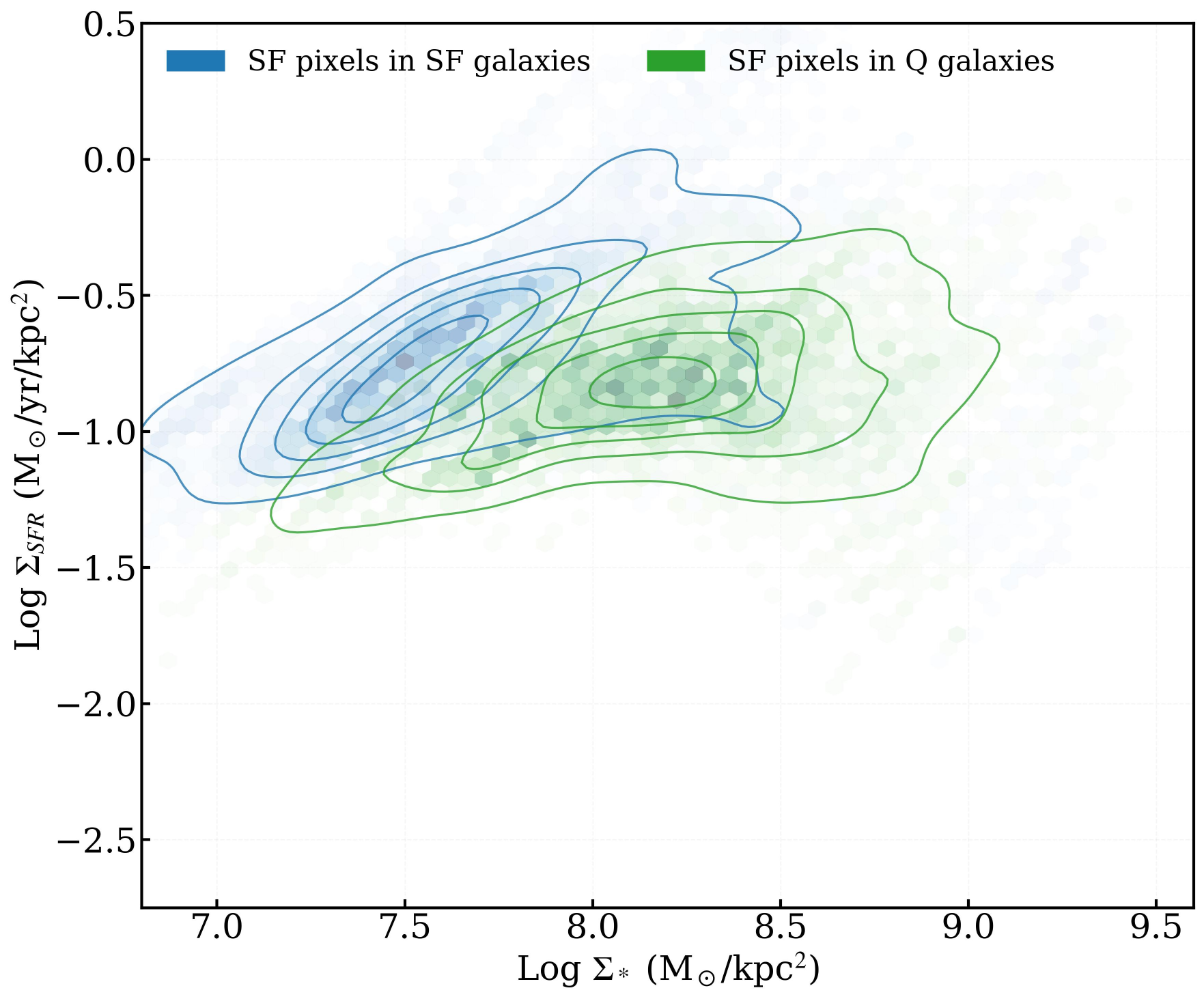}
\hfill
\includegraphics[width=0.49\textwidth]{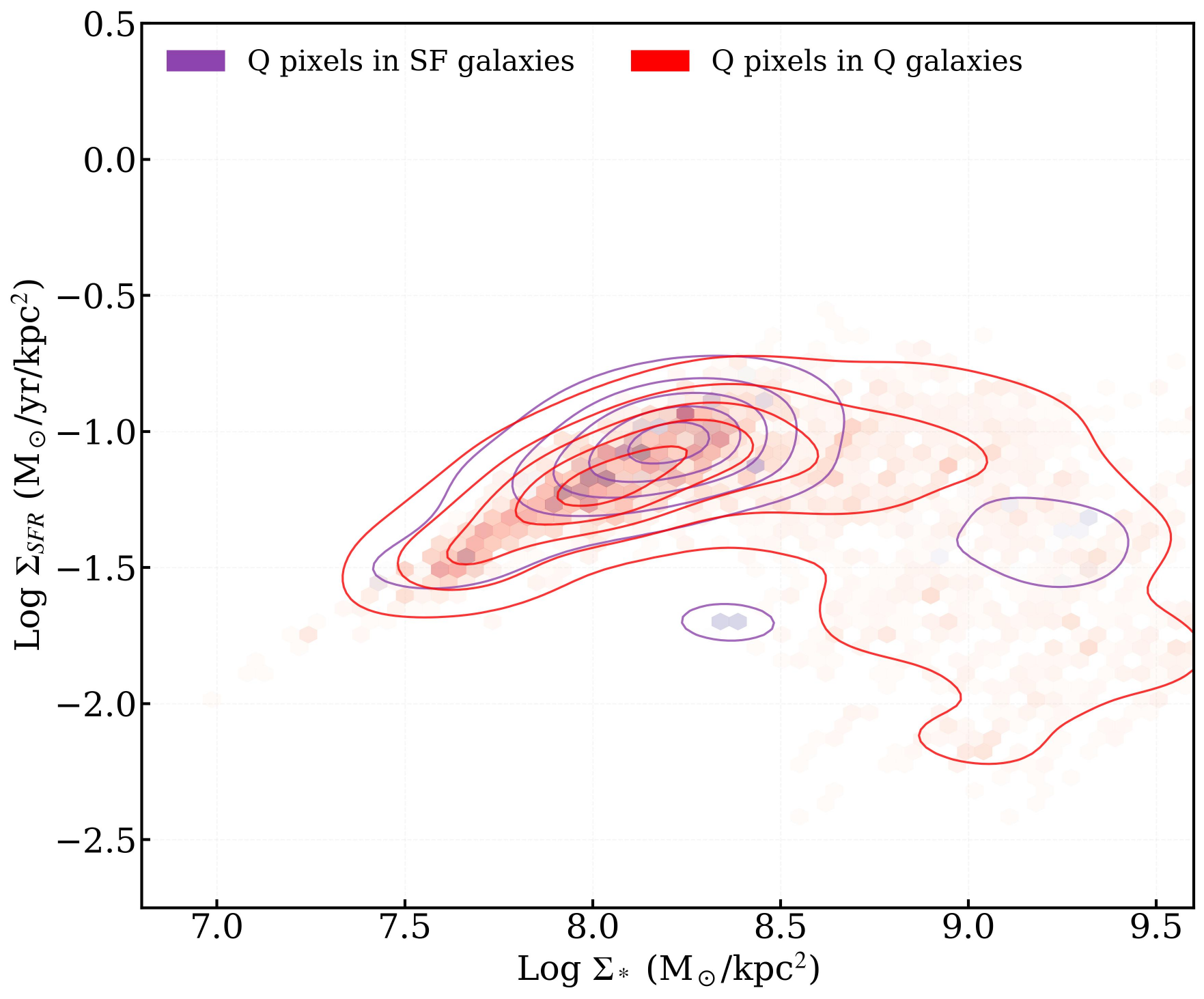}
\caption{Surface density relationships for pixels classified by UVJ colors and host galaxy type. \textbf{Left:} Star-forming pixels in star-forming galaxies (blue) and star-forming pixels in quiescent galaxies (green). \textbf{Right:} Quiescent pixels in star-forming galaxies (purple) and quiescent pixels in quiescent galaxies (red). Density contours were computed using Gaussian kernel density estimation and represent normalized density levels of 0.1, 0.3, 0.5, 0.7, and 0.9 relative to the peak density for each population. Star-forming pixels show quantitatively different distributions depending on host galaxy type, while quiescent pixels exhibit statistically similar properties regardless of host population.}
\label{fig:uvj_surface_densities}
\end{figure*}

The relationship between $\Sigma_*$ and $\Sigma_{\rm SFR}$ reveals quantitatively distinct patterns when examined separately for different pixel classifications and host galaxy types, as shown in Figure \ref{fig:uvj_surface_densities}. Our analysis encompasses 19,981 individual pixels across 38 galaxies, providing robust statistical sampling for each population.

Star-forming pixels in quiescent host galaxies exhibit systematically higher stellar mass surface densities compared to star-forming pixels in star-forming galaxies, with median values of $8.21 \pm 0.43$ versus $7.67 \pm 0.50$ log M$_\odot$ kpc$^{-2}$.  These star-forming regions in quiescent hosts also demonstrate systematically lower SFR surface densities ($-0.85 \pm 0.25$ versus $-0.74 \pm 0.34$ log M$_\odot$ yr$^{-1}$ kpc$^{-2}$), demonstrating that star-forming regions exhibit different properties depending on their host galaxy type.

Conversely, quiescent pixels demonstrate remarkably similar surface density distributions regardless of their host galaxy type. Quiescent pixels in star-forming and quiescent galaxies show nearly identical median stellar mass surface densities of $8.22 \pm 0.40$ and $8.39 \pm 0.56$ log M$_\odot$ kpc$^{-2}$ respectively, with corresponding SFR surface densities of $-1.13 \pm 0.22$ and $-1.23 \pm 0.31$ log M$_\odot$ yr$^{-1}$ kpc$^{-2}$. While quiescent pixels comprise only a small minority within star-forming galaxies (typically $<$15\%), they represent the dominant component in quiescent galaxies ($>$60\% in most cases). Despite this disparity in abundance, their surface density properties remain statistically indistinguishable between host galaxy types, indicating that quiescent regions exhibit consistent characteristics independent of their global galaxy classification.

\subsection{Morphological Dependence of Star Formation: Sérsic Index Analysis} \label{sec:sersic_analysis}

The spatially resolved UVJ analysis reveals systematic differences in star formation properties between different galaxy populations and within individual systems. To further explore the physical drivers underlying these variations, we examine the role of galaxy morphology in regulating local star formation processes. We investigate the correlation between Sérsic index and spatially resolved star formation properties to assess how galaxy structure influences star formation activity independently of the global star-forming versus quiescent classification.
The Sérsic index provides a quantitative measure of light concentration, with low values ($n < 2$) characteristic of disk-dominated systems and high values ($n > 4$) indicative of bulge-dominated morphologies. This analysis enables assessment of how galaxy structure influences local star formation processes independently of the global star-forming versus quiescent classification.

\begin{figure}
\centering
\includegraphics[width=\linewidth]{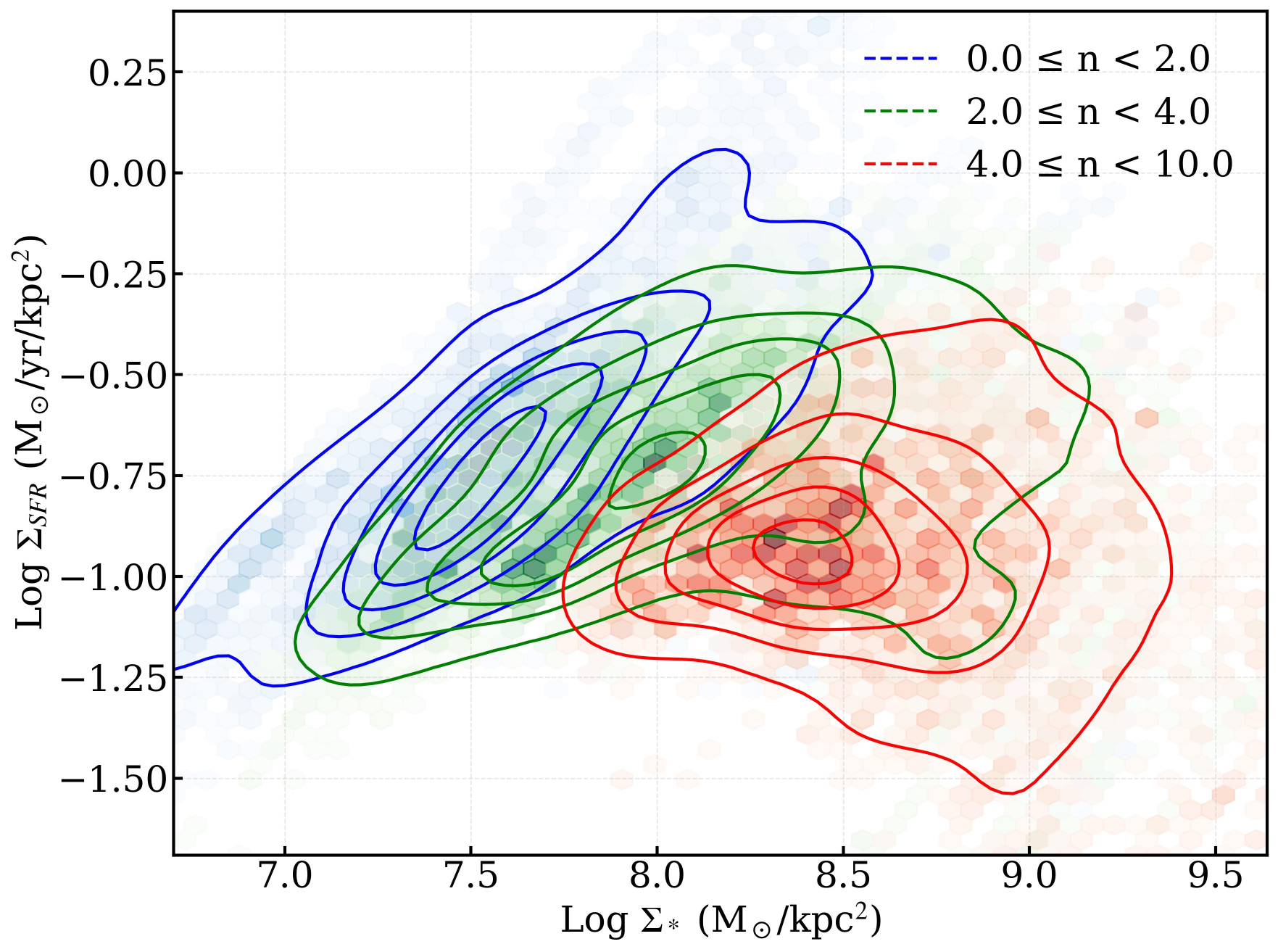}
\caption{Surface density relationships segregated by Sérsic index morphological classification. Contour levels show the distribution of $\Sigma_{\rm SFR}$ versus $\Sigma_*$ for three Sérsic index bins: $0.0 \leq n < 2.0$ (blue, disk-dominated), $2.0 \leq n < 4.0$ (green, intermediate), and $4.0 \leq n < 10.0$ (red, bulge-dominated). Density contours were computed using Gaussian kernel density estimation and represent normalized density levels of 0.1, 0.3, 0.5, 0.7, and 0.9 relative to the peak density for each population. The clear separation between morphological types reveals systematic differences in star formation efficiency and demonstrates the connection between galaxy structure and local star formation processes.}
\label{fig:sersic_surface_densities}
\end{figure}

Figure \ref{fig:sersic_surface_densities} displays the $\Sigma_{\text{SFR}}$-$\Sigma_*$ relationship segregated by Sérsic index into three morphological bins. Disk-dominated systems with low Sérsic indices ($0.0 \leq n < 2.0$, blue contours) occupy a distinct region of parameter space characterized by enhanced star formation efficiency at fixed stellar mass surface density. These systems exhibit a positive slope between surface densities, indicating sustained star formation activity across a range of stellar mass concentrations. The distribution extends to relatively high $\Sigma_{\text{SFR}}$ values, reflecting the active star formation characteristic of disk-dominated morphologies.

Bulge-dominated galaxies with high Sérsic indices ($4.0 \leq n < 10.0$, red contours) demonstrate markedly different behavior, with a negative correlation slope between $\Sigma_{\text{SFR}}$ and $\Sigma_*$. This population is characterized by systematically lower star formation surface densities at fixed stellar mass surface density, indicating reduced star formation efficiency in regions of high stellar concentration. The distribution is concentrated toward higher $\Sigma_*$ and lower $\Sigma_{\text{SFR}}$ values, consistent with the quenched central regions typical of bulge-dominated systems.

Galaxies with intermediate Sérsic indices ($2.0 \leq n < 4.0$, green contours) display transitional behavior between these two extremes, occupying an intermediate region in the $\Sigma_{\text{SFR}}$-$\Sigma_*$ plane. This morphological sequence suggests a systematic change in local star formation physics as galaxy structure evolves from disk-dominated to bulge-dominated configurations, with implications for understanding the physical mechanisms linking morphology and star formation regulation \citep{Martig_et_al_2009}.

\begin{figure}
\includegraphics[width=\linewidth]{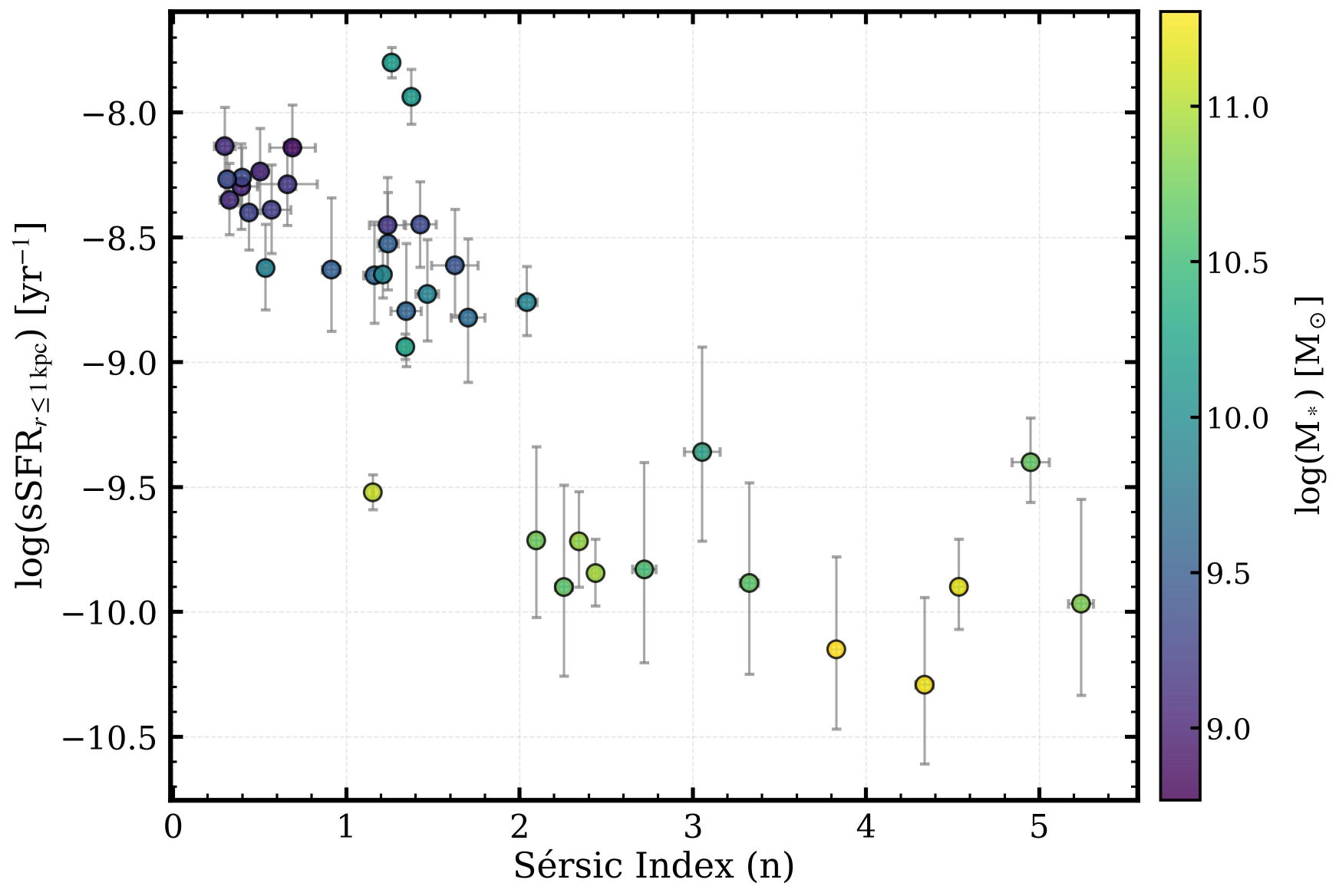}
\caption{Central sSFR (measured within 1 kpc) as a function of Sérsic index for the protocluster galaxy sample. Each point represents an individual galaxy, with colors indicating stellar mass as shown by the colorbar. Error bars represent Sérsic index fitting uncertainties and Monte Carlo-propagated sSFR uncertainties. The plot demonstrates a strong anti-correlation between structural concentration and central star formation activity.}
\label{fig:sersic_central_ssfr}
\end{figure}

The relationship between Sérsic index and central star formation activity provides additional insight into the morphology-quenching connection, as illustrated in Figure \ref{fig:sersic_central_ssfr}. We observe a strong anti-correlation between Sérsic index and sSFR measured within the central 1 kpc, with star formation rates decreasing systematically as structural concentration increases. Points color-coded by stellar mass (ranging from $\log(M_*/M_\odot) = 8.8$ to 11.3) reveal a fundamental coupling between morphology and mass: low Sérsic index galaxies are predominantly lower-mass systems ($\log(M_*/M_\odot) = 8.8-9.9$), while high Sérsic index galaxies are systematically more massive ($\log(M_*/M_\odot) = 10.6-11.2$). Error bars represent Sérsic index fitting uncertainties and sSFR uncertainties derived from Monte Carlo error propagation using asymmetric 16th and 84th percentile confidence intervals. This demonstrates that morphological structure and galaxy mass are intimately connected in regulating star formation in galaxy centers.

Quantitatively, low Sérsic index systems ($n < 1$) maintain $\log(\text{sSFR}) \approx -8.3$ yr$^{-1}$, while high Sérsic index systems ($n > 4$) exhibit values as low as $\log(\text{sSFR}) \approx -10.3$ yr$^{-1}$. The underlying mass dependence is quantified by a highly significant Spearman rank correlation between stellar mass and central sSFR ($r = -0.83$, $p < 0.001$). Low-mass systems ($\log(M_*/M_\odot) < 10$) show median values of -8.5 yr$^{-1}$, while high-mass systems ($\log(M_*/M_\odot) > 11$) exhibit median values of -10.0 yr$^{-1}$.

This morphology-mass-sSFR coupling reveals that structural concentration and mass growth work in concert to suppress star formation in galaxy centers. The stellar mass distribution (indicated by the color coding) highlights this fundamental relationship: high-mass systems ($\log(M_*/M_{\odot}) > 11$) systematically exhibit both high Sérsic indices ($n > 3.5$) and suppressed activity ($\log(\text{sSFR}) < -10$), while lower-mass systems ($\log(M_*/M_{\odot}) < 10$) display disk-dominated morphologies ($n < 2$) and elevated activity ($\log(\text{sSFR}) > -9$). This tight connection suggests that the physical processes responsible for building dense stellar cores are fundamentally linked to the mechanisms that quench star formation.

The morphological dependence of central quenching, combined with the strong mass-sSFR correlation, suggests that galaxy structure and mass evolution follow correlated evolutionary pathways. The systematic variation of both surface density relationships and central sSFR with Sérsic index indicates that the transition from disk-dominated to bulge-dominated morphology correlates with both mass growth and significant changes in the physical conditions governing star formation. These observed correlations provide important observational constraints on theoretical models of galaxy structure, mass assembly, and quenching mechanisms, though distinguishing between simultaneous versus sequential physical processes requires further investigation.

\section{Discussion}

Our spatially resolved analysis of galaxies in the Spiderweb protocluster at $z \sim 2.16$ provides new observational evidence for the physical processes governing galaxy evolution during cosmic noon. Through pixel-by-pixel SED fitting and comprehensive analysis of stellar mass surface density profiles, sSFR, mass-to-light ratios, stellar ages, UVJ color distributions, and morphological properties, we reveal clear signatures of inside-out quenching operating in massive protocluster galaxies, coupled with strong correlations between galaxy structure and star formation activity.

\subsection{Inside-Out Quenching in High-Mass Protocluster Galaxies}

Our most significant finding is the clear demonstration of inside-out quenching within the Spiderweb protocluster, particularly evident in our quiescent galaxy population. These systems exhibit systematic radial gradients in sSFR, with central regions showing sSFR values approximately an order of magnitude lower than outer regions. The mean sSFR increases systematically from {$9.4 \times 10^{-11}$} yr$^{-1}$ at $R/R_{\mathrm{e}} = 0.5$ to {$1.4 \times 10^{-9}$} yr$^{-1}$ at $R/R_{\mathrm{e}} = 3.0$, providing direct observational evidence for the suppression of central star formation while outer regions maintain residual star formation activity. The mass-to-light (M/L) ratio profiles provide independent confirmation: quiescent galaxies exhibit systematically elevated M/L ratios (M/L $\sim 3.4~M_{\odot}/L_{\odot}$ in centers, decreasing to $\sim 1.9~M_{\odot}/L_{\odot}$ at outer regions). This structural configuration, characterized by redder, higher-M/L centers, is consistent with the negative M/L gradients observed in local galaxies \citep{Garcia-Benito_et_al_2019} and the negative color gradients ($r_{e,\mathrm{mass}} < r_{e,\mathrm{light}}$) that characterize the quiescent population at lower redshifts \citep{Suess_et_al_2019a}. Notably, while field galaxies at $z \gtrsim 2$ typically exhibit flat color gradients \citep{Suess_et_al_2019a}, our protocluster quiescent galaxies display well-established negative gradients, suggesting accelerated structural evolution in this dense environment. Despite these strong sSFR and M/L gradients, mass-weighted stellar age profiles remain flat across all radii, indicating that quenching occurred relatively recently and rapidly after a concentrated epoch of mass assembly.

This radial sSFR gradient is remarkably consistent with the seminal observations of \citet{Tacchella_et_al_2015}, who first demonstrated inside-out quenching in massive star-forming galaxies at $z \sim 2.2$ using ground-based adaptive optics observations. Their work established that the most massive systems at cosmic noon can sustain high SFR at large radii while hosting already quenched central regions, with quenching timescales of less than 1 Gyr in inner regions.  Our observations confirm that inside-out quenching proceeds similarly in the protocluster environment, consistent with its interpretation as an internal mass-driven process.

The contrast with our lower-mass star-forming sample ($\log(M_*/M_{\odot}) < 10.5$) is particularly striking. These star-forming galaxies show relatively flat sSFR profiles with minimal radial variation, indicating sustained star formation activity across their entire extent. However, they exhibit substantially lower M/L ratios (M/L $\sim 0.37~M_{\odot}/L_{\odot}$ in centers decreasing to $\sim 0.17~M_{\odot}/L_{\odot}$ at outskirts) with negative M/L and mass-weighted age gradients that trace inside-out mass assembly, where central regions formed stars earlier but ongoing star formation maintains consistently low M/L values throughout \citep{Nelson_et_al_2016}. Detailed analysis of inside-out mass assembly in star-forming galaxies will be presented in future work. While our sample exhibits a strong correlation between evolutionary state and stellar mass, with quiescent galaxies predominantly found at $\log(M_*/M_{\odot}) \geq 10.5$ and most of the star-forming galaxies at $\log(M_*/M_{\odot}) \leq 10.5$, the primary distinction driving the observed radial sSFR patterns is evolutionary state rather than stellar mass alone. This evolutionary state-dependent behavior aligns with observational evidence that quenching processes become increasingly efficient above critical stellar mass thresholds \citep{Contini_et_al_2020}, and with theoretical predictions from simulations \citep{Zolotov_et_al_2015, Tacchella_et_al_2016, Nelson_et_al_2021} that inside-out quenching primarily affects systems where feedback mechanisms become dominant over other star formation regulation processes.

Recent spatially resolved studies have similarly found evidence for inside-out stellar mass buildup and quenching in massive disk galaxies \citep{Gonzalez_Delgado_et_al_2016, Abdurro_Akiyama_2018, Nelson_et_al_2021}. Particularly relevant is the work by \citet{Spilker_et_al_2019}, who used ALMA CO observations to directly trace inside-out quenching through molecular gas depletion in a compact galaxy at z=2.2, and \citet{Bluck_et_al_2020}, who demonstrated that central galaxies exhibit inside-out quenching through steeply rising sSFR radial profiles. Our high-redshift observations provide some of the clearest evidence for this process operating in a protocluster environment during the peak epoch of cosmic star formation.

The inside-out quenching signature we observe is also consistent with recent JWST observations of rapid quenching at cosmic noon. \citet{Park_et_al_2024} found that massive quiescent galaxies at $z \sim 2$ show evidence of recent rapid quenching accompanied by central starbursts, with AGN activity driving multiphase gas outflows. Similarly, \citet{Belli_et_al_2024} detected powerful neutral gas outflows in a rapidly quenching galaxy at z = 2.445, with a mass outflow rate sufficient to quench star formation, providing direct evidence for ejective AGN feedback. The enhanced AGN fraction in the Spiderweb protocluster established by \citet{Tozzi_et_al_2022b} suggests similar rapid feedback processes may be operating in our sample \citep{Perez_et_al_2023, Shimakawa_et_al_2024, Perez_et_al_2025,Shimakawa_et_al_2025, Travascio_et_al_2025}.

Our pixel-by-pixel observations at $z \sim 2.16$ are consistent with local integral field spectroscopy studies. \citet{Cano_Diaz_et_al_2019} analyzed $\sim$2000 MaNGA galaxies at $z \sim 0$, finding that the spatially resolved $\Sigma_{\rm SFR}$-$\Sigma_*$ diagram exhibits clear bimodality: star-forming areas follow a tight local SFMS (slope $\approx 0.94$, scatter $\approx 0.27$ dex), while retired areas lie $\sim 1.5$ dex below. Critically, the local SFMS segregates by morphology, with star-forming regions in earlier-type galaxies showing systematically lower $\Sigma_{\rm SFR}$ at fixed $\Sigma_*$ compared to late-type systems. This morphological segregation parallels our observation that high-Sérsic index galaxies exhibit suppressed $\Sigma_{\rm SFR}$ at fixed $\Sigma_*$. While absolute sSFR values are approximately an order of magnitude lower at $z \sim 0$ (as expected from cosmic star formation evolution), the pattern persists: positive sSFR gradients in quiescent/early-type systems versus flat profiles in star-forming/late-type galaxies, demonstrating that inside-out quenching operates consistently from cosmic noon to the present. \citet{Gonzalez_Delgado_et_al_2016} extended this analysis across the full Hubble sequence using CALIFA, finding that sSFR(R) increases radially outward with steeper slopes in the inner regions, suggesting galaxies are quenched inside-out with this process proceeding faster in central bulge-dominated regions than in disks. Their result that spirals show declining $\Sigma_{\rm SFR}$(R) profiles with values at $R_{\rm e}$ of $\sim 20~M_{\odot}~{\rm Gyr}^{-1}~{\rm pc}^{-2}$ (within a factor of two across Hubble types) provides a quantitative anchor for the local SFMS that, when compared to our observed central suppression in massive quiescent systems, confirms the inside-out nature of the quenching process across $>$10 Gyr of cosmic history (see also \citealt{Sanchez_2020} for a comprehensive review of spatially resolved properties from CALIFA and MaNGA).

\subsection{Physical Mechanisms and Environmental Context}

Our analysis reveals systematic relationships between galaxy structure, stellar mass, and star formation activity that provide observational constraints on quenching processes. We observe a strong anti-correlation between S\'ersic index and central specific SFR, with galaxies exhibiting higher S\'ersic indices (bulge-dominated systems) systematically showing lower central sSFR values. This demonstrates that morphological concentration is linked to star formation suppression \citep{Bluck_et_al_2014, Teimoorinia_et_al_2016, Bluck_et_al_2020}.

The systematic differences in the $\Sigma_{\mathrm{SFR}}$-$\Sigma_*$ relationship as a function of Sérsic index provide observational evidence for morphology-dependent star formation physics. Disk-dominated systems show enhanced $\Sigma_{\rm SFR}$ at fixed $\Sigma_*$, while bulge-dominated systems exhibit systematically reduced $\Sigma_{\rm SFR}$. These results are consistent with the morphological quenching framework proposed by \citet{Martig_et_al_2009}, where the growth of a stellar bulge can stabilize gas disks against fragmentation and suppress star formation. However, given that bulge mass tightly correlates with supermassive black hole mass \citep{Haring_et_al_2004, Kormendy_et_al_2013}, the observed morphological dependencies likely reflect both disk stabilization effects and the underlying role of AGN feedback. \citet{Fang_et_al_2013} demonstrated that dense central cores are necessary but not sufficient for complete quenching, consistent with a two-step process involving both bulge-driven stabilization and AGN feedback. These results complement extensive work on $\Sigma_*$ thresholds for quenching \citep{Barro_et_al_2017, Whitaker_et_al_2017}.

The mass threshold of $\log(M_*/M_{\odot}) \geq 10.5$ where we predominantly observe inside-out quenching aligns with the transition where feedback mechanisms become increasingly important in massive galaxies \citep{Croton_et_al_2006, Schawinski_et_al_2007, Fabian_2012}. While our sample exhibits strong correlations between stellar mass and evolutionary state, both factors appear to influence the presence of inside-out quenching signatures. Notably, two massive star-forming galaxies (ID 335 and ID 551) also exhibit clear radial sSFR gradients, suggesting they may represent systems in early phases of the quenching process. The systematic differences between star-forming and quiescent populations - flat sSFR profiles in lower-mass galaxies and strong radial gradients in massive systems—aligns with theoretical predictions that different physical mechanisms dominate in different mass regimes \citep{Bower_et_al_2006,Dubois_et_al_2013}. At the lower mass range, stellar feedback appears sufficient to maintain uniform star formation across galactic disks. Above this threshold, additional feedback mechanisms become increasingly important, leading to the inside-out quenching patterns we observe.

Recent theoretical work using the TNG50 simulation \citep{Nelson_et_al_2021} has shown that inside-out quenching is caused by AGN feedback, with remarkable agreement between simulated and observed sSFR profiles. The consistency between our observations and these state-of-the-art simulations provides strong support for AGN-driven inside-out quenching as a dominant mechanism in massive galaxies during cosmic noon. \citet{Sanchez_et_al_2018} demonstrated that MaNGA AGN hosts at $z \sim 0$ exhibit similar radial sSFR gradients with centrally suppressed star formation accompanied by molecular gas deficits and decreased star formation efficiency, confirming AGN feedback as a persistent quenching mechanism across $>10$ Gyr of cosmic history. Recent JWST observations have demonstrated that massive quiescent galaxies at $z \sim 2$ show evidence of rapid quenching accompanied by powerful outflows \citep{Park_et_al_2024, Belli_et_al_2024}.

An alternative physical mechanism capable of producing inside-out quenching is the compaction process proposed by \citet{Dekel_Burkert_2014}, in which violent disk instabilities drive gas inflows that lead to the formation of a compact stellar core accompanied by an intense central starburst. This compaction phase, examined in detail through simulations by \citet{Tacchella_et_al_2016b}, results in the rapid consumption of the central gas reservoir through the starburst itself, leading to the gradual cessation of star formation in the galaxy center while the outer disk may continue forming stars. The compaction scenario naturally produces inside-out quenching as a consequence of the centrally concentrated gas consumption and subsequent feedback from the compact stellar core and central AGN activity. The high S\'ersic indices and concentrated stellar mass distributions we observe in our quiescent galaxy population are consistent with galaxies that have undergone such compaction, though our current observations cannot definitively distinguish between compaction-driven quenching and other mechanisms that produce similar morphological and star formation signatures. The combination of central star formation suppression, elevated stellar mass surface densities, and bulge-dominated morphologies in our massive quiescent systems is consistent with the expected outcomes of the compaction process operating during the peak epoch of cosmic star formation.

While our analysis does not reveal strong dependence on distance from the protocluster center, likely due to the relatively compact projected scale of our current JWST observations, the enhanced quiescent fraction and AGN activity in the Spiderweb system suggest that the dense environment plays an important role in facilitating the physical processes that drive quenching. \citet{naufalRevealingQuiescentGalaxy2024} reported a quiescent fraction of $\sim 50\%$ for massive galaxies in the Spiderweb protocluster, significantly higher than field galaxies at similar redshifts. Recent molecular gas observations of $H\alpha$ emitters \citep{Perez_et_al_2025}  identified a sharp transition in gas fractions with most massive galaxies hosting AGN activity and exhibiting depletion timescales of 1--3 Gyr that coincide with the expected timeline for red sequence establishment. The recent detection of hot intracluster gas through the thermal Sunyaev-Zel'dovich effect \citep{Di_Mascolo_et_al_2023} confirms the presence of environmental conditions that can influence galaxy evolution.

Environmental effects in protoclusters operate differently than in fully virialized clusters. Recent observations suggest that environmentally-specific processes such as ram pressure stripping do not operate efficiently in high-redshift protoclusters \citep{Dannerbauer_et_al_2017}. \cite{Tadaki_et_al_2019} demonstrated that environmental effects on gas properties in protoclusters are mass-dependent, with gas accretion through cosmic filaments being accelerated in less massive galaxies while suppressed in the most massive systems \citep{Perez_et_al_2023}. Our UVJ pixel analysis supports this picture, showing that star-forming regions in quiescent massive galaxies exhibit different surface density properties compared to those in star-forming galaxies, suggesting modified star formation conditions rather than complete gas removal.
Future observations covering larger projected areas will be essential to fully characterize the radial dependence of these environmental processes across the protocluster structure.

The combination of rapid quenching timescales, enhanced AGN activity, and morphological transformation observed in this protocluster represents a crucial phase in the evolution of massive galaxies. Our results demonstrate that the physical processes responsible for creating the red sequence of massive elliptical galaxies were already operating efficiently at $z \sim 2$, during the peak epoch of cosmic star formation. However, quantifying the specific contribution of the protocluster environment to these processes requires a detailed comparison with field galaxies at similar redshifts and stellar masses, which is beyond the scope of the current analysis. Future work will address this critical question by conducting parallel pixel-by-pixel SED fitting analysis of field galaxies to isolate the environmental effects on galaxy quenching and morphological evolution.

\subsection{Spatially Resolved UVJ Analysis and Quenching Heterogeneity}

Our spatially resolved UVJ color analysis, based on the selection criteria of \citet{williamsRestFrameUltravioletColors2009}, reveals important insights into the heterogeneous nature of star formation cessation within individual galaxies. The finding that star-forming pixels within quiescent host galaxies occupy distinct regions of the $\Sigma_*$ versus $\Sigma_{\rm SFR}$ parameter space compared to star-forming pixels in globally star-forming galaxies indicates that the local physical conditions governing star formation differ significantly between these two systems.

This result suggests that regions maintaining residual star formation activity in otherwise quiescent galaxies operate under fundamentally different physical conditions than those in actively star-forming systems. The systematically lower star formation efficiency observed in star-forming pixels within quiescent hosts may reflect the influence of the evolved internal galactic conditions, including reduced gas availability, altered gas thermodynamics, or modified feedback processes following the initial quenching event.

Conversely, our observation that quiescent pixels exhibit remarkably similar surface density properties regardless of their host galaxy type provides important constraints on quenching physics. This similarity implies that once a region achieves quiescent colors through local star formation cessation, its subsequent evolution becomes largely independent of the global galaxy state. This finding supports theoretical models in which quenching represents a threshold-driven phase transition rather than a gradual process dependent on global galaxy properties \citep{Peng_et_al_2010,Peng_et_al_2012}.

The UVJ-based classification of pixels also reveals the spatial structure of the quenching process, with quiescent galaxies showing strong radial gradients in pixel classifications that mirror the sSFR trends. The systematic increase in star-forming pixel fractions from $\sim 10\%$ in galaxy centers to $\sim 50\%$ in outer regions provides independent confirmation of inside-out quenching patterns and demonstrates the utility of color-based diagnostics for tracing spatially resolved star formation histories \citep{Breda_et_al_2020, Nelson_et_al_2021}

\section{Summary and Conclusions} \label{sec:conclusions}

We have conducted a comprehensive pixel-by-pixel spectral energy distribution analysis of 38 galaxies in the Spiderweb protocluster at $z \sim 2.16$ using deep multi-wavelength imaging from \textit{JWST} NIRCam and \textit{HST} ACS/WFC3. Our spatially resolved investigation spanning rest-frame UV to near-infrared wavelengths provides new observational insights into the physical processes governing galaxy evolution during cosmic noon, the epoch of peak cosmic star formation. The main findings and conclusions of this study are summarized below.

Inside-Out Quenching: We find clear evidence for inside-out quenching primarily in quiescent galaxies, which in our sample are predominantly found at $\log(M_*/\mathrm{M}_{\odot}) \geq 10.5$. These systems exhibit systematic radial gradients in sSFR, characterized by central sSFR suppression by approximately one order of magnitude relative to their outer regions, with mean values increasing from $9.4 \times 10^{-11}$ yr$^{-1}$ at $R/R_{\mathrm{e}} = 0.5$ to $1.4 \times 10^{-9}$ yr$^{-1}$ at $R/R_{\mathrm{e}} = 3.0$. Notably, two massive star-forming galaxies also exhibit inside-out quenching signatures, suggesting they represent systems in transition to quiescence. In contrast, star-forming galaxies show relatively flat sSFR profiles with minimal radial variation, indicating sustained star formation activity across their entire extent.

Morphological Regulation of Star Formation: Our analysis reveals a fundamental anti-correlation between Sérsic index and central star formation activity. However, given that bulge mass tightly correlates with supermassive black hole mass, AGN feedback emerges as the more practical mechanism for quenching star formation rather than morphological quenching alone. Modern simulations demonstrate that AGN feedback is essential to reproduce abundant high-redshift quiescent galaxies, with the observed morphological dependencies potentially reflecting the underlying physical processes that regulate star formation in massive systems.

Spatially Resolved Color Evolution: Through UVJ color analysis at the pixel level, we identify heterogeneous star formation states within individual galaxies. Star-forming pixels in quiescent host galaxies occupy systematically different regions of parameter space compared to those in globally star-forming systems, while quiescent pixels show similar properties regardless of host galaxy type. This asymmetry provides new constraints on the multi-phase nature of galaxy quenching processes.

The combination of inside-out quenching patterns, morphological dependencies, and enhanced AGN activity points toward a scenario where multiple physical processes work in concert to regulate star formation. Our results support models where AGN feedback drives central quenching in systems undergoing the transition to quiescence, with this process becoming increasingly prevalent above $M_* \sim 10^{10.5} M_{\odot}$, while morphological quenching through disk stabilization provides a complementary mechanism linking galaxy structure to star formation cessation.

The spatially resolved approach employed in this study demonstrates that sub-galactic analysis is essential for understanding galaxy evolution processes that would be obscured in integrated measurements. Our pixel-by-pixel methodology, enabled by the exceptional spatial resolution and depth of \textit{JWST} observations, reveals the complex spatial structure of star formation regulation that underlies global galaxy properties.

Our results demonstrate that cosmic noon represents a critical transition period when the fundamental processes shaping modern galaxy populations were actively operating. The clear evidence for inside-out quenching, morphological regulation of star formation, and environmental facilitation of rapid evolution provides new observational constraints on galaxy formation theory and establishes the Spiderweb protocluster as an exceptional laboratory for studying galaxy evolution during the peak epoch of cosmic star formation.

\section{Data Availability}
Some of the data presented in this article were obtained from the Mikulski Archive for Space Telescopes (MAST) at the Space Telescope Science Institute. The specific observations analyzed can be accessed via \dataset[10.17909/qt4t-dz57]{https://doi.org/10.17909/qt4t-dz57}.

\begin{acknowledgments}
This work was supported by JSPS KAKENHI grant No. J23H01219 and JSPS Core-to-Core Program (grant No.: JPJSCCA20210003). TK acknowledges financial support from JSPS KAKENHI Grant Numbers 24H00002 (Specially Promoted Research by T. Kodama et al.), 22K21349 (International Leading Research by S. Miyazaki et al.), and JSPS Core-to-Core Program (JPJSCCA20210003; M. Yoshida et al.). HD and JMPM acknowledge support from the Agencia Estatal de Investigación del Ministerio de Ciencia, Innovación y Universidades (MCIU/AEI) under grant (Construcción de cúmulos de galaxias en formación a través de la formación estelar oscurecida por el polvo) and the European Regional Development Fund (ERDF) with reference (PID2022-143243NB-I00/10.13039/501100011033). P.G.P.-G. acknowledges support from grant PID2022-139567NB-I00 funded by Spanish Ministerio de Ciencia e Innovaci\'on MCIN/AEI/10.13039/501100011033, FEDER {\it Una manera de hacer Europa}. We thank Dr. Abdurro'uf for insightful discussions. 
\end{acknowledgments}

\appendix
\section{Point Spread Function Matching and Spatial Resolution Homogenization}
\label{app:psf}

\begin{figure*}
\centering
\includegraphics[width=0.95\textwidth]{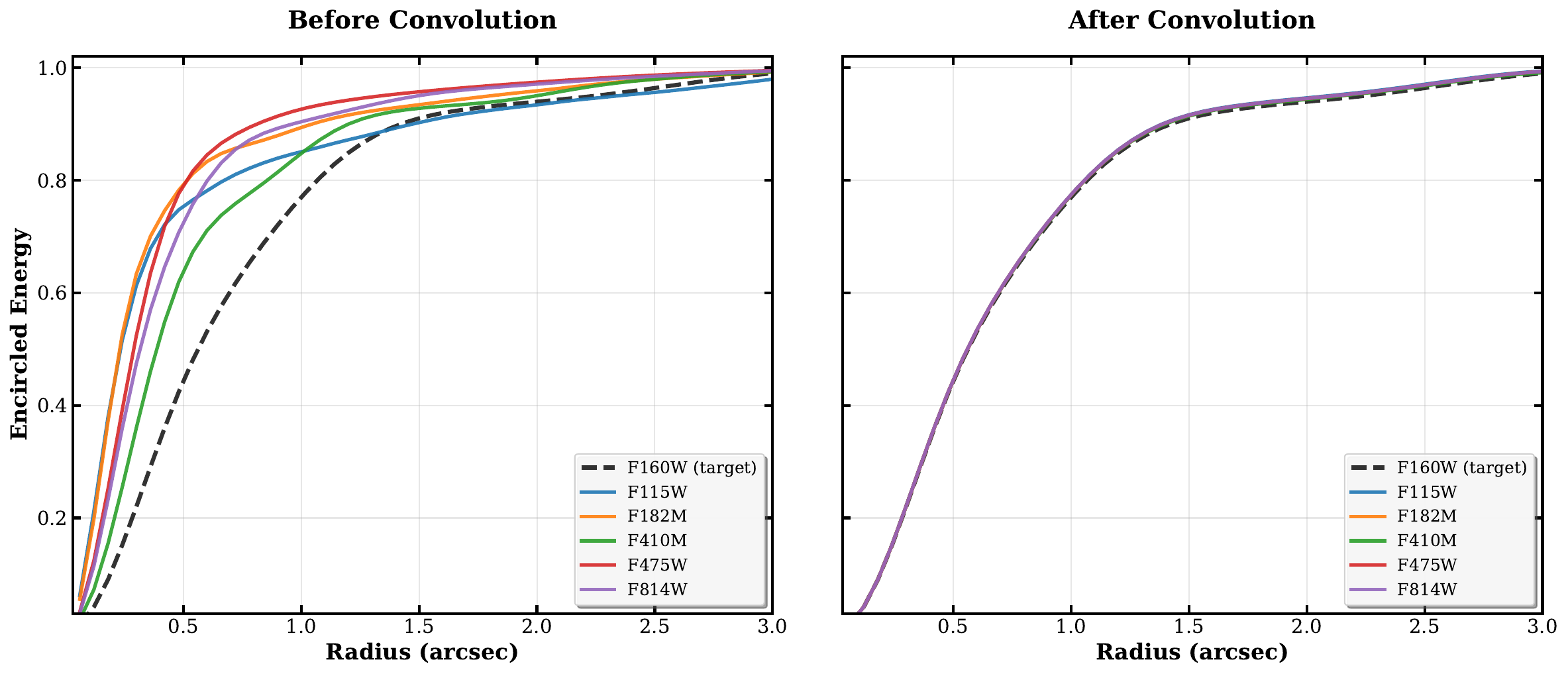}
\caption{Point Spread Function (PSF) convolution analysis for HST and JWST multi-wavelength observations of the Spiderweb protocluster. The left panel shows the original PSF profiles as a function of radius for our six-filter dataset (HST/ACS F475W, F814W; HST/WFC3 F160W; JWST/NIRCam F115W, F182M, F410M), demonstrating the instrumental and wavelength-dependent variations in spatial resolution. The right panel displays the result after PSF matching, where all bands have been convolved to match the  -seeing filter (HST/WFC3 F160W), ensuring consistent spatial resolution across all wavelengths. This PSF homogenization is essential for accurate pixel-by-pixel SED fitting, as it eliminates systematic biases that would otherwise arise from different effective beam sizes when comparing fluxes across multiple bands. The successful convergence of all profiles to a single curve in the right panel confirms the effectiveness of our PSF matching procedure, enabling robust spatially resolved analysis of galaxy properties throughout the protocluster environment at $z \approx 2.1$.}
\label{fig:psf_convolution}
\end{figure*}

The pixel-by-pixel SED fitting analysis requires careful treatment of the Point Spread Function (PSF) variations across different HST and JWST filters to ensure accurate photometric measurements and robust parameter derivation. Figure~\ref{fig:psf_convolution} illustrates our PSF matching procedure, which is critical for maintaining consistent spatial resolution across all wavelengths used in the analysis. The left panel demonstrates the inherent variations in the PSF profiles for our multi-instrument dataset (HST/ACS F475W, F814W; HST/WFC3 F160W; JWST/NIRCam F115W, F182M, F410M), with different instruments and wavelengths exhibiting varying profile shapes due to the diffraction-limited nature of JWST and the optical characteristics of HST. To eliminate systematic biases in our spatially resolved analysis, we convolved all images to match the worst-seeing filter (HST/WFC3 F160W), as shown in the right panel where all PSF profiles converge to a single curve. This PSF homogenization ensures that flux measurements at each pixel position represent the same physical scale across all bands, thereby enabling accurate SED fitting and reliable derivation of stellar population properties. Without this correction, the varying spatial resolution would introduce significant uncertainties in the derived stellar masses, star formation rates, and other physical parameters, particularly in regions where sharp gradients in galaxy properties exist.

\section{Stellar mass surface and SFR density map}
\label{app:map}

We present spatially resolved stellar mass surface density ($\Sigma_*$) and star formation rate surface density ($\Sigma_{\rm SFR}$) maps derived from our pixel-by-pixel SED fitting analysis for galaxies in the Spiderweb protocluster. The maps are generated for both quiescent and star-forming galaxy populations, with galaxy classifications and identifications based on \citet{naufalRevealingQuiescentGalaxy2024}. 

\begin{figure*}
\centering
% Row 1: Galaxies 258 and 369
\includegraphics[width=0.48\textwidth]{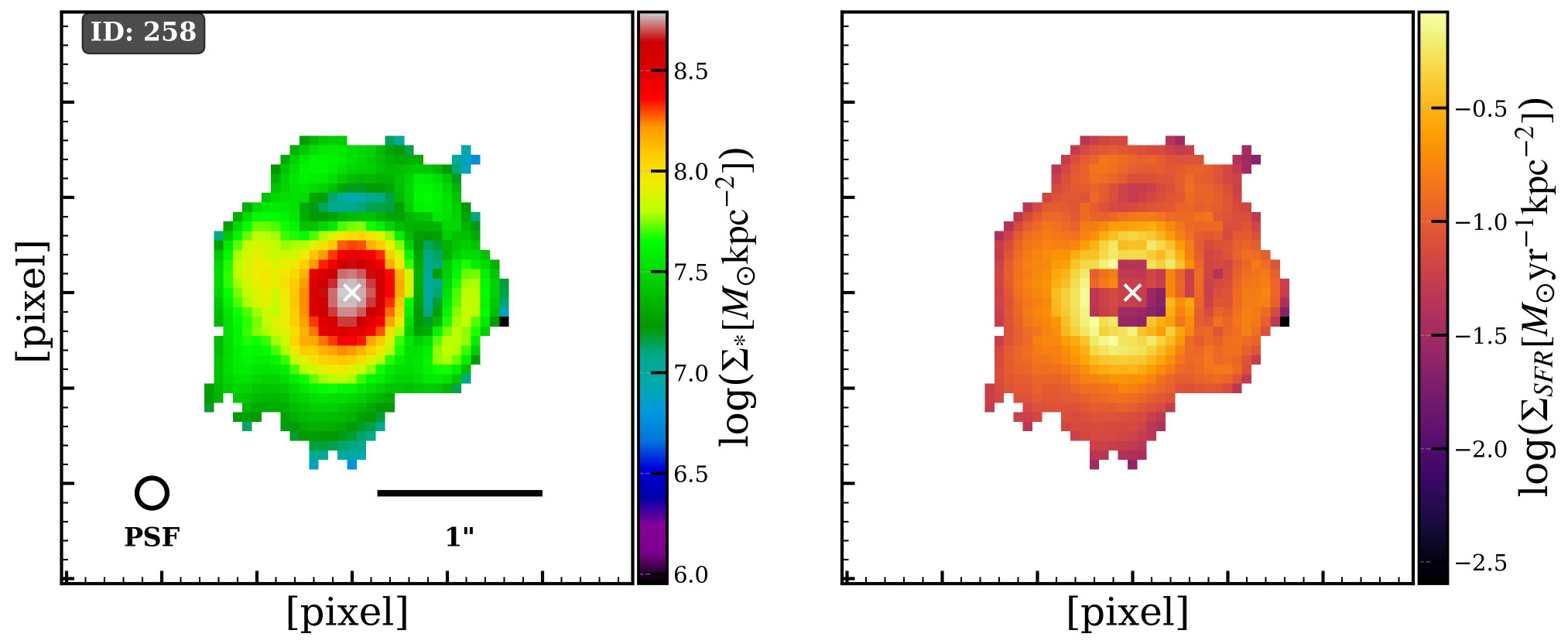}
\hfill
\includegraphics[width=0.48\textwidth]{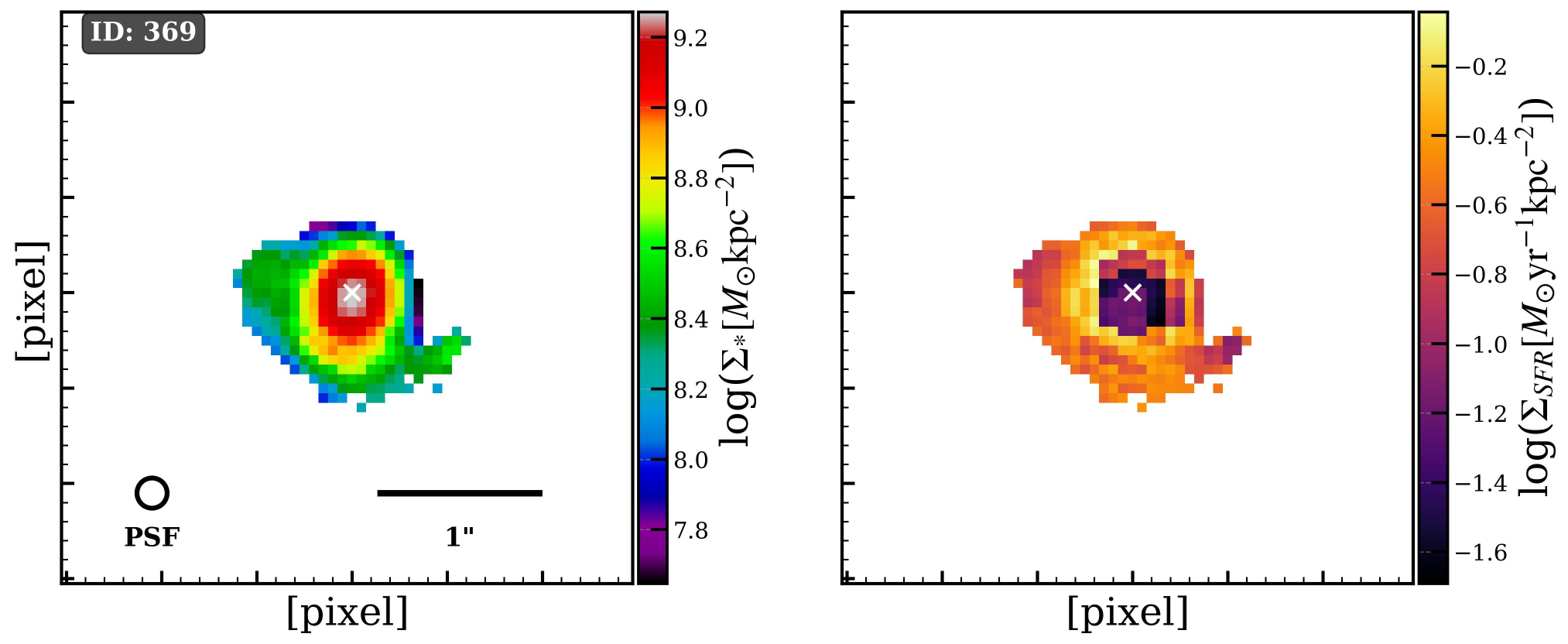}

\vspace{0.3cm}

% Row 2: Galaxies 412 and 440
\includegraphics[width=0.48\textwidth]{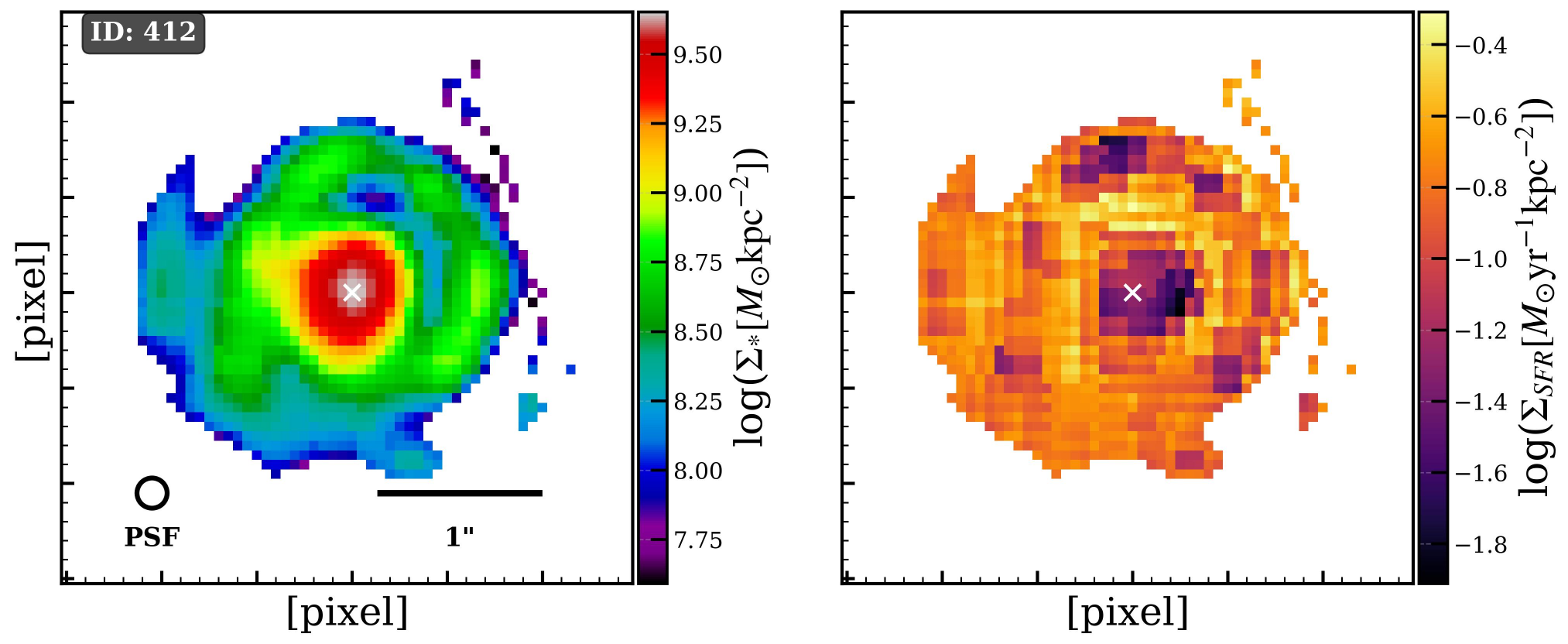}
\hfill
\includegraphics[width=0.48\textwidth]{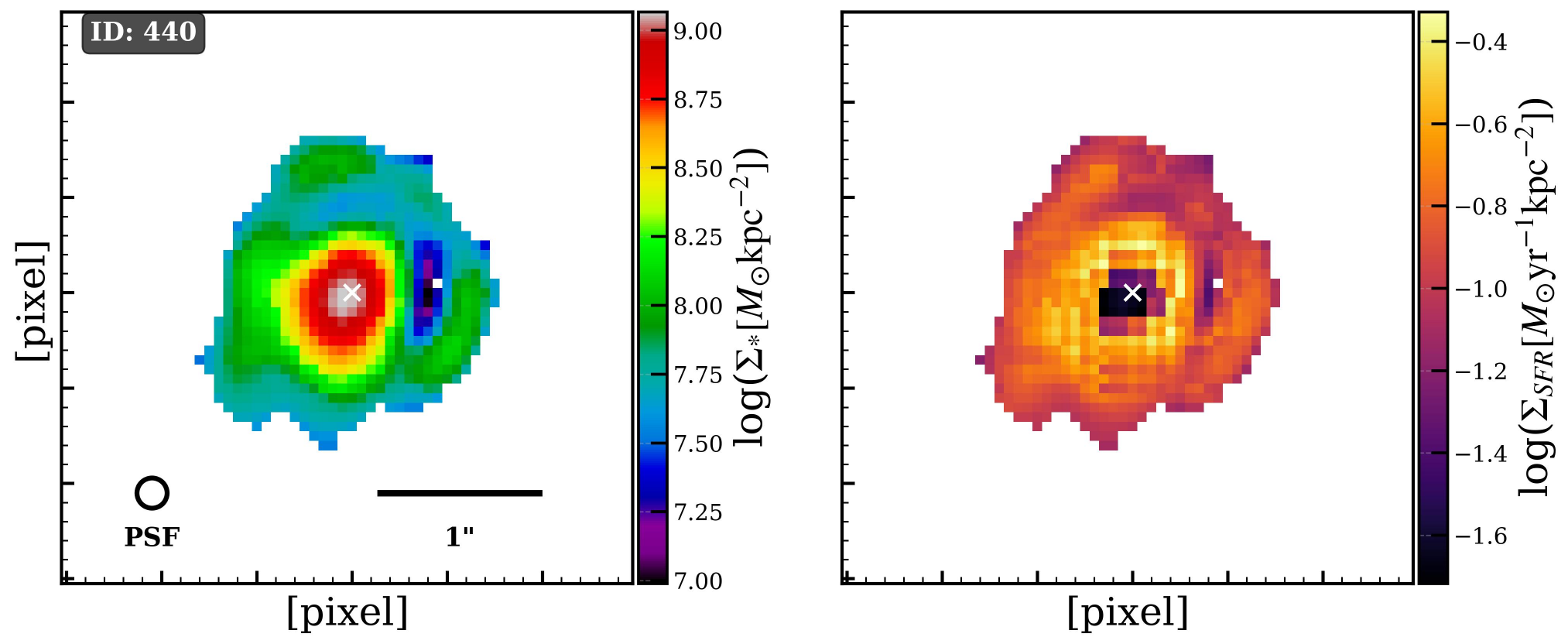}

\vspace{0.3cm}

% Row 3: Galaxies 443 and 467
\includegraphics[width=0.48\textwidth]{galaxy_maps_443_Q.pdf}
\hfill
\includegraphics[width=0.48\textwidth]{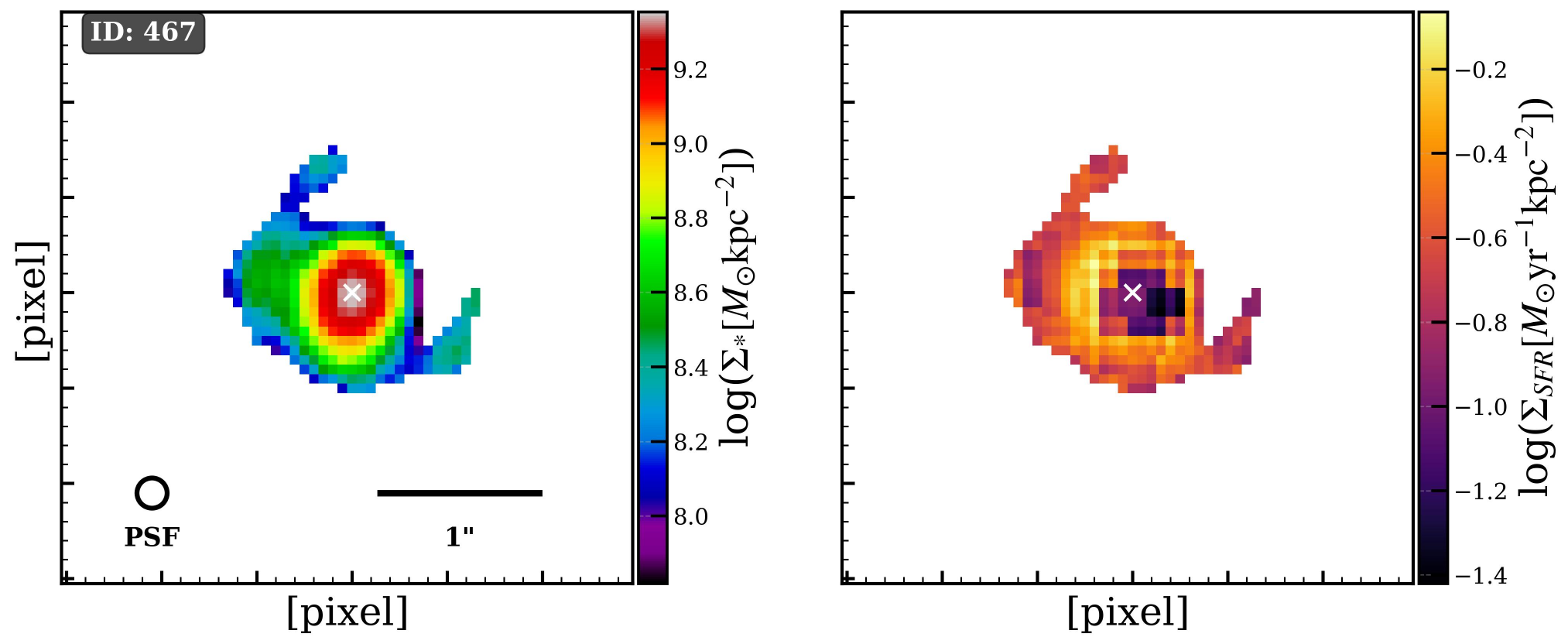}

\vspace{0.3cm}

% Row 4: Galaxies 557 and 569
\includegraphics[width=0.48\textwidth]{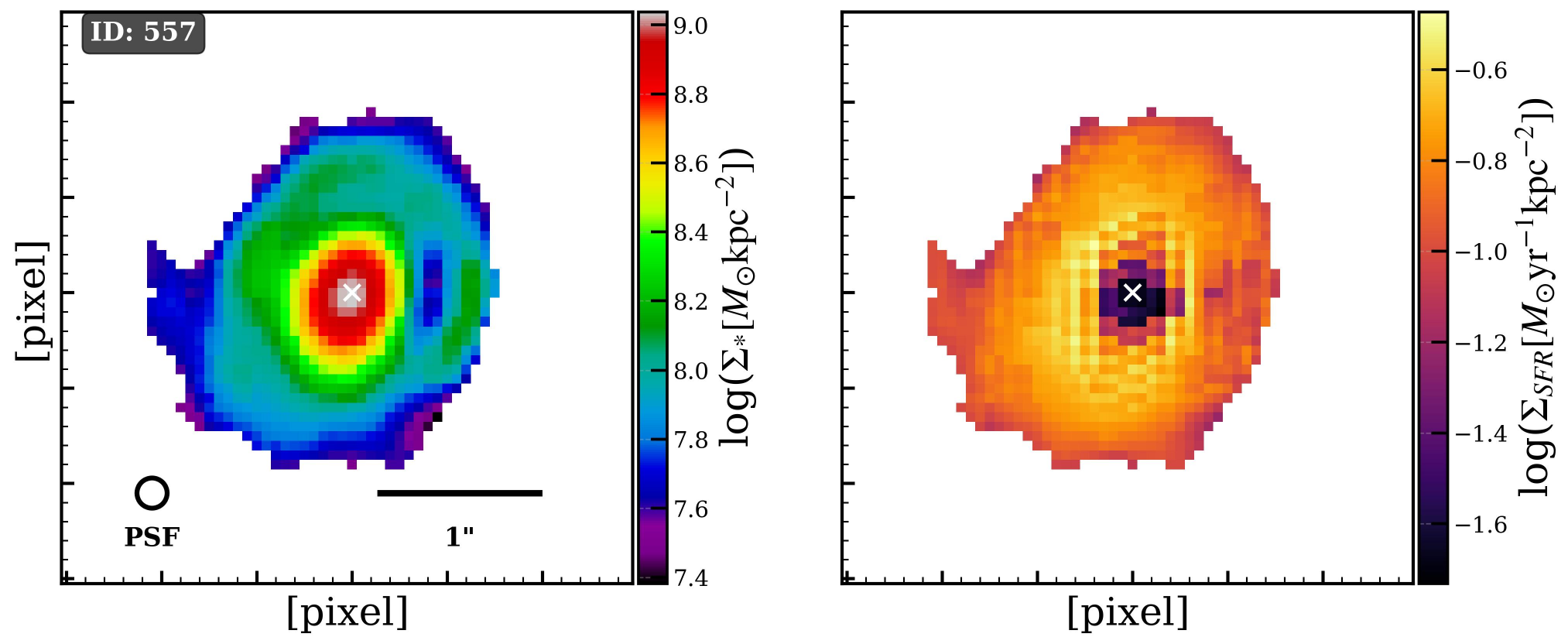}
\hfill
\includegraphics[width=0.48\textwidth]{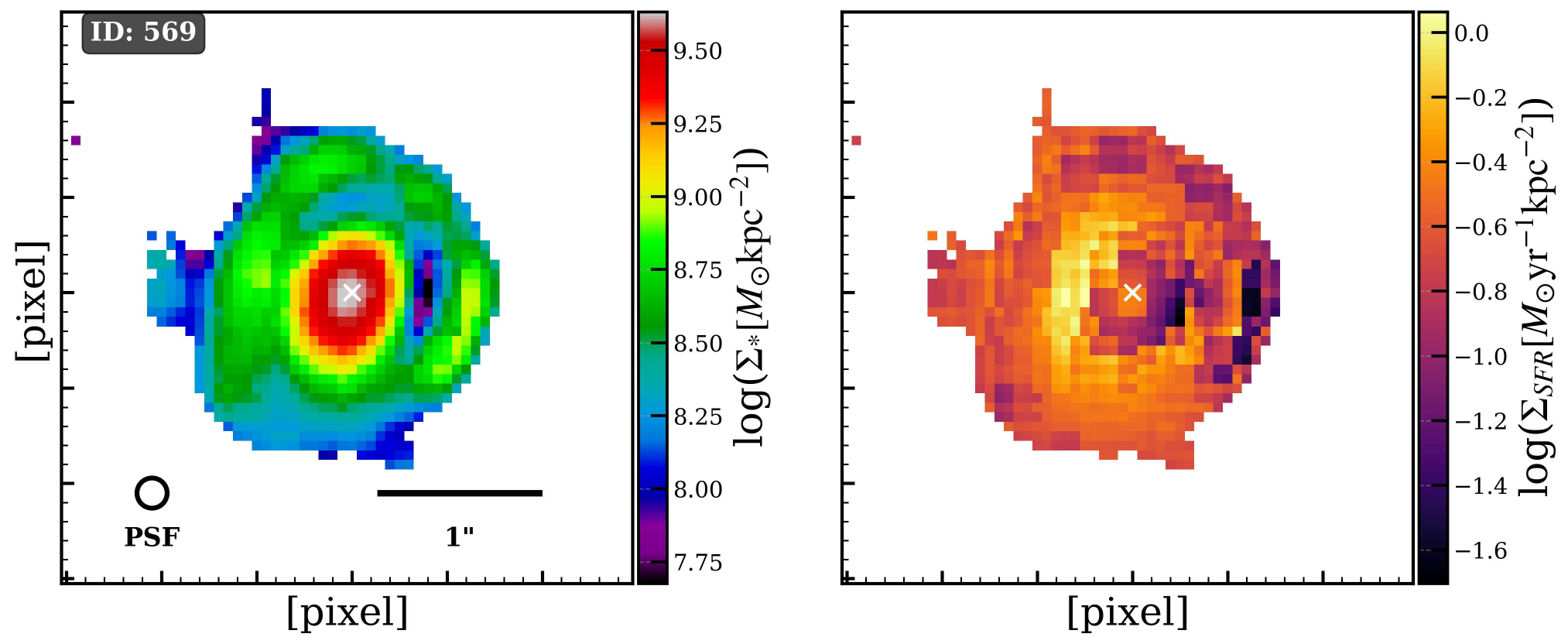}

\vspace{0.3cm}

% Row 5: Galaxies 588 and 642
\includegraphics[width=0.48\textwidth]{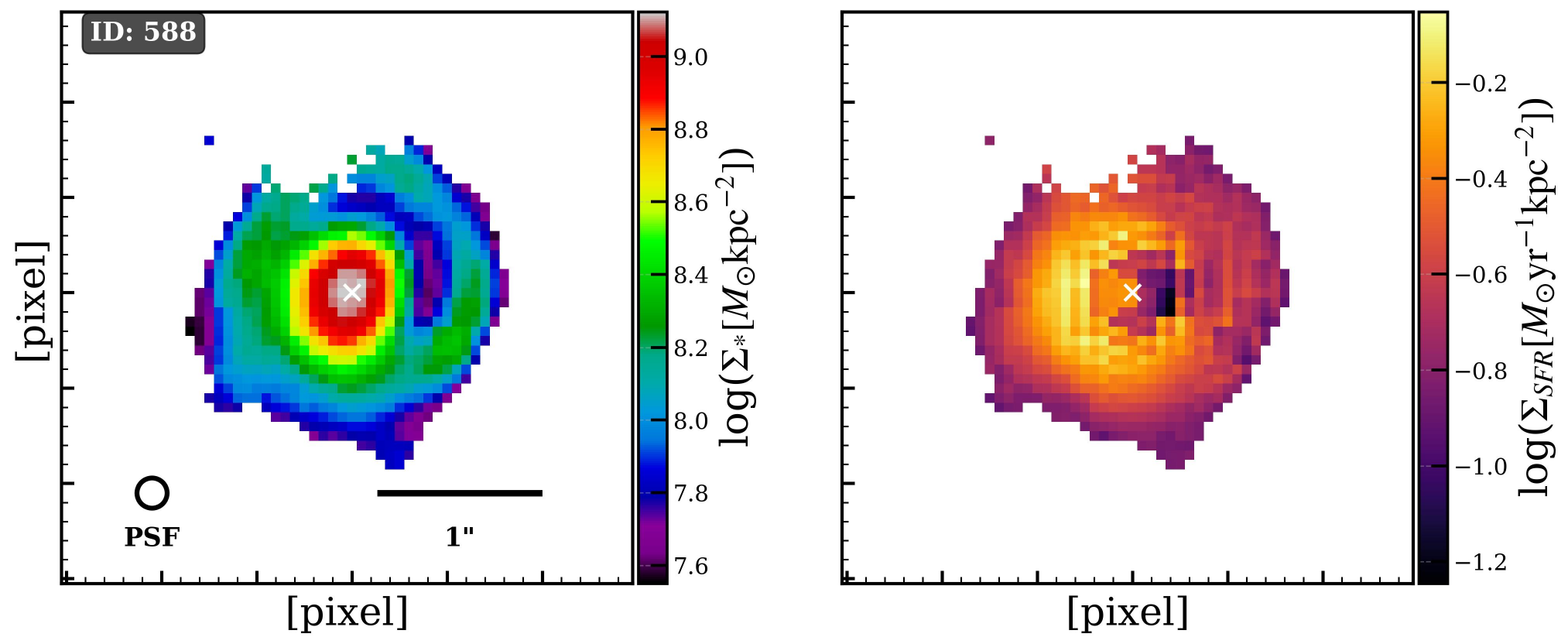}
\hfill
\includegraphics[width=0.48\textwidth]{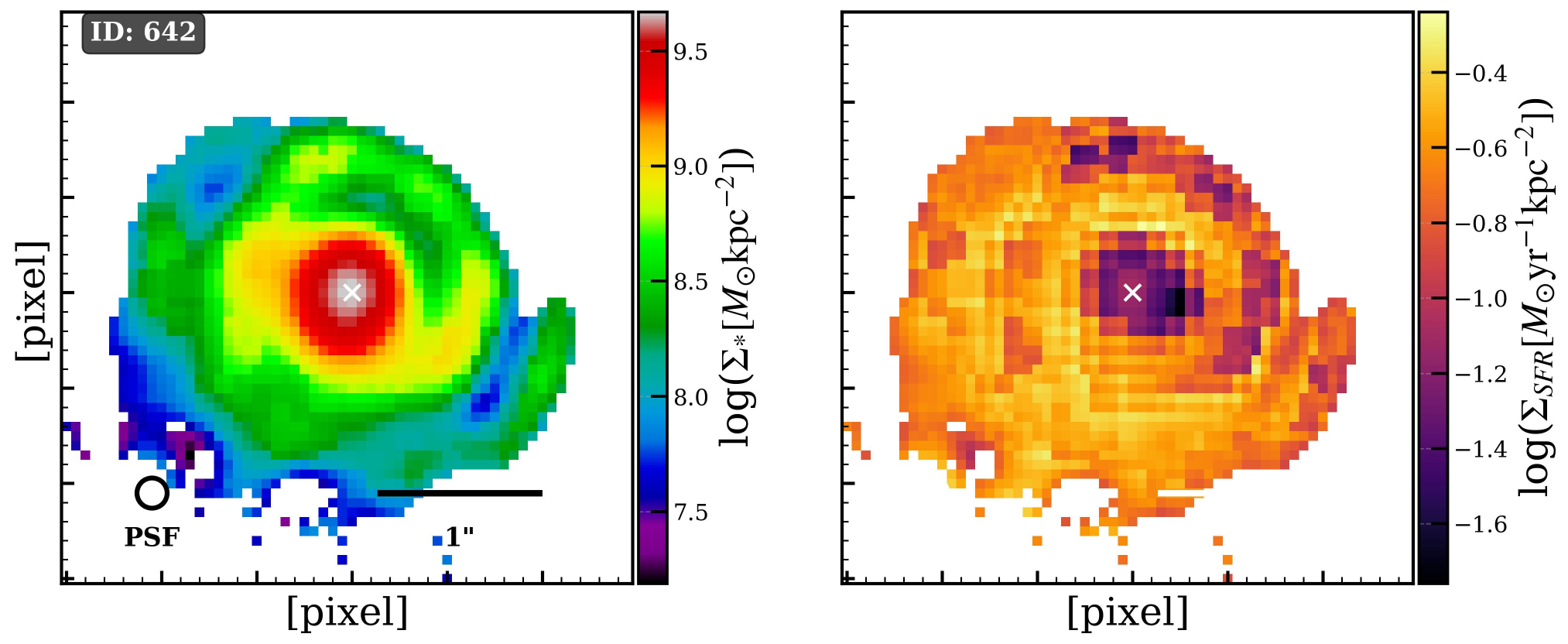}

\vspace{0.3cm}
% Row 6: Galaxy 654 (single galaxy, centered)
\begin{center}
\includegraphics[width=0.48\textwidth]{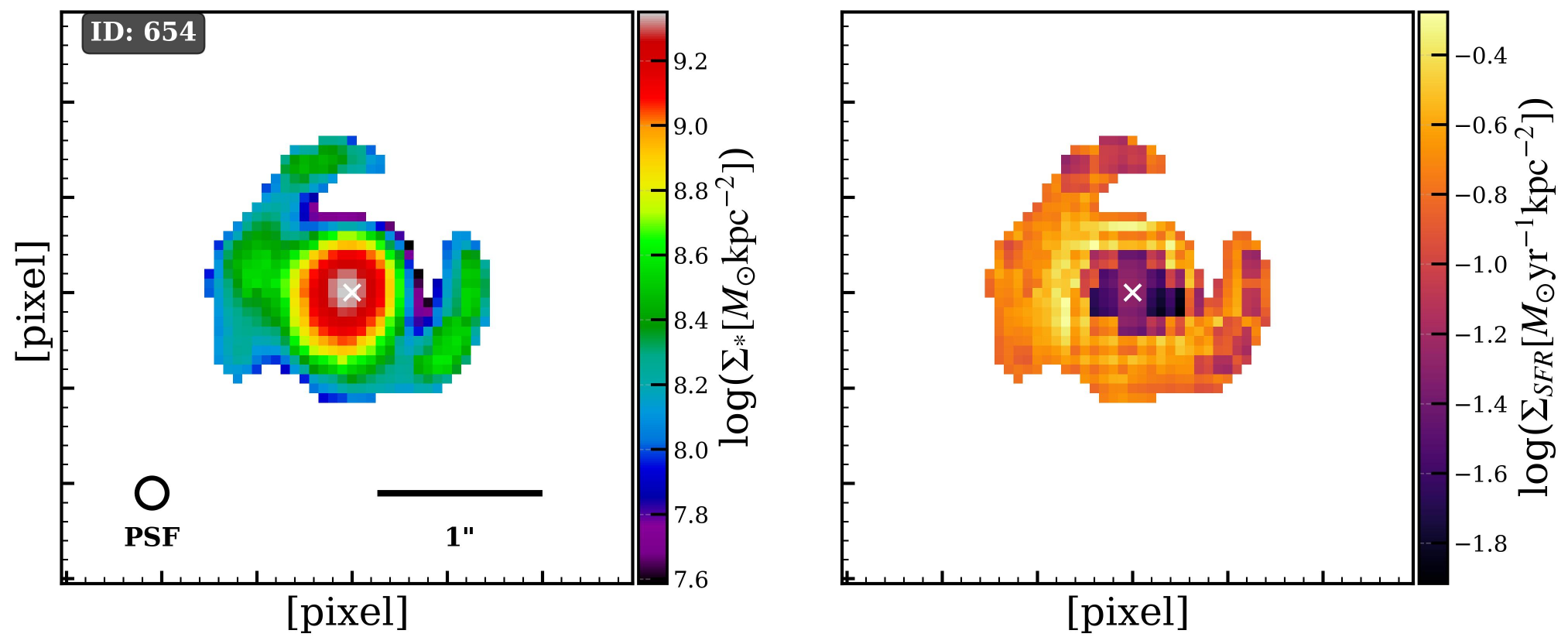}
\end{center}

\caption{$\Sigma_*$ (left panels) and $\Sigma_{\rm SFR}$ (right panels) maps for 11 quiescent galaxies. Each galaxy panel shows a 40×40 pixel region (±20 pixels from center) centered on the galaxy. The galaxy IDs correspond to those from \citet{naufalRevealingQuiescentGalaxy2024}}

\label{fig:quiescent_galaxy_maps}

\end{figure*}

Figure~\ref{fig:quiescent_galaxy_maps} shows the derived property maps for 11 quiescent galaxies, while Figures~\ref{fig:sf_galaxy_maps} present the corresponding maps for star-forming galaxies in the sample. The stellar mass surface density maps reveal the underlying stellar distribution and structural properties of each galaxy, while the SFR surface density maps trace the spatial distribution of ongoing star formation activity. These spatially resolved measurements provide insights into the internal structure and star formation patterns within individual galaxies at $z \sim 2$, enabling detailed analysis of the relationship between stellar mass distribution and star formation activity across different galaxy types in the dense protocluster environment.

%%%%%%%%%%%%%%%%%%%%%%

% Figure 8 - Part 1
\begin{figure*}
\figurenum{13}
\centering
% Row 1: Galaxies 210 and 211
\includegraphics[width=0.48\textwidth]{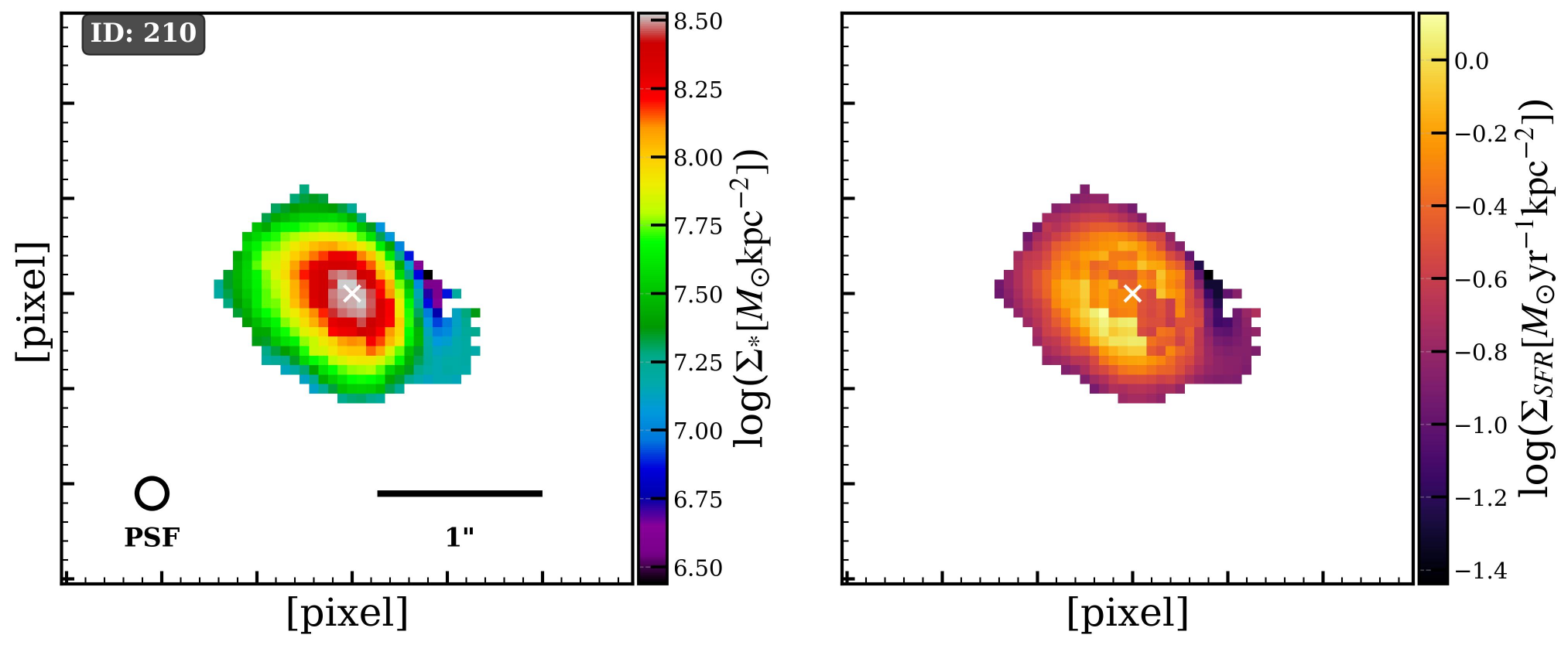}
\hfill
\includegraphics[width=0.48\textwidth]{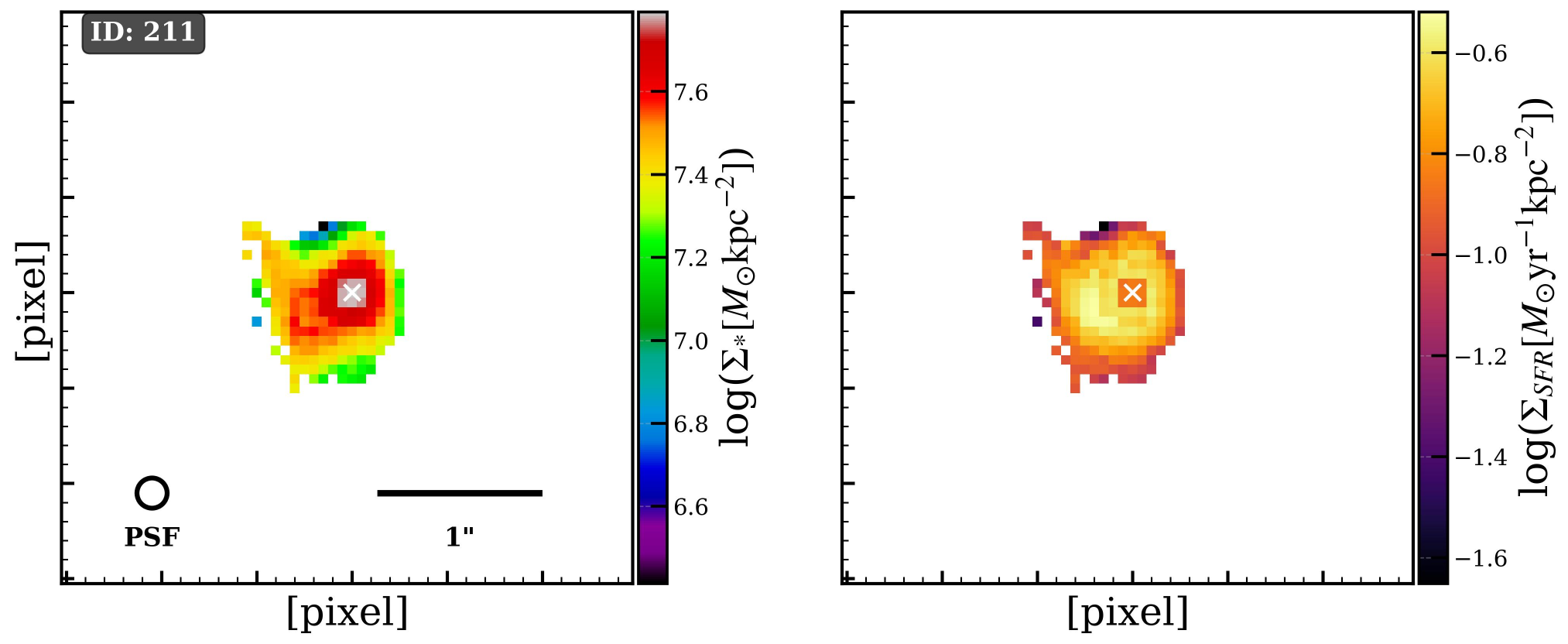}
\vspace{0.3cm}
% Row 2: Galaxies 266 and 296
\includegraphics[width=0.48\textwidth]{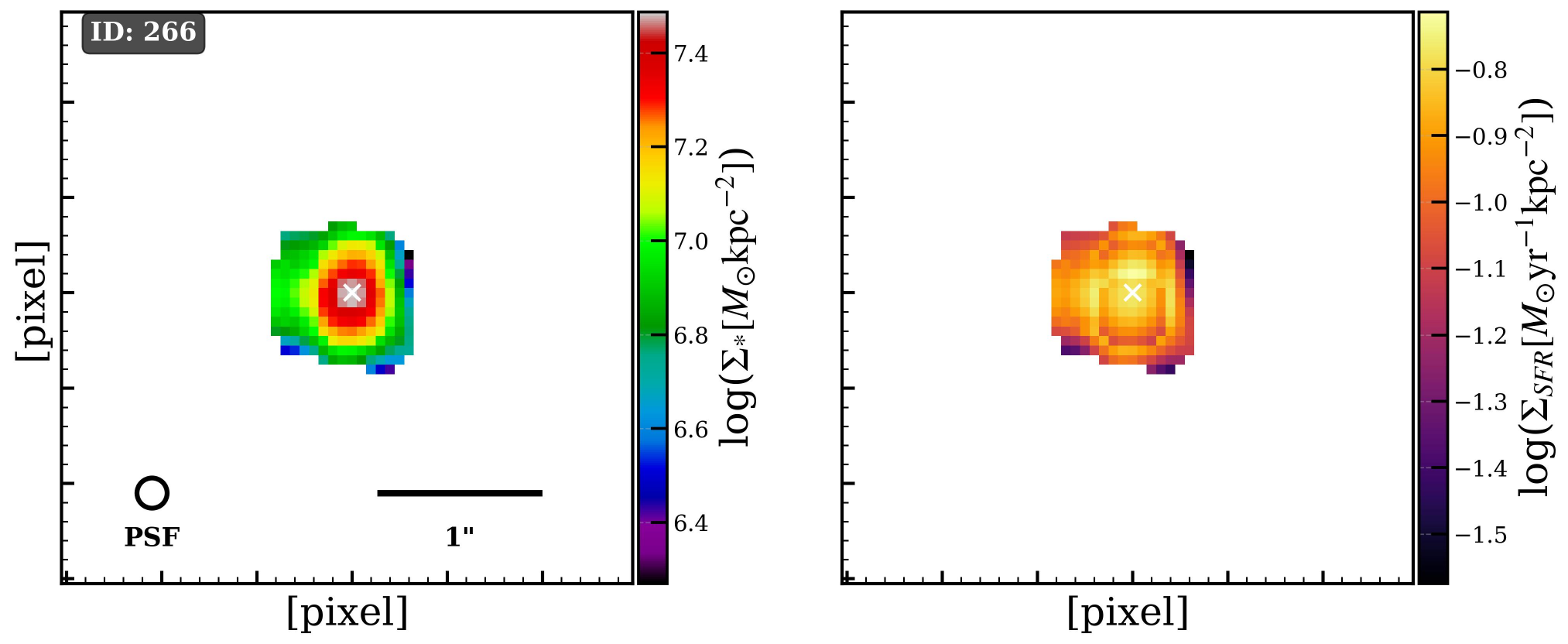}
\hfill
\includegraphics[width=0.48\textwidth]{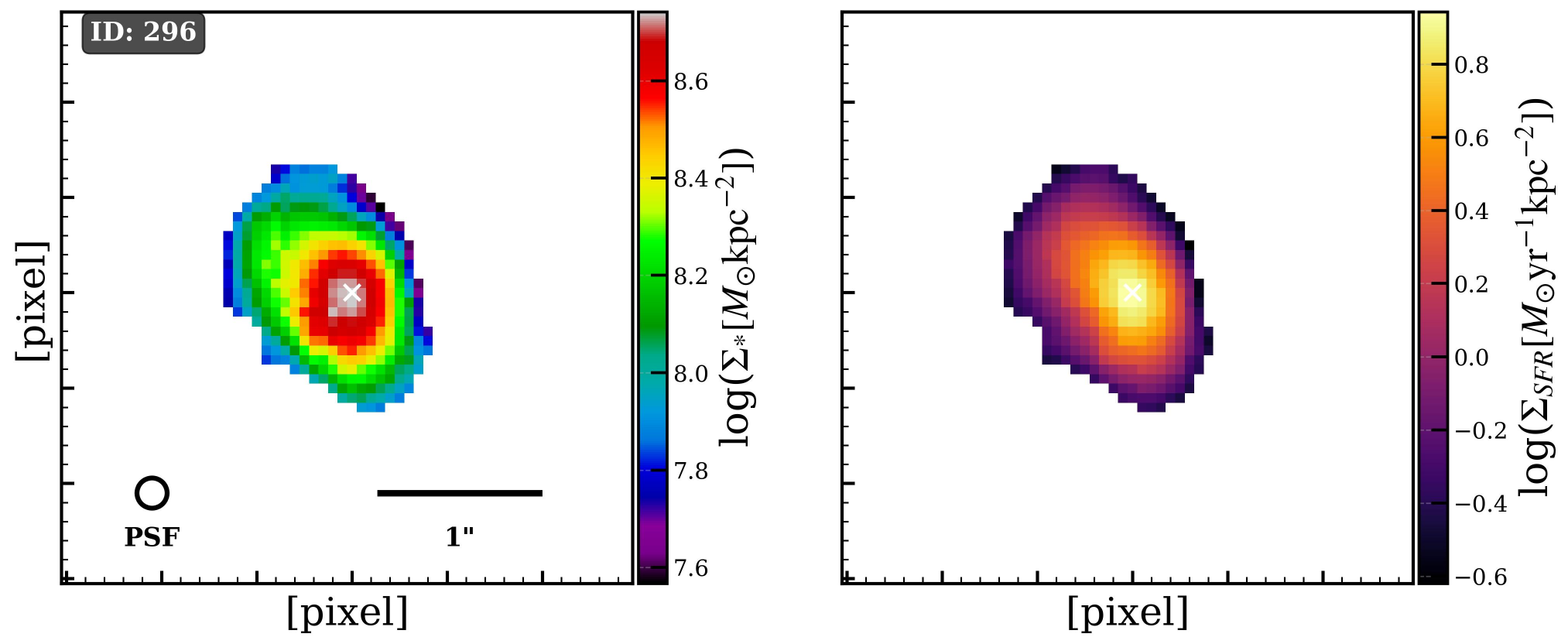}
\vspace{0.3cm}
% Row 3: Galaxies 297 and 335
\includegraphics[width=0.48\textwidth]{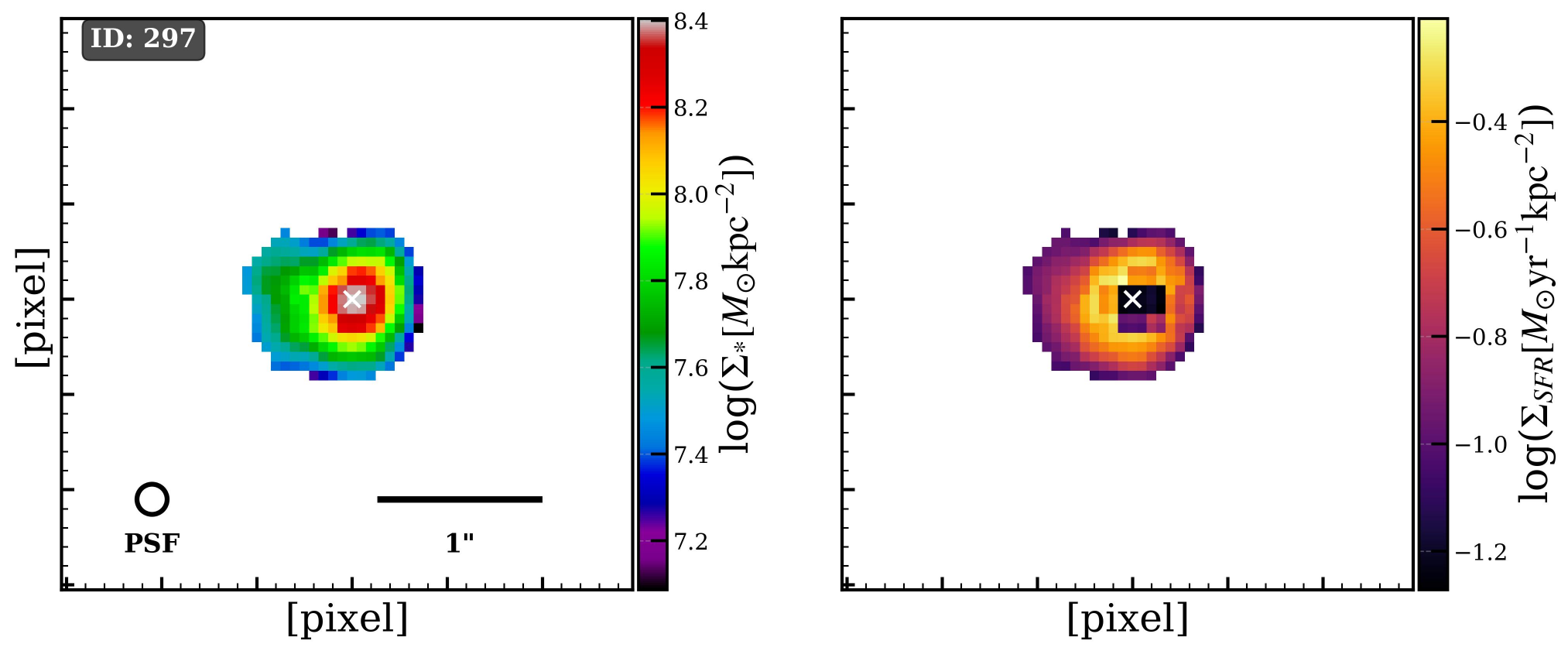}
\hfill
\includegraphics[width=0.48\textwidth]{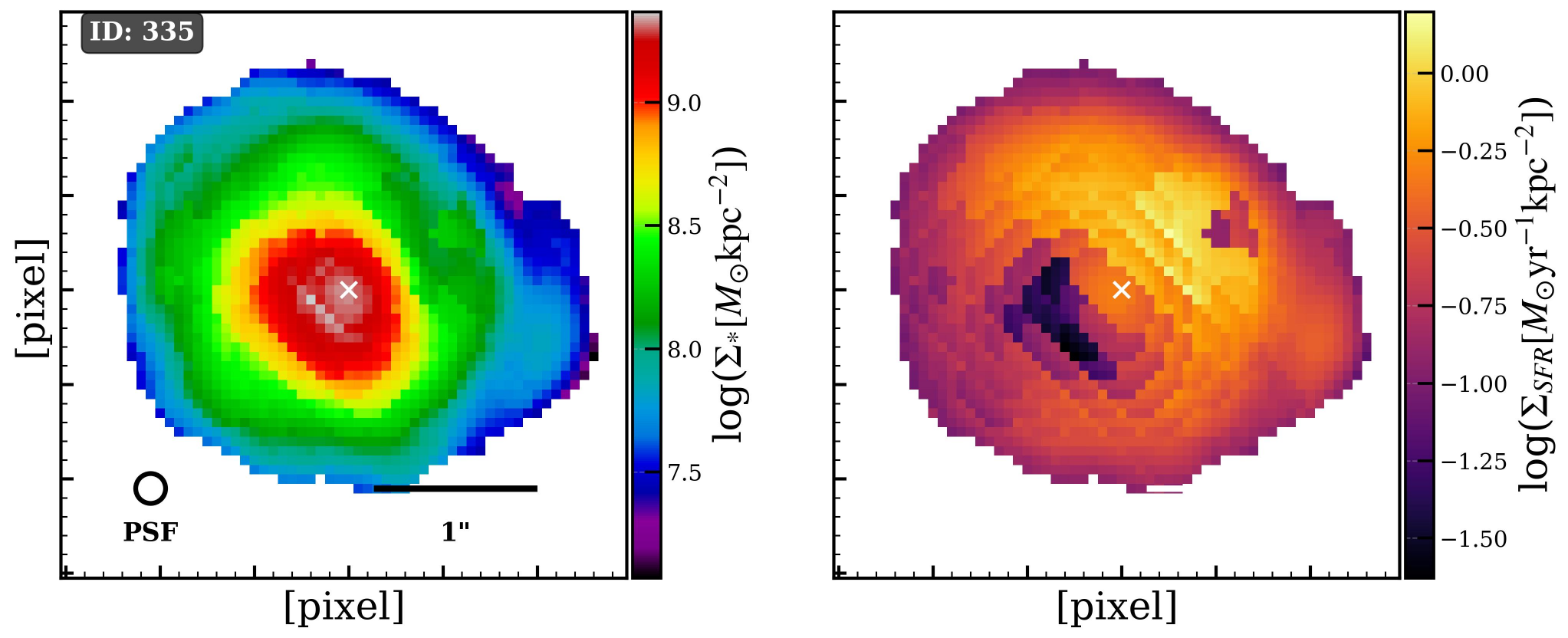}
\vspace{0.3cm}
% Row 4: Galaxies 364 and 386
\includegraphics[width=0.48\textwidth]{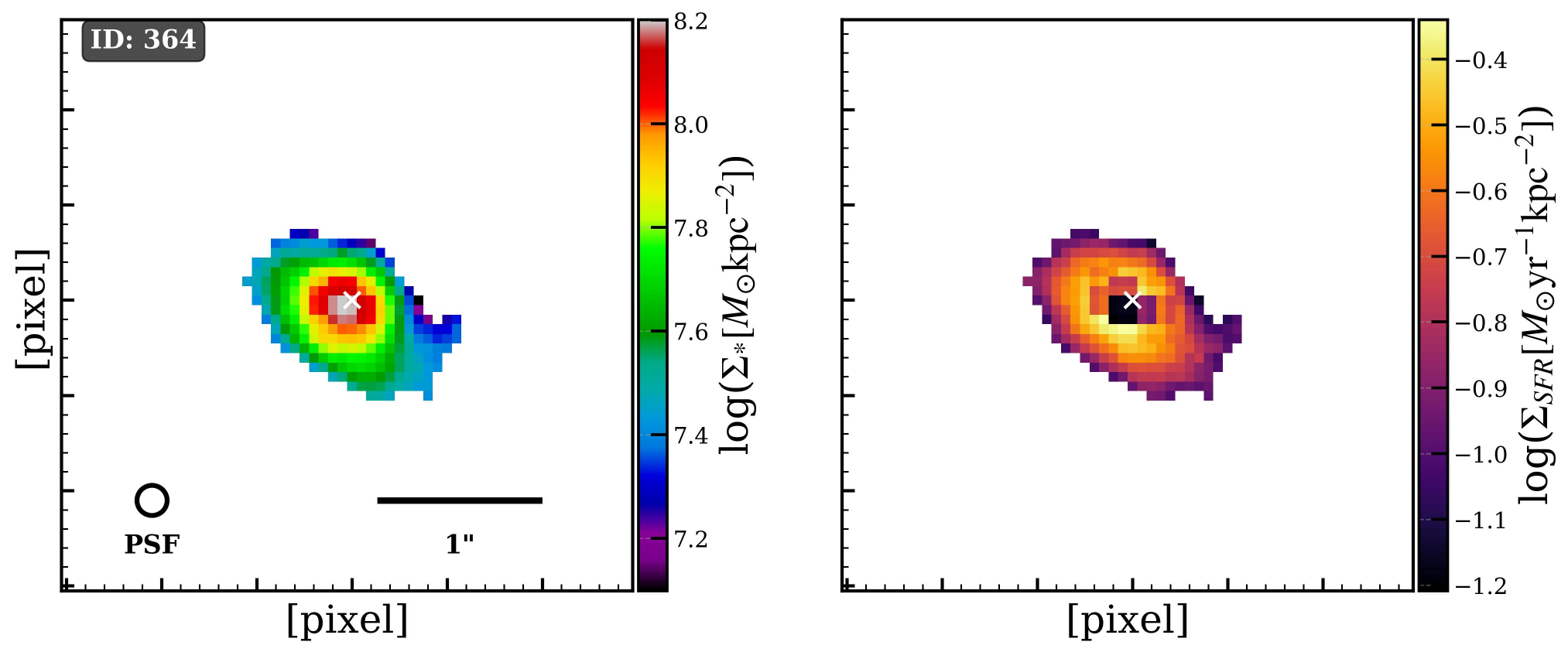}
\hfill
\includegraphics[width=0.48\textwidth]{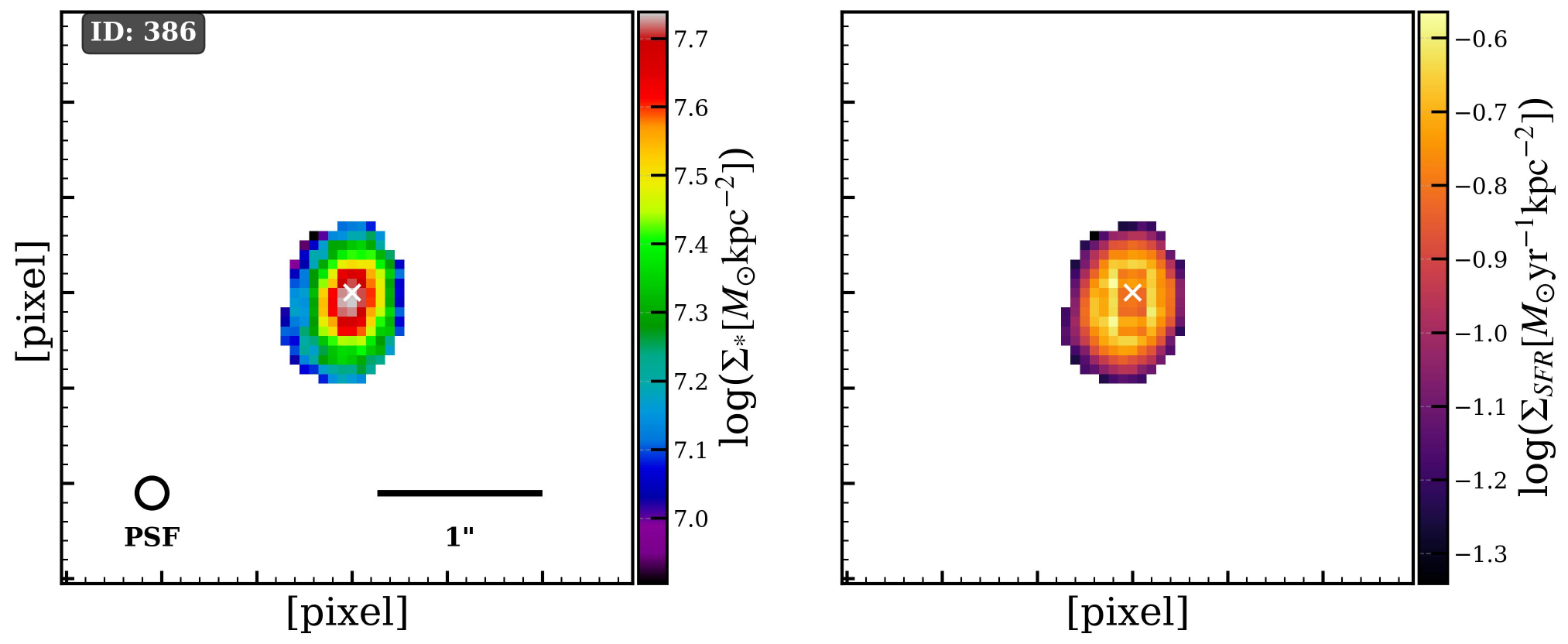}
\vspace{0.3cm}
% Row 5: Galaxy 392 (single, centered)
\begin{center}
\includegraphics[width=0.48\textwidth]{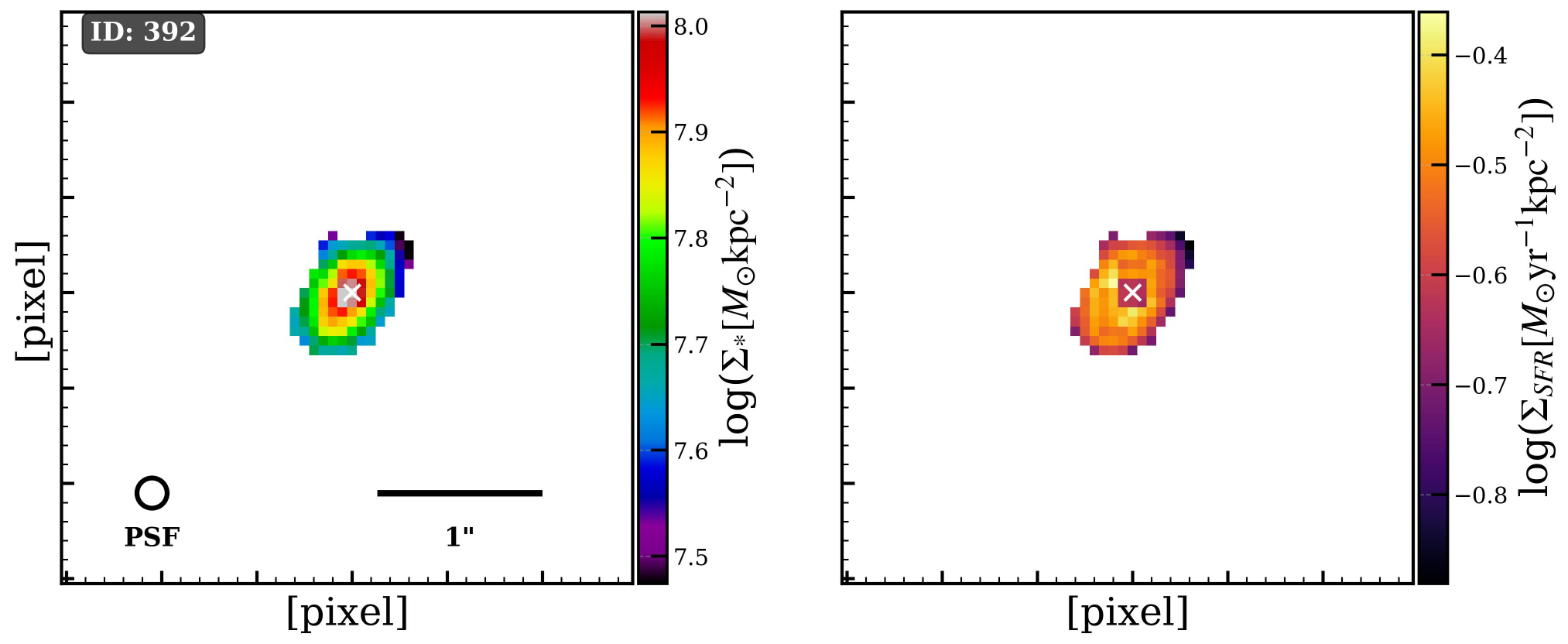}
\hfill
\includegraphics[width=0.48\textwidth]{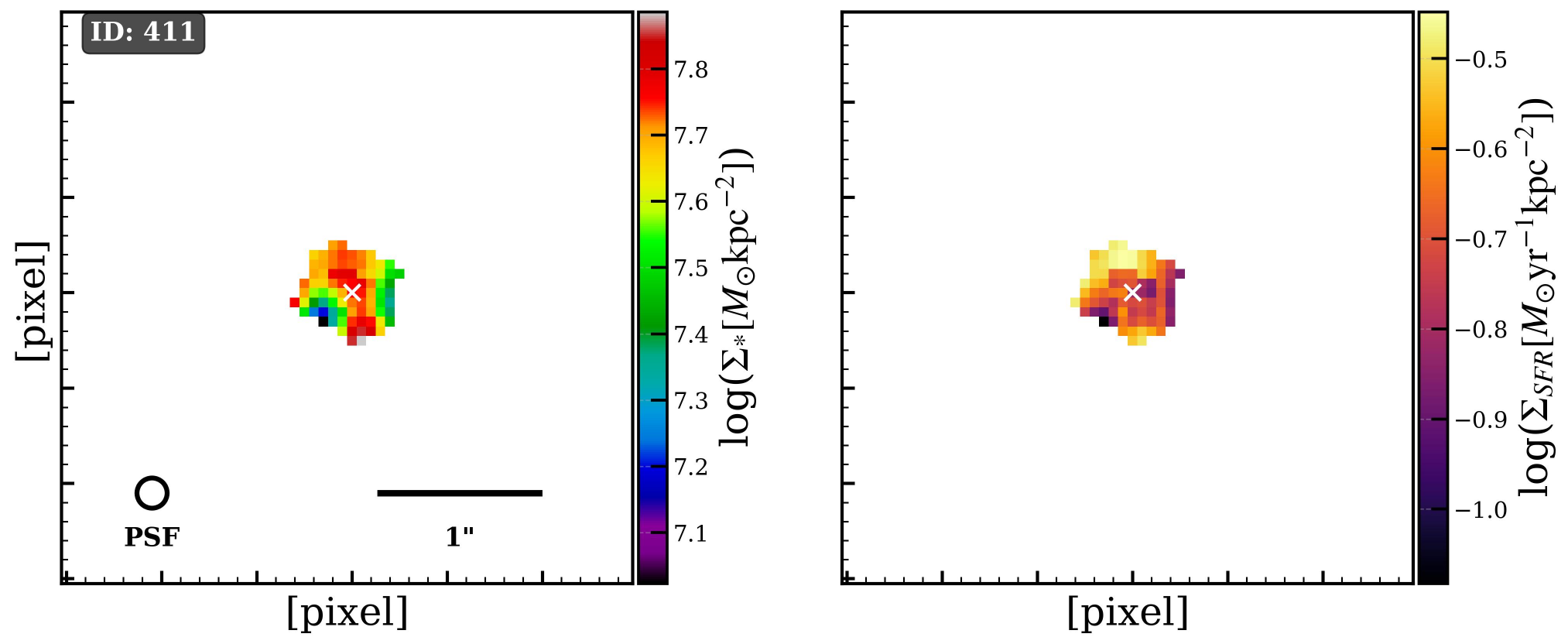}
\end{center}
\caption{$\Sigma_*$ (left panels) and $\Sigma_{\rm SFR}$ (right panels) maps for star-forming galaxies. Each galaxy panel shows a 40×40 pixel region (±20 pixels from center) centered on the galaxy. Galaxy IDs correspond to those from \citet{naufalRevealingQuiescentGalaxy2024}.}
\label{fig:sf_galaxy_maps}
\end{figure*}

% Figure 8 - Part 2 (continues on next page)
\begin{figure*}
\figurenum{13}
\centering
% Row 1: Galaxies 418 and 429
\includegraphics[width=0.48\textwidth]{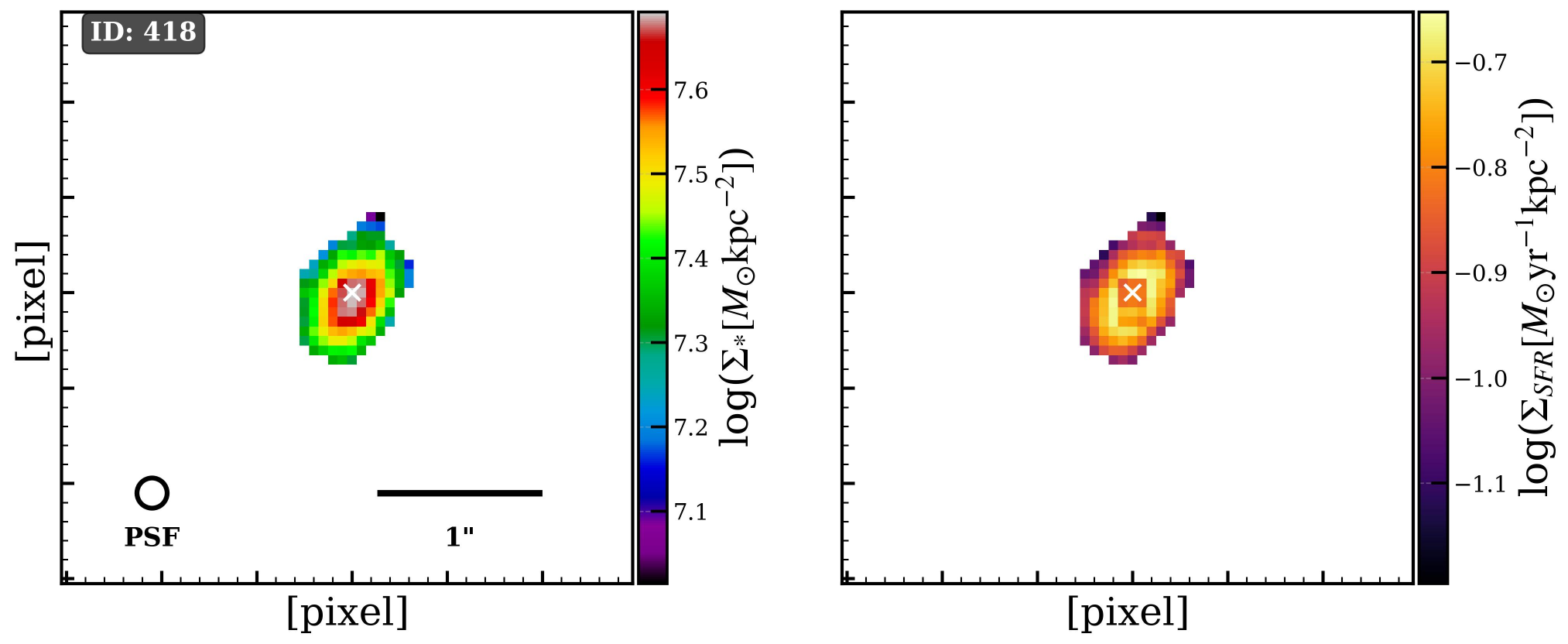}
\hfill
\includegraphics[width=0.48\textwidth]{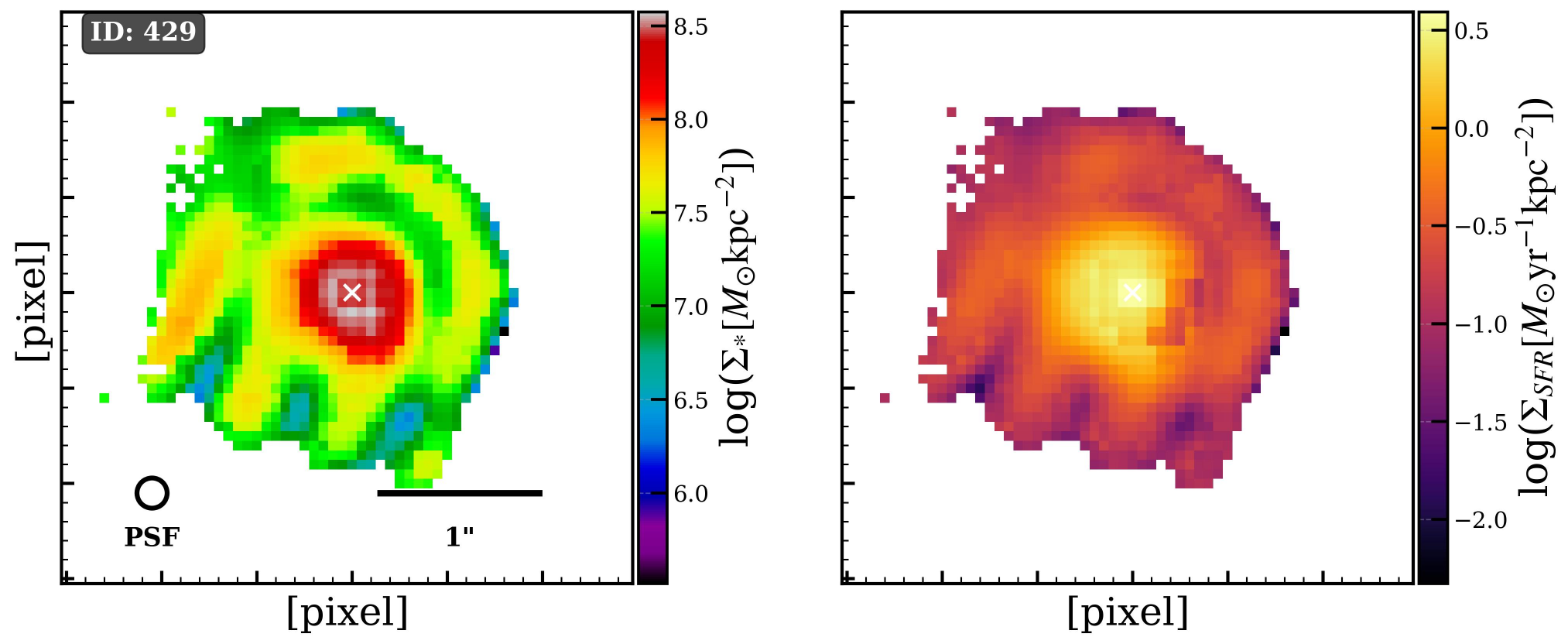}
\vspace{0.3cm}
% Row 2: Galaxies 432 and 459
\includegraphics[width=0.48\textwidth]{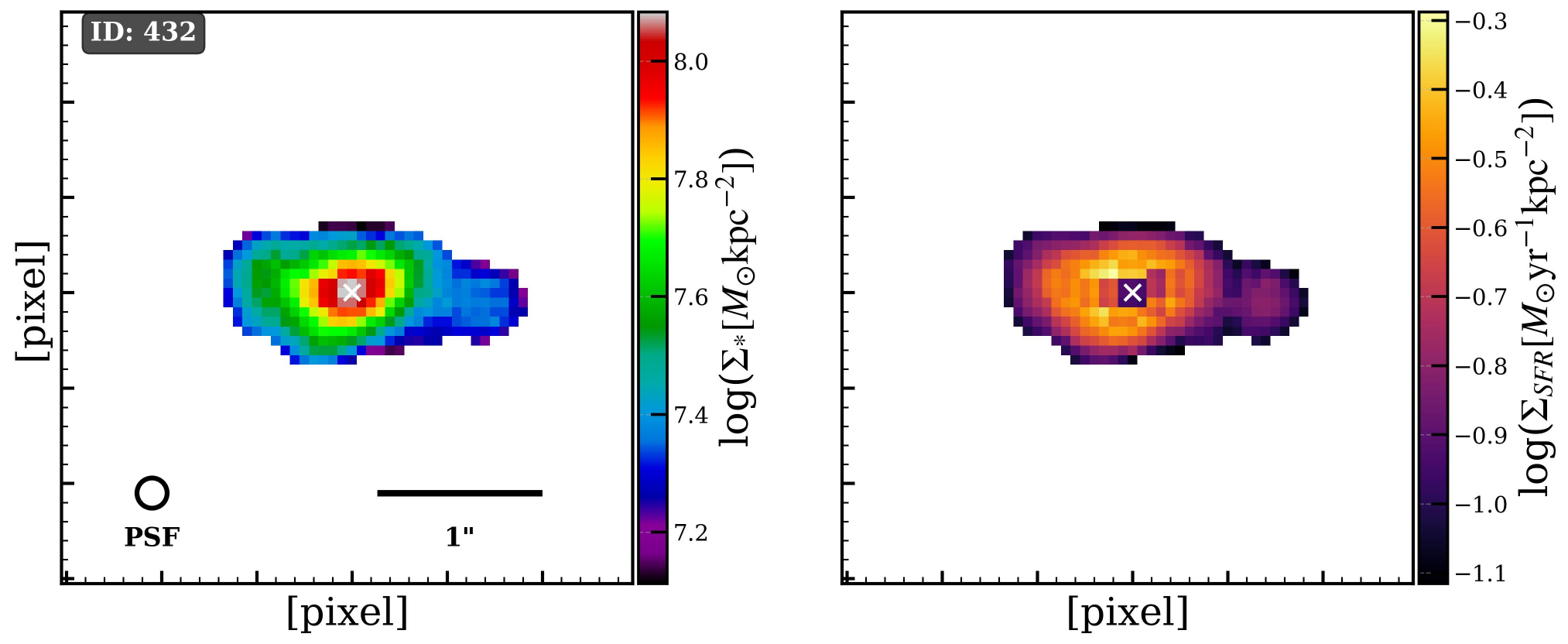}
\hfill
\includegraphics[width=0.48\textwidth]{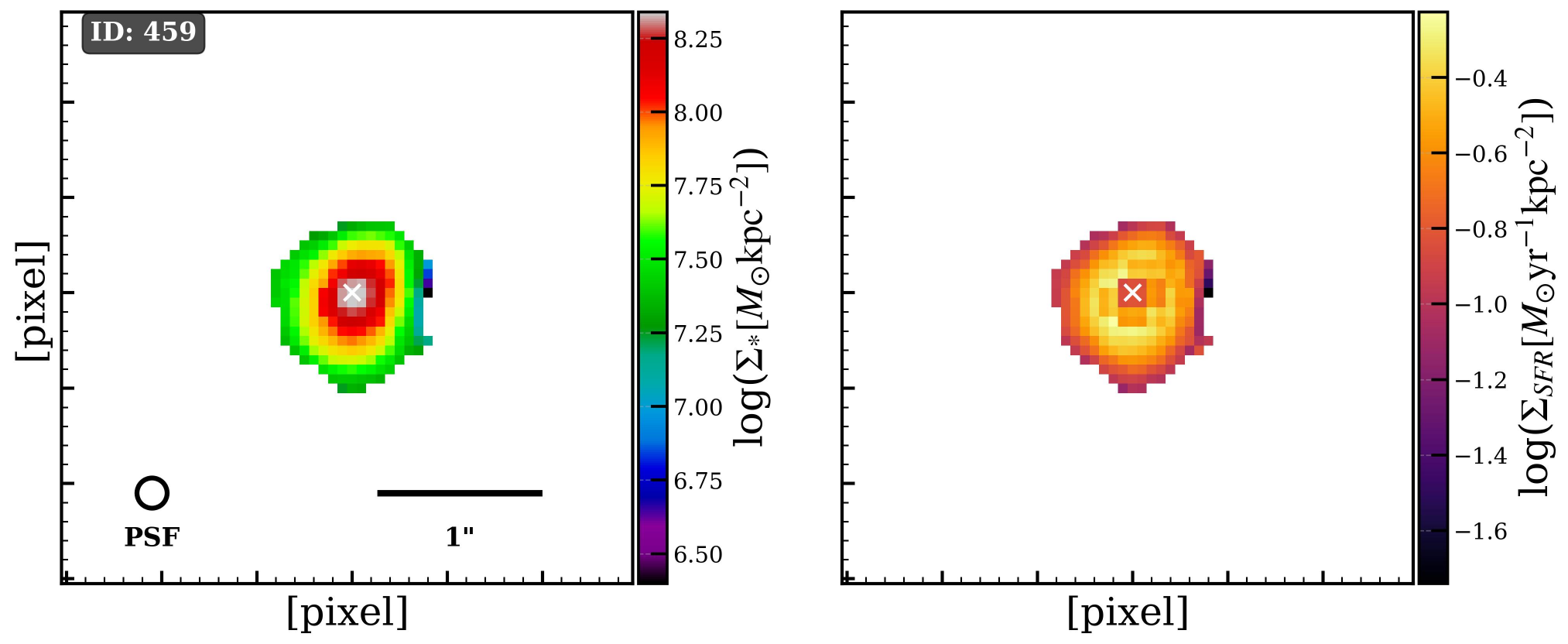}
\vspace{0.3cm}
% Row 3: Galaxies 459 and 461
\includegraphics[width=0.48\textwidth]{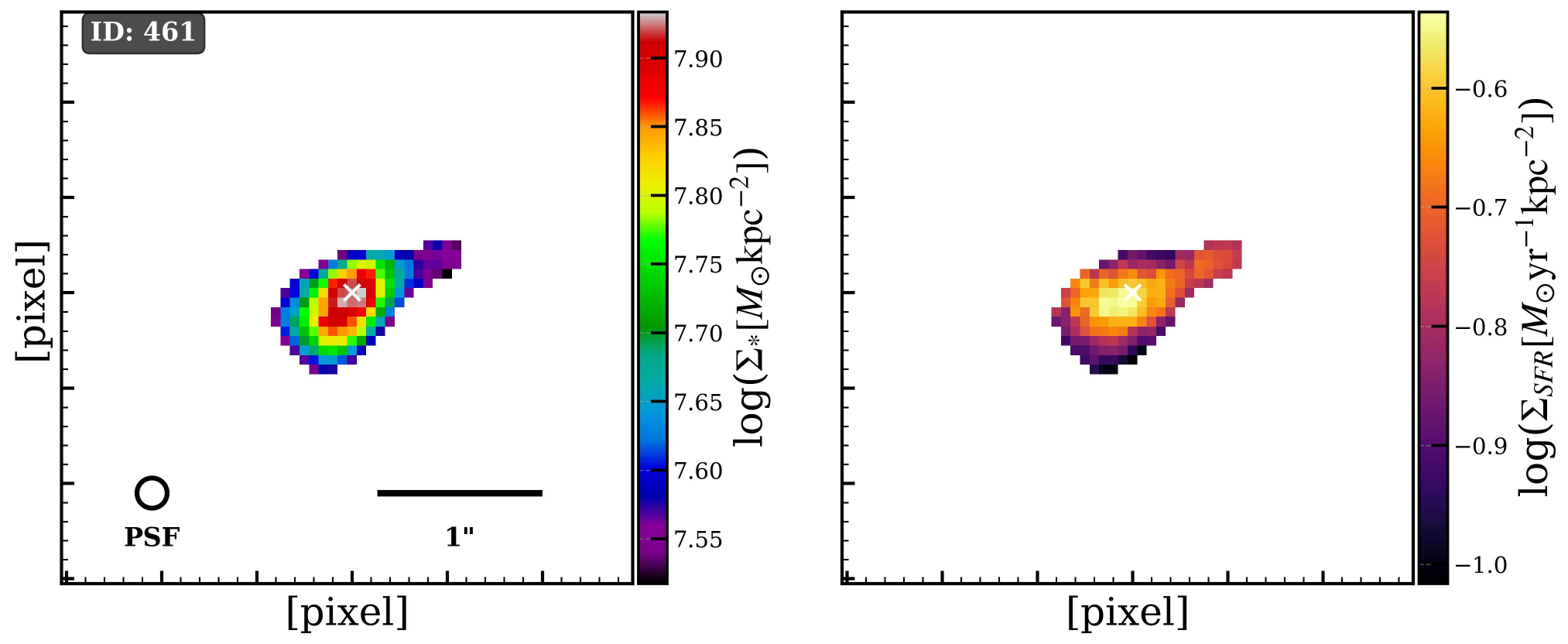}
\hfill
\includegraphics[width=0.48\textwidth]{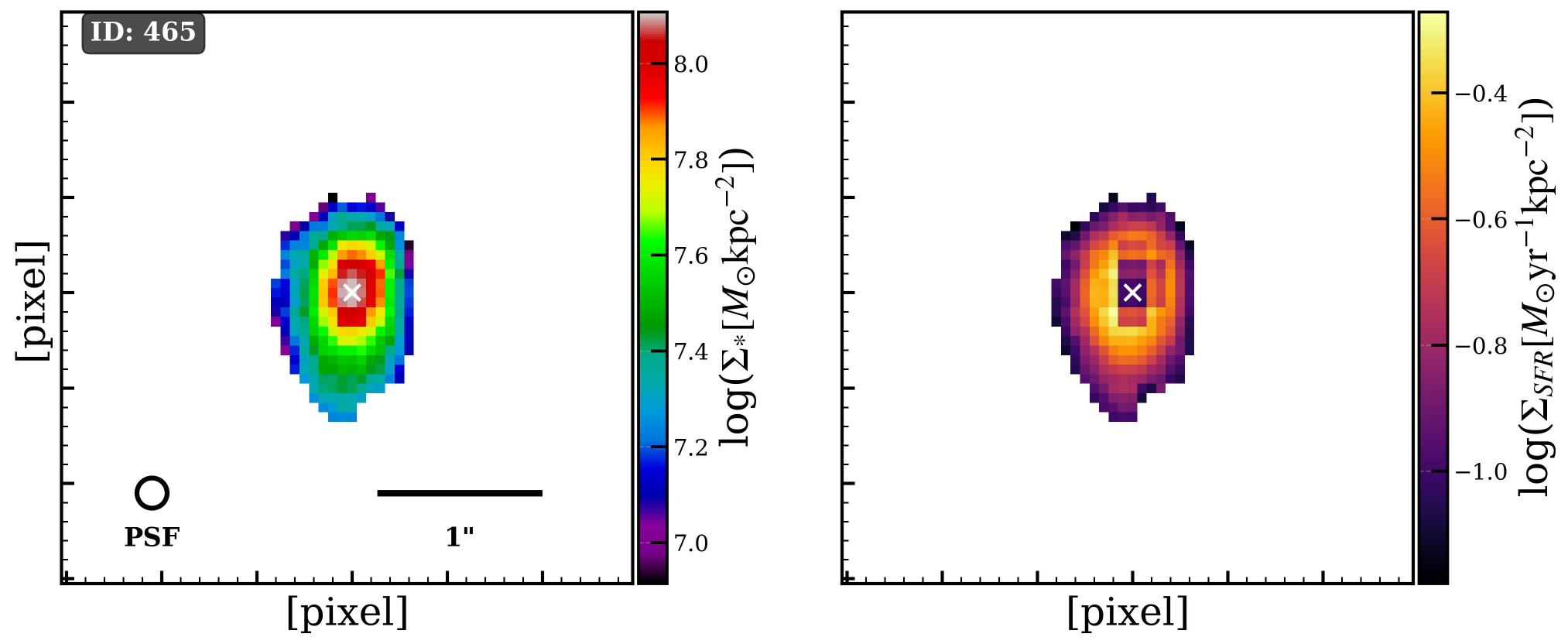}
\vspace{0.3cm}
% Row 4: Galaxies 465 and 479
\includegraphics[width=0.48\textwidth]{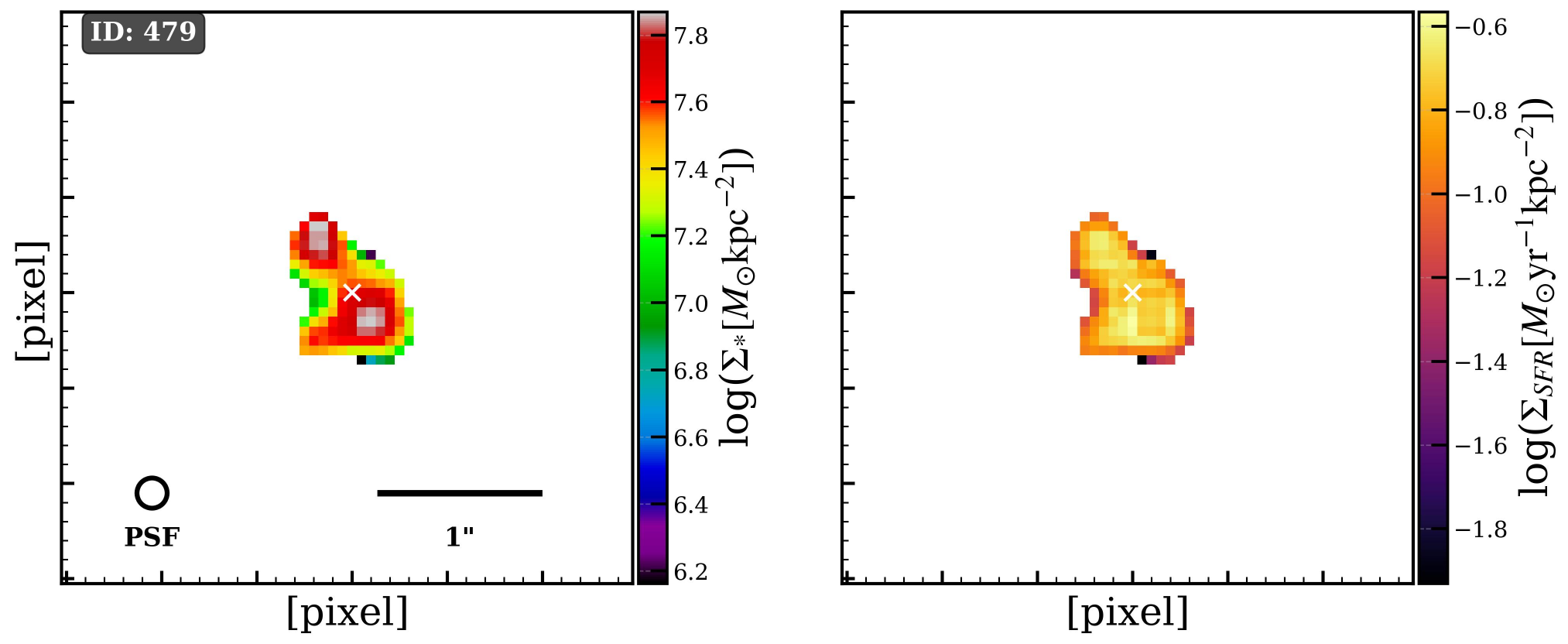}
\hfill
\includegraphics[width=0.48\textwidth]{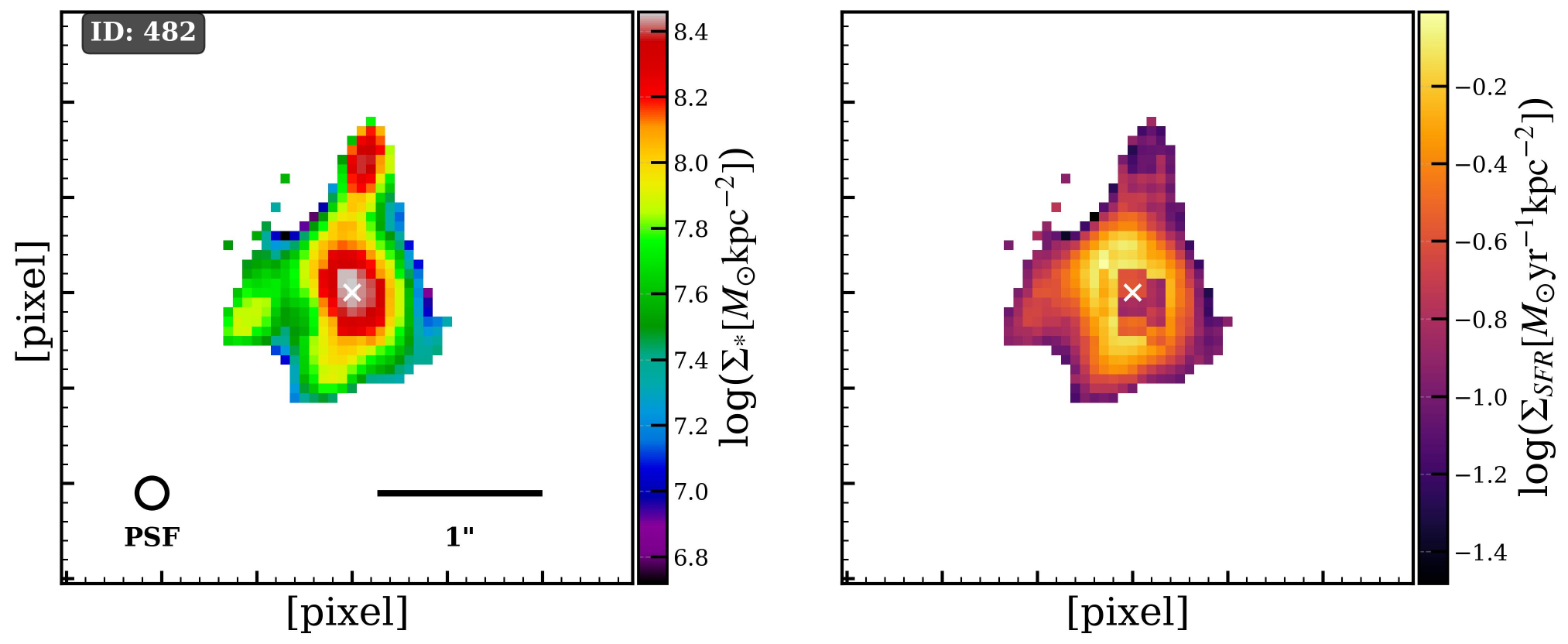}
\vspace{0.3cm}
% Row 5: Galaxy 482 (single, centered)
\includegraphics[width=0.48\textwidth]{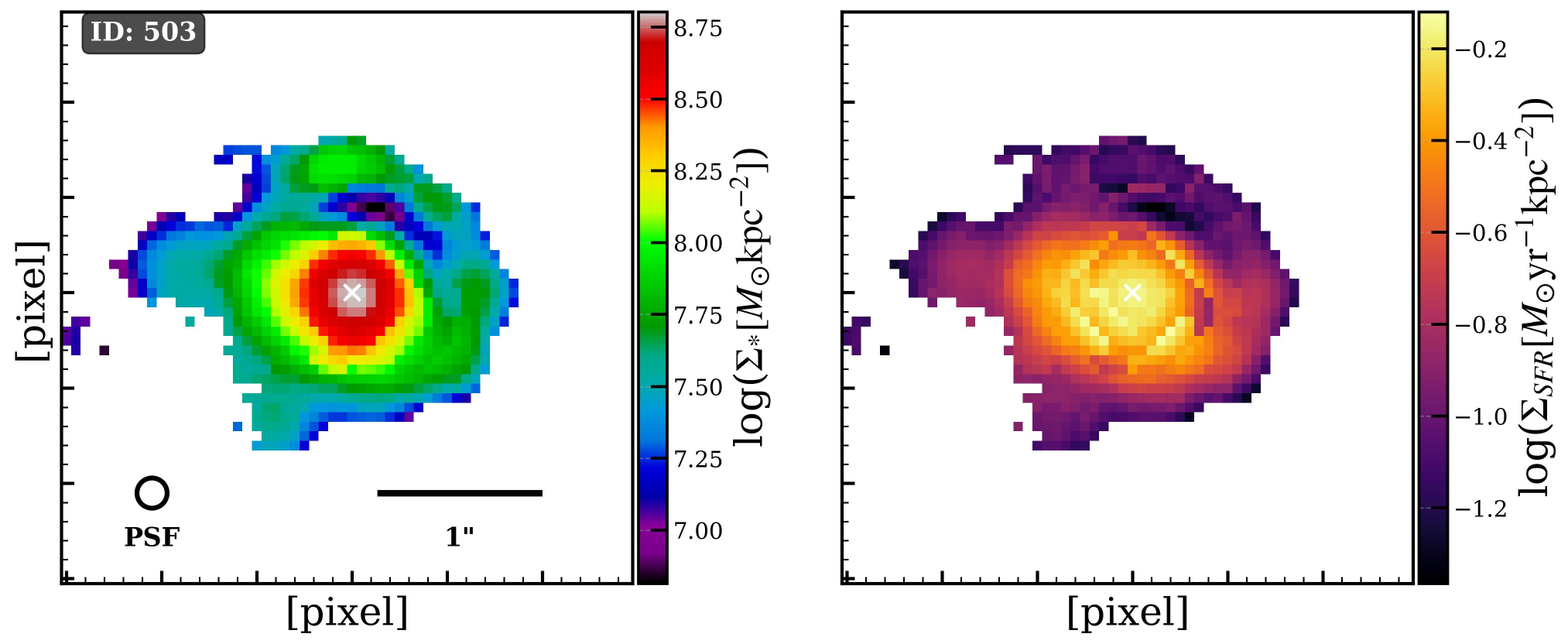}
\hfill
\includegraphics[width=0.48\textwidth]{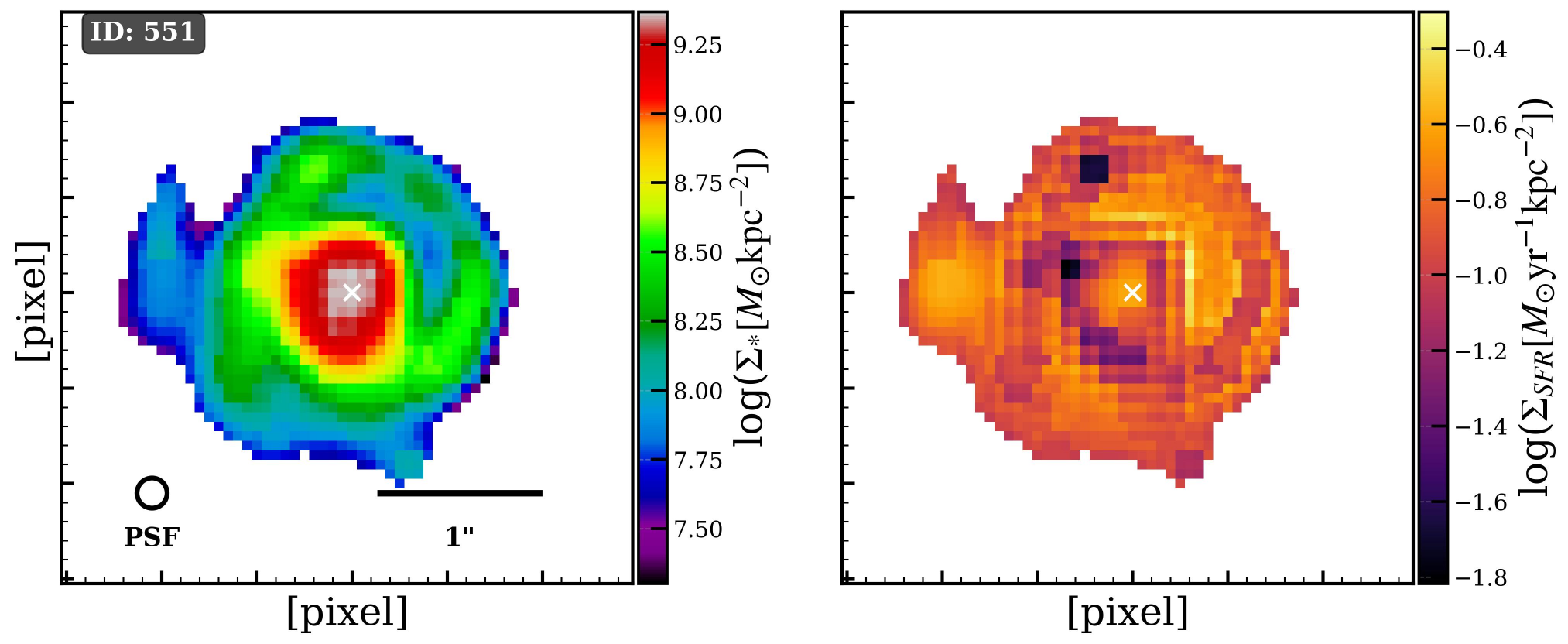}
\caption{continued.}
\end{figure*}

% Figure 8 - Part 3 (continues on next page)
\begin{figure*}
\figurenum{13}
\centering
% Row 1: Galaxies 551 and 558
\includegraphics[width=0.48\textwidth]{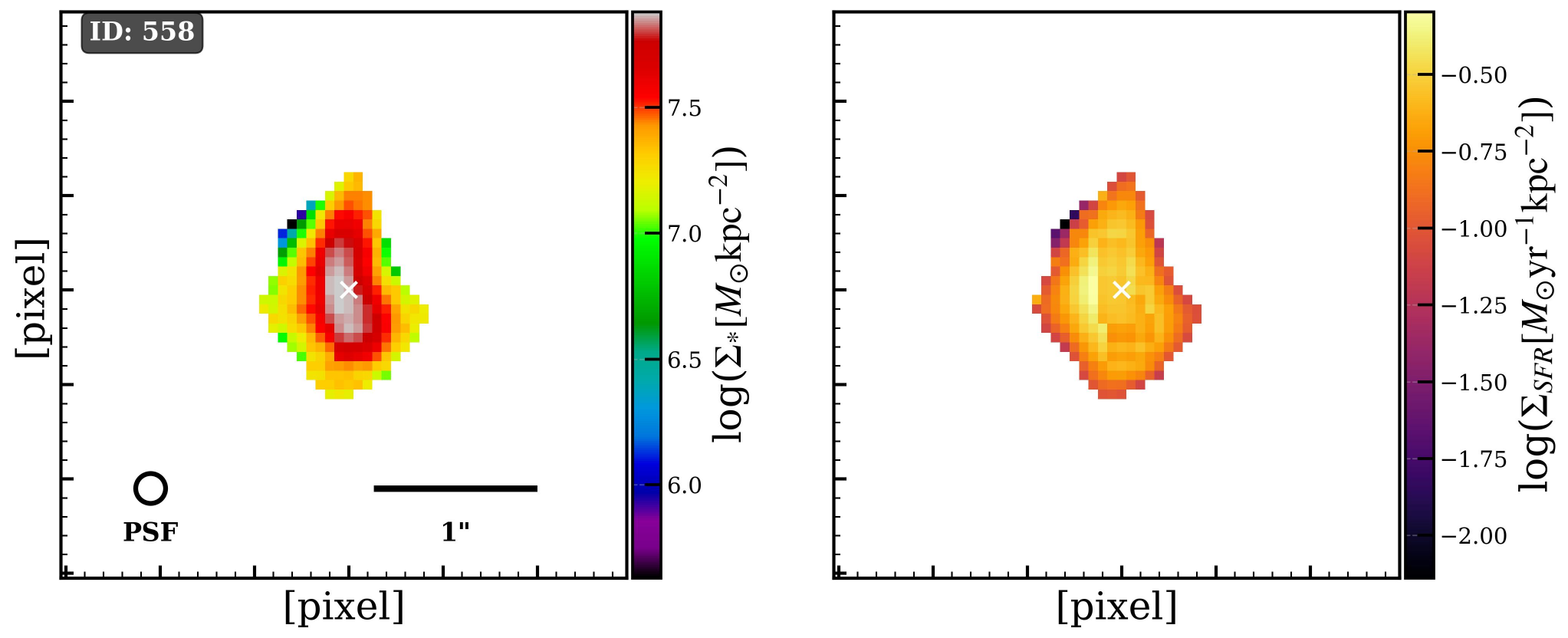}
\hfill
\includegraphics[width=0.48\textwidth]{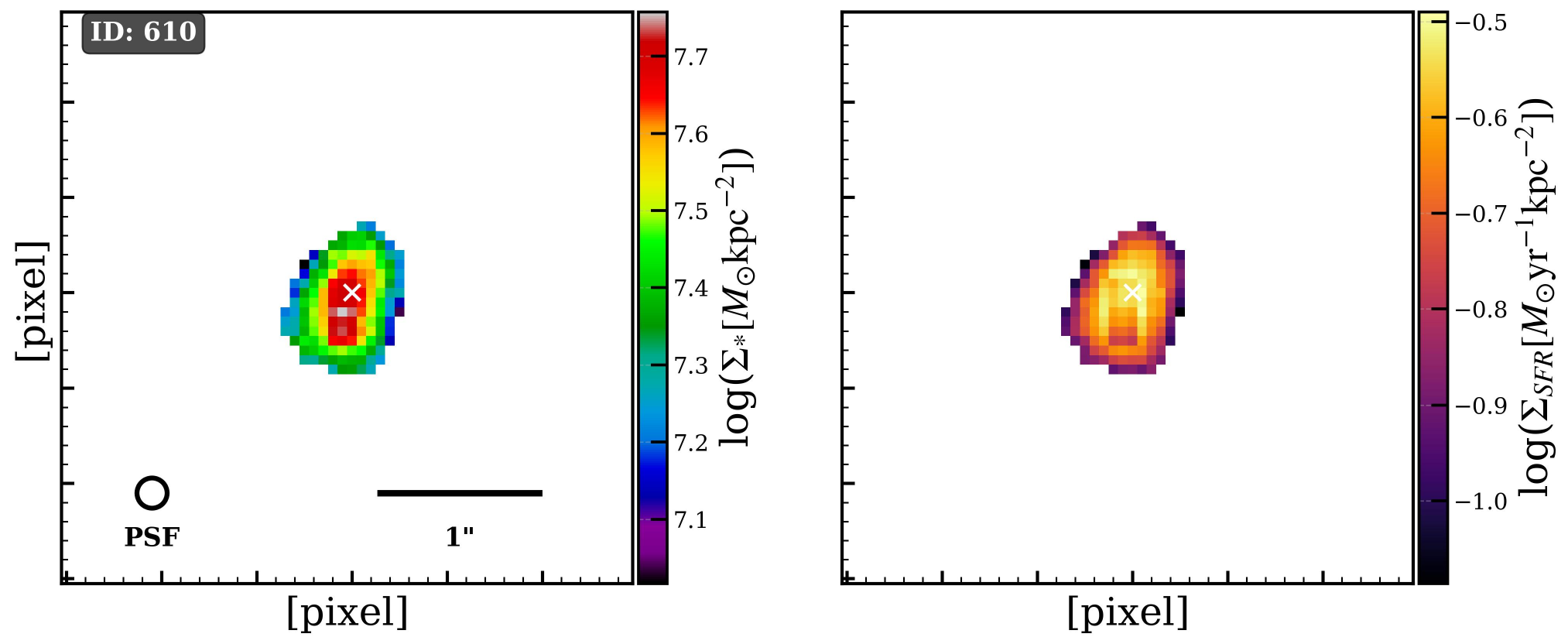}
\vspace{0.3cm}
% Row 2: Galaxies 610 and 624
\includegraphics[width=0.48\textwidth]{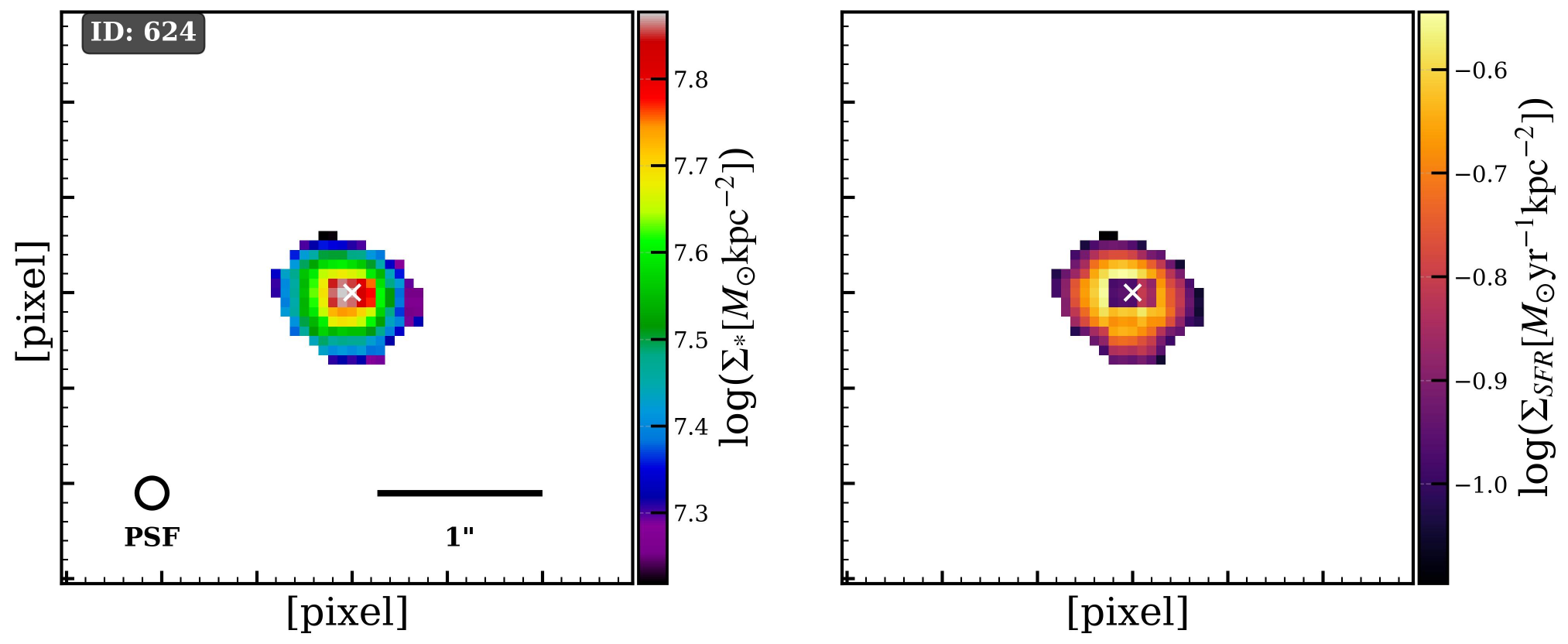}
\hfill
\includegraphics[width=0.48\textwidth]{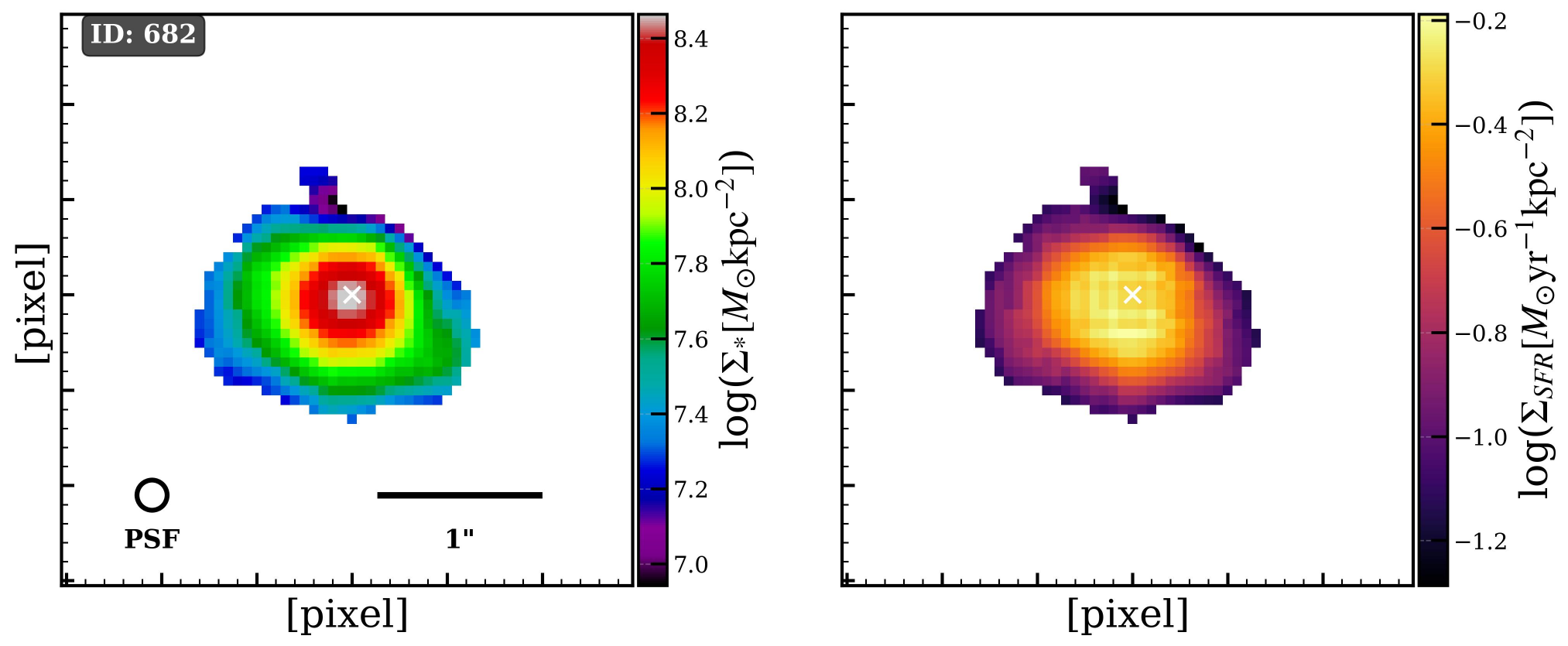}
\vspace{0.3cm}
% Row 3: Galaxies 682 and 686
\includegraphics[width=0.48\textwidth]{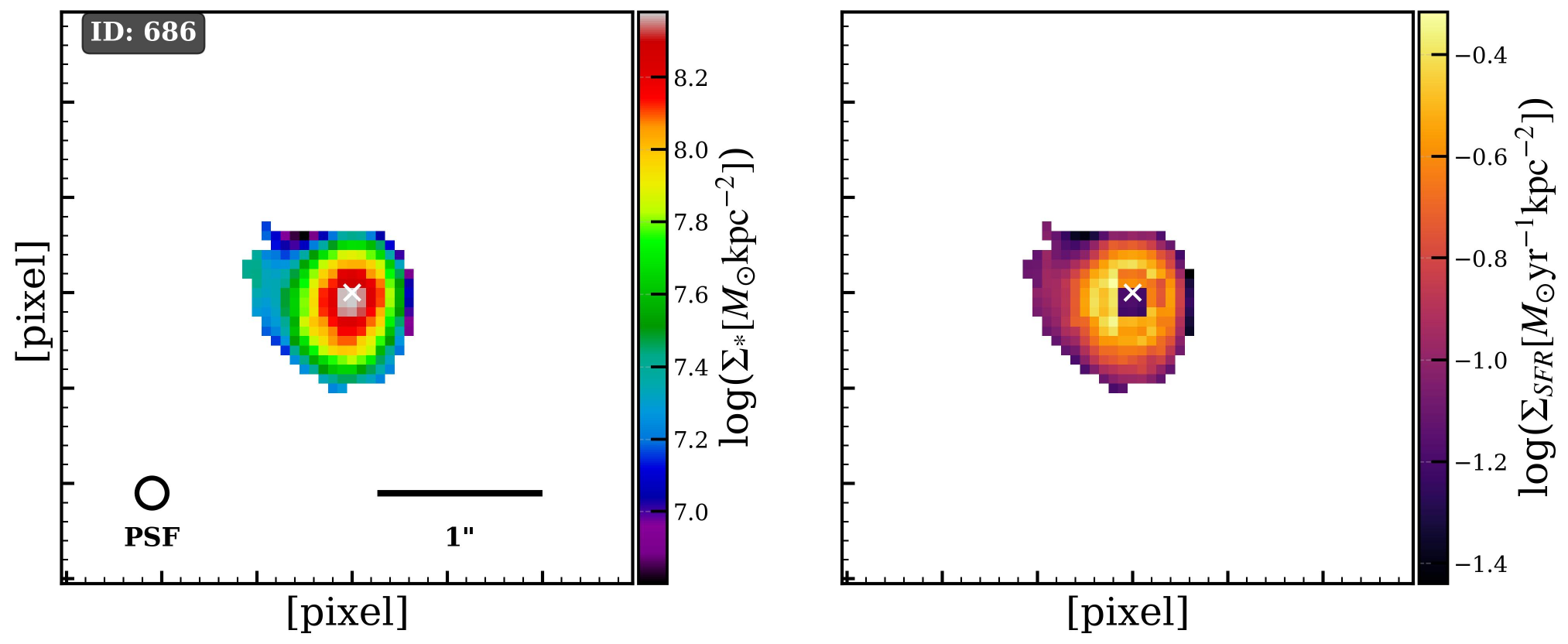}
\hfill
\includegraphics[width=0.48\textwidth]{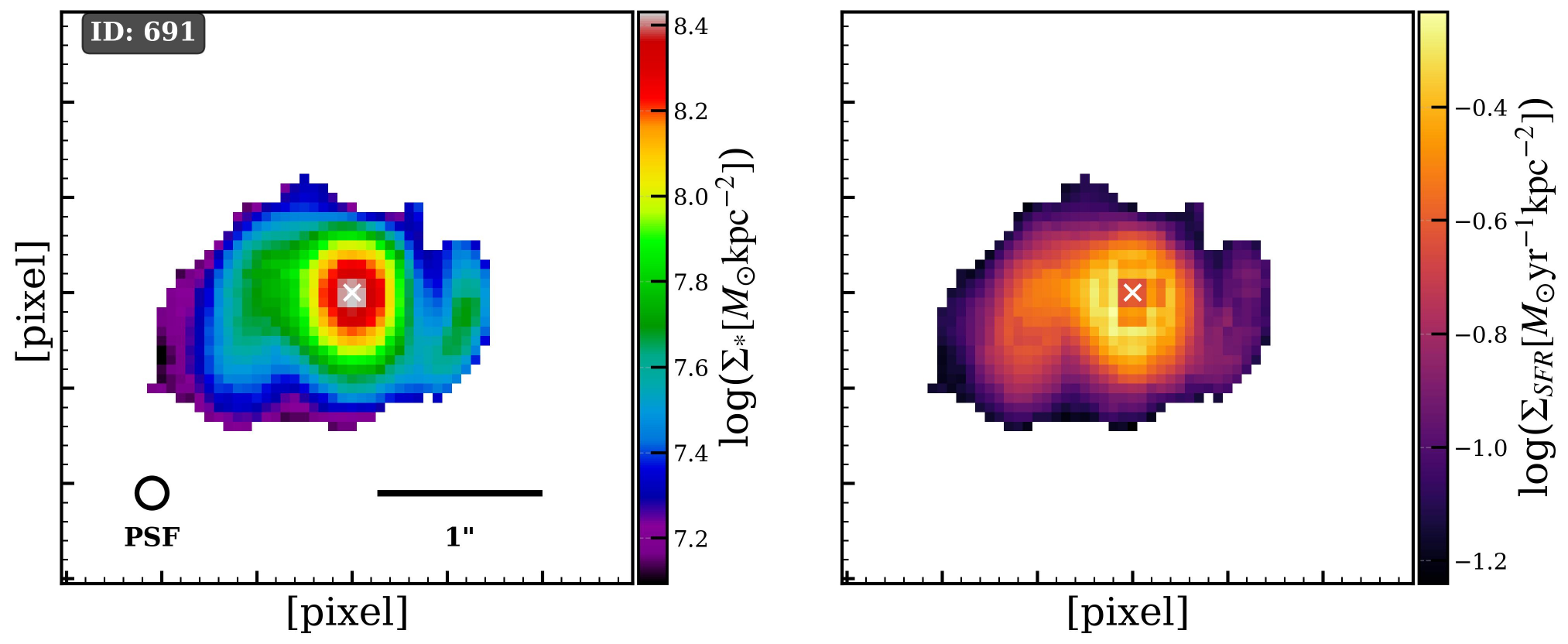}
\vspace{0.3cm}
% Row 5: Galaxies 770
\begin{center}
\includegraphics[width=0.48\textwidth]{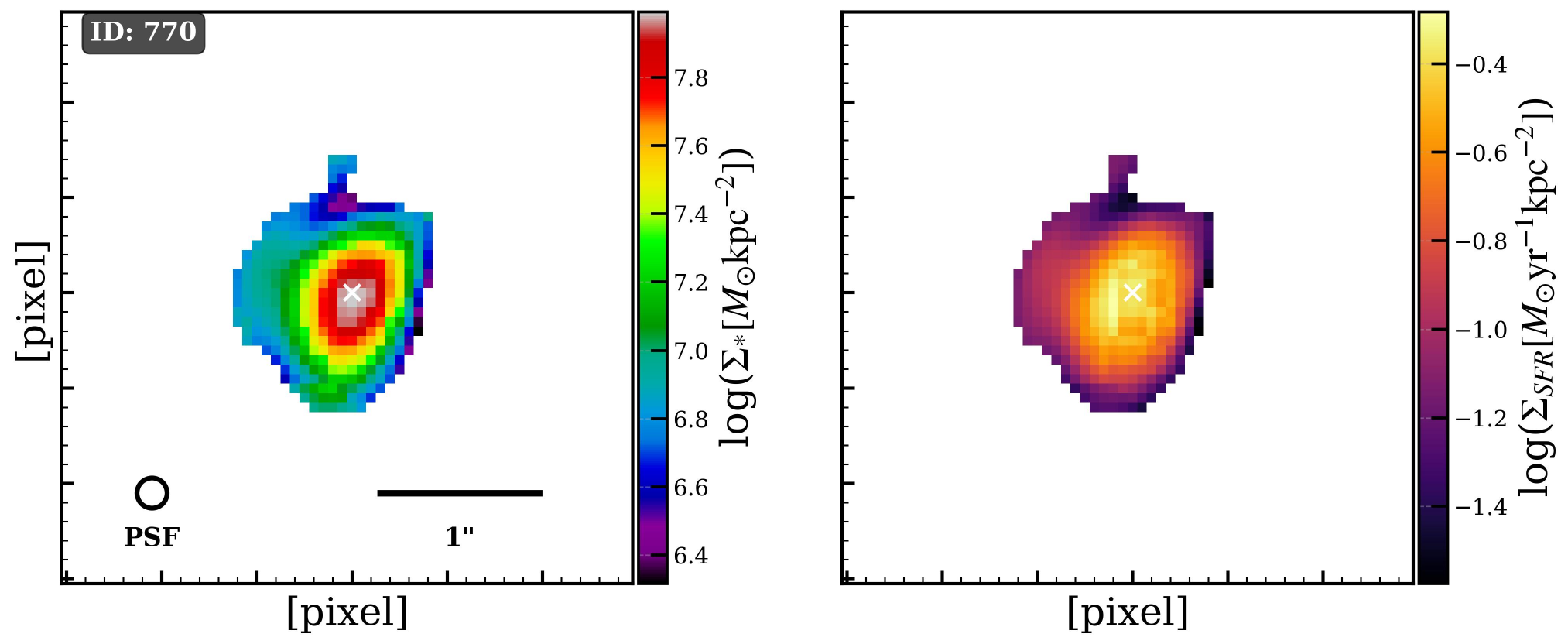}
\end{center}
\caption{continued.}
\end{figure*}

\bibliography{sample631}{}
\bibliographystyle{aasjournal}

\end{document}